\documentclass[usenatbib,useAMS]{mnras}
\usepackage{graphicx}
\usepackage{ctable}
\usepackage{url}
\usepackage{times,rotating,float}
\usepackage{xspace,subfig}

\newcommand{\msun}{~\mathrm{M_{\odot}}}

\newcommand{\sunrise}{\textsc{sunrise}\xspace}

\newcommand{\gadgetthree}{\textsc{gadget-3}\xspace}

\newcommand{\magphys}{\textsc{magphys}\xspace}

\newcommand{\acknowledgments}{\begin{small}\section*{Acknowledgments}\end{small}}

\newcommand{\gtsim}{\mbox{{\raisebox{-0.4ex}{$\stackrel{>}{{\scriptstyle\sim}}$}}}}
\newcommand{\ltsim}{\mbox{{\raisebox{-0.4ex}{$\stackrel{<}{{\scriptstyle\sim}}$}}}}

\newcommand\sref[1]{\hyperref[#1]{Section~\ref*{#1}}}
\newcommand\fref[1]{\hyperref[#1]{Fig.~\ref*{#1}}}

\title[SED modelling of spatially resolved galaxies]{
Panchromatic SED modelling of spatially-resolved galaxies 
}

\author[D.~J.~B. Smith \&\ C.~C. Hayward]{
\parbox[t]{\textwidth}{
Daniel J.~B.~Smith$^1$\thanks{E-mail: d.j.b.smith@herts.ac.uk}
and 
Christopher C. Hayward$^{2,3}$
}
\vspace*{6pt} \\
$^1$Centre for Astrophysics, Science \& Technology Research Institute, University of Hertfordshire, Hatfield, Herts, AL10 9AB, UK\\
$^2$Center for Computational Astrophysics, Flatiron Institute, 162 Fifth Avenue, New York, NY 10010, USA\\
$^3$Harvard--Smithsonian Center for Astrophysics, 60 Garden Street, Cambridge, MA 02138, USA
}

\begin{document}

\date{\today; MNRAS {\it in press}}

\pagerange{\pageref{firstpage}--\pageref{lastpage}} \pubyear{2018}

\maketitle

\label{firstpage}

\begin{abstract}
We test the efficacy of the energy-balance spectral energy distribution (SED) fitting code \magphys\ for recovering the spatially-resolved properties of a simulated isolated disc galaxy, for which it was not designed. We perform 226,950 \magphys\ SED fits to regions between 0.2\,kpc and 25\,kpc in size across the galaxy's disc, viewed from three different sight-lines, to probe how well \magphys\ can recover key galaxy properties based on 21 bands of UV--far-infrared model photometry. \magphys\ yields statistically acceptable fits to $> 99$ per cent of the pixels within the $r$-band effective radius and between 59 and 77 per cent of pixels within 20\,kpc of the nucleus. \magphys\ is able to recover the distribution of stellar mass, star formation rate (SFR), specific SFR, dust luminosity, dust mass, and $V$-band attenuation reasonably well, especially when the pixel size is $\ga 1$ kpc, whereas non-standard outputs (stellar metallicity and mass-weighted age) are recovered less well. Accurate  recovery is more challenging in the smallest sub-regions of the disc (pixel scale $\la 1$ kpc), where the energy balance criterion becomes increasingly incorrect. Estimating integrated galaxy properties by summing the recovered pixel values, the true integrated values of all parameters considered except metallicity and age are well recovered at all spatial resolutions, ranging from 0.2\,kpc to integrating across the disc, albeit with some evidence for resolution-dependent biases. These results must be considered when attempting to analyse the structure of real galaxies with actual observational data, for which the `ground truth' is unknown. 
\end{abstract}


\section{Introduction} \label{S:intro}

Contemporary observational astrophysics is extremely data rich; we have never had a larger range of survey data, covering wavelengths ranging from the ultraviolet to the far-infrared and sub-millimetre. Whatever the wavelength, the data have never been more sensitive, or sharper, whether obtained from the ground \citep[e.g.][]{hartley13,jarvis13,tanaka17}, using interferometers (such as LOFAR: \citealt{hardcastle16}, or ALMA: \citealt{dunlop17}) or from space (e.g. with {\it Herschel}: \citealt{kennicutt11}, or the {\it Hubble Space Telescope}: \citealt{illingworth13}). This means that we are becoming able to go beyond modelling the integrated multi-wavelength emission from galaxies, and beyond modelling individual components \citep[performing bulge\slash disc decomposition, e.g.][]{vika14,kennedy16,johnston17}, to modelling individual pixels at all wavelengths simultaneously, building up unprecedented understanding of the physics of galaxy formation and evolution.
One of the many benefits of this is that the widely-used panchromatic modelling techniques such as \magphys\ \citep{dacunha08} and {\sc Cigale} \citep{burgarella05,noll09} use energetic arguments to build a consistent model of all these data simultaneously, bringing the starlight that dominates the luminosity at optical\slash near-infrared wavelengths on to an even footing with the dust emission that dominates at mid- and far-infrared wavelengths. In the real Universe this is complicated by the potentially very different resolutions afforded by observations at different wavelengths (e.g. optical versus far-infrared data), though techniques exist to deconvolve the data and perhaps prevent observational effects from biasing the results \citep[see e.g.][for examples]{viero13,merlin15,hurley17,wright16}. 

The aforementioned \magphys\ code uses a physically motivated model to consistently model the multi-wavelength properties of galaxies from the rest-frame ultra-violet to the far-infrared. By balancing the energy absorbed by dust at optical\slash near-infrared wavelengths with the far infrared luminosity, consistently modelling the emission from the stars and dust simultaneously can give a greater degree of understanding than considering either wavelength range alone \citep[e.g.][]{smith12}, since around half of the photons ever emitted by stars and AGN have been absorbed and re-radiated by dust \citep[e.g.][]{dole06}. Only by studying all of the available data simultaneously can we hope to build up a full picture of the mass growth through star formation in galaxies, their dust mass evolution, and the role of environment for example. As a result, \magphys\ has been widely used throughout the literature \citep[e.g.][]{dacunha10,smith12,berta13,lanz13,rowlands14a,rowlands14b,viaene14,HS15,SH15,dariush16,viaene16,miettinen17,gurkan18}, and the same is true of {\sc Cigale} \citep[e.g.][]{giovannoli11,johnston15,wylezalek16,pearson17}. 
One key advantage of \magphys\ over other SED fitting codes is that the required SED libraries are pre-defined and publicly available, which means that results produced by different groups are directly comparable since they use the same priors, stellar population library, star formation histories, and dust law. This ensures that the results obtained by any group are repeatable given the same input photometry. 

Particular among the \magphys\ investigations, \citet{HS15} performed an idealised test of the extent to which \magphys\ is able to recover reliable properties of galaxies by analysing a suite of smoothed-particle-hydrodynamical (SPH) simulations on which dust radiative transfer was performed in post-processing to compute UV--mm spectral energy distributions (SEDs) of the simulated galaxies. The advantages of testing \magphys\ in this way are wide-ranging: since the true properties (e.g. stellar mass, star formation rate, etc) of the galaxies being modelled are known, and since we can eliminate uncertainties due to e.g. the uncertainty in the attenuation law of high redshift galaxies \citep[e.g.][]{reddy16} by making the same assumptions in the radiative transfer as in \magphys. \citet{HS15} used simulations of an isolated disc galaxy and major galaxy merger to demonstrate that \magphys\ can reliably recover properties of galaxies virtually independent of viewing angle and evolutionary stage, even in cases in which merger-induced AGN activity contributes as much as 25 per cent of the system's total bolometric luminosity.

Building on this initial study, \citet{SH15} modified \magphys\ to provide statistical estimates of galaxy star formation histories, by marginalising over the default input libraries, and used similar simulations from \citet{lanz13} to test how well the known simulated star formation histories could be recovered. \citet{SH15} showed that \magphys\ is able to derive realistic marginalised estimates of the star formation histories of isolated disc simulations, but found that the best-fit SFHs were unreliable, varying with viewing angle, and showing spurious bursts that were not present in the actual SFH. \citet{SH15} also found that \magphys\ was unable to recover realistic SFHs for major mergers, attributing this to the parametrised SFHs (exponentially declining, with bursts superposed) in the input libraries.

At the same time, there is a growing amount of work in the literature based on spatially resolved galaxy modelling and SED fitting, and the natural sources to begin with are bright, local galaxies which cover a large sky area. For example, the spatially-resolved SED of the nearby galaxy M51 was studied by \citet{mentuch12}  without using an energy balance model. More recently, \citet{viaene14} used a version of the \magphys\ code (with a non-standard library including a slightly broadened range of dust temperatures) to model the SED of the nearby Andromeda galaxy \citep[M31, originally studied using \magphys\ by][to investigate the source of dust heating]{groves12}. This was done both for the galaxy as a whole and by decomposing the photometry into individual structural components (including the bulge, the ring, and the inner disc) as well as into individual resolution elements as small as $140\times 610$\,pc. \citet{viaene14} produced resolved maps of various parameters produced by the model and compared the parameters derived from the global photometry and with those obtained via the pixel-by-pixel fitting. Though the values derived in each case were comparable, a robust and direct comparison between the pixel-by-pixel values and the integrated properties is complicated, since for a variety of practical observational reasons, the area used to sum over the pixel-by-pixel values differs for each parameter (individual resolution elements with large parameter uncertainties were disregarded), and are therefore always smaller than the area used for the global photometry. 

The recent work by \citet{sorba15} highlighted the alarming prospect of a resolution-dependent systematic bias in stellar mass estimates for a sample of 67 nearby galaxies using just six bands of photometry from GALEX and SDSS (i.e. in the absence of far-infrared data), in the sense that integrated stellar masses were underestimated relative to the sum of the masses of the individual components.  \citet{sorba15} attributed this discrepancy to an ``outshining" bias, in which the youngest stars dominate the integrated emergent spectrum, and loss of information about the star formation history, and therefore mass-to-light ratio, in which the numerator is dominated by the less luminous old stellar population. This result is highly alarming, since if it is confirmed, it implies that the vast majority of known galaxies (i.e. those for which we are unable to decompose the photometry into individual resolution elements) have had their stellar mass content underestimated. Other works have found tentative evidence for similar resolution-dependent bias in dust masses derived for the Large Magellanic Cloud \citep{galliano11}, and for NGC 628 and NGC 6946 \citep{aniano12}. Elsewhere, \citet{amblard17} attempted a spatially-resolved SED decomposition of two local early type galaxies (IC1459 and NGC2768) on sub-kpc scales from rest-frame UV to the far-infrared (including wavelengths up to 160\,$\mu$m). Most recently, \citet{jung17} studied the difference in stellar mass, SFR, UV dust obscuration, and mass-weighted age recovered whether the integrated or resolved UV to near-IR photometry is used for parameter estimation (using a Monte-Carlo Markov Chain sampler based on CB07 models), in order to investigate the radial dependence of the specific star formation rate in high-redshift galaxies. The recent work by \citet{cibinel15} suggested that identifying merging galaxies at high-$z$ according to their SED-fitting derived stellar mass estimates is more robust than using single-band photometry, and there are many other works that have attempted to decompose the SEDs of spatially-resolved galaxies  \citep[e.g.][]{baes10,delooze12,delooze14}.  In addition, recent studies of spatially resolved lensed galaxies \citep[e.g. with ALMA;][]{dye15,swinbank15} suggest that similar analyses may be possible out to the highest redshifts using forthcoming facilities.  

Age gradients in the stellar populations of galaxies from integral field spectroscopy can used to critically differentiate between different formation mechanisms of early type galaxies (e.g. whether gas stripping, stellar feedback or major mergers were responsible for truncating their star formation). Given the vastly greater observational expense of spectroscopy relative to photometry, it is clearly desirable to reliably estimate the same properties from panchromatic photometry alone, since photometry will always be available for more galaxies than can be observed using integral field spectroscopy (even with new, wide-field, high-throughput spectrographs such as MUSE, MaNGA and WEAVE; see \citealt{bacon10}, \citealt{bundy15} and \citealt{dalton12,dalton14}, respectively).

Simulated data are ideal for investigating how well \magphys\ can perform with spatially resolved data, since they present an idealised proving ground, free from nuisance observational effects (e.g. noisy photometry, different spatial resolution at different wavelengths, etc), and where the true distributions of e.g. star formation, stellar mass and dust mass are known. In this paper, we test the ability of \magphys\ to produce reliable estimates of galaxy properties (including stellar mass, SFR, sSFR, dust luminosity, stellar metallicity, $A_V$, $M_{\mathrm{dust}}$ and mass-weighted age) at a range of different spatial scales varying from the integrated photometry to pixels 0.2\,kpc on a side \citep[similar to the highest resolution sampled in the nearby galaxy M31 by][]{viaene14}. Although \citet{HS15} found that \magphys\ was able to infer global properties of simulated galaxies, such as the total SFR, from their integrated SEDs
reasonably well, it is by no means guaranteed that a similar level of success will be achieved when employing \magphys\ on a pixel-by-pixel basis. For example, the energy-balance criterion employed by \magphys\ breaks down if the dust emission in a given pixel is predominantly powered by emission from stars located in other pixels. This must occur at some arbitrarily high resolution, but whether this assumption holds on the scales relevant for pixel-by-pixel SED modelling of galaxies is an open question.

In this work, we extend the analysis of \citet{HS15} by performing pixel-by-pixel SED modelling of a simulated galaxy to test how well \magphys\ can recover the simulated galaxy's \emph{spatially resolved} physical properties. In a companion
paper (Hayward \& Smith {\it in prep}), we compare the \magphys\ results with those obtained using simpler tracers (e.g. H$\alpha$ luminosity-based SFR maps). The remainder of this paper is organised as follows.
In section \ref{S:methods} we describe the \magphys\ model, alongside the simple modifications we have made to the fitting code. Section \ref{S:results} contains our results, in which we compare the output produced by \magphys\ with the known true parameters of the simulated galaxy as a function of the spatial scales that we probe. We discuss these results in section \ref{S:discussion}, while section \ref{S:conclusions} contains some concluding remarks. Throughout we adopt a standard cosmology with $H_0 = 71$ km s$^{-1}$ Mpc$^{-1}$, $\Omega_M = 0.27$ and $\Omega_\Lambda = 0.73$.

\section{Methods} \label{S:methods}

\subsection{Synthetic images of a simulated galaxy}

We employ galaxy images generated by performing dust radiative transfer on a 3-D hydrodynamical simulation of an isolated disc galaxy.
The approach has been described in detail in many previous works \citep[e.g.][]{pj10,H11,H12,HS15}, so we will only summarise it
briefly here.

First, we generate initial conditions for an isolated disc galaxy following \citet{springel05}. The galaxy is composed of a dark matter halo,
exponential stellar and gas discs, a stellar bulge, and a central supermassive black hole (BH). We analyse a single model
galaxy, the `M3' model presented in \citet{Lanz14}, whose properties (e.g. gas fraction and disc scalelength given its stellar mass)
are designed to be representative of local galaxies (see \citealt{Cox08}). For this work, we employ eight times the number of
particles used in \citet{Lanz14} to provide better sampling of young stars. 
The simulated galaxy's virial, stellar disc, gas disc, stellar bulge, and BH masses are
$1.16 \times 10^{12}$, $4.2 \times 10^{10}$, $8 \times 10^9$, $9 \times 10^9$, and $10^5 \msun$, respectively. All other
properties are identical to those of the `G3' model of \citet{Cox08}. The number of particles
used to represent the dark matter halo, stellar disc, gas disc, and stellar bulge is $9.6 \times 10^5$, $4 \times 10^5$, $4 \times 10^5$
and $8 \times 10^4$, respectively. The gravitational softening lengths are 50 (200) pc for the gas and stellar (dark matter) particles.

Initial conditions in hand, we then evolve the simulated galaxy using a non-public version of the \gadgetthree smoothed particle
hydrodynamics (SPH) code \citep{Springel05gadget}.\footnote{Although we employ the traditional density-entropy formulation
of SPH, this is not a concern, because \citet{H14arepo} demonstrated that the results of idealised, non-cosmological simulations of isolated
galaxies and galaxy mergers performed using SPH are almost identical to those performed using the more-accurate moving-mesh
hydrodynanics code {\sc arepo} \citep{springel10}.} The code directly accounts for the effects of gravity, hydrodynamics, and radiative
heating and cooling. We account for the unresolved structure of the ISM, including the effects
of supernova feedback, using the model of \citet{sh03}. Gas particles with a density greater than a threshold of $n \sim 0.1$ cm$^{-3}$
are assumed to form stars at a rate set by the volume-density-dependent form of the Schmidt--Kennicutt relation \citep{schmidt59,kennicutt98}.
BH accretion and feedback is included following \citet{springel05} (modified Eddington-limited Bondi-Hoyle accretion and thermal
`quasar-mode' feedback), but this is unimportant for the simulation studied here. The initial stellar ages (which matter only for the radiative
transfer calculations; see below) and stellar and gas metallicities are assigned following \citet{Lanz14}. Aging of the stellar population and
enrichment of the ISM is followed self-consistently as the system evolves.

\begin{figure*}
\centering 
\fbox{\subfloat[Camera 0]{\includegraphics[width=0.27\columnwidth]{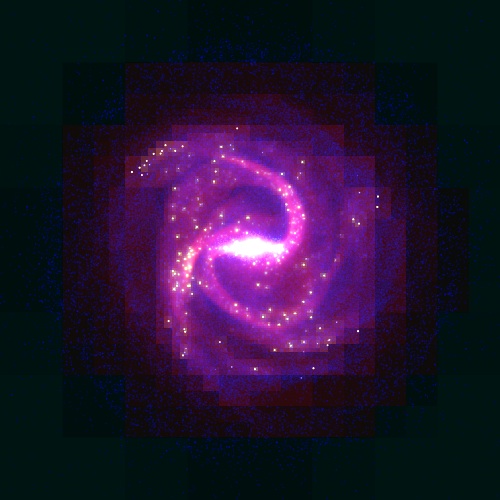}}}
\subfloat[Camera 1]{\includegraphics[width=0.27\columnwidth]{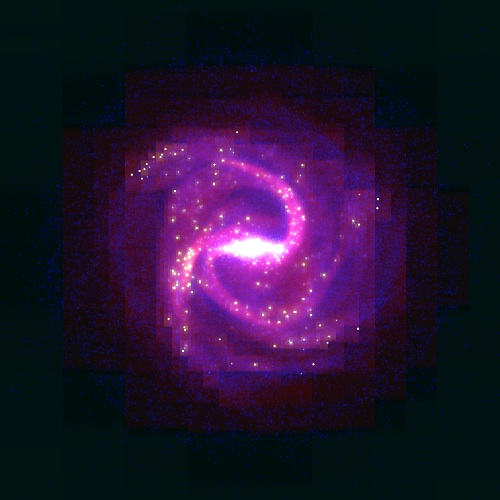}}
\hspace{0.02cm}
\subfloat[Camera 2]{\includegraphics[width=0.27\columnwidth]{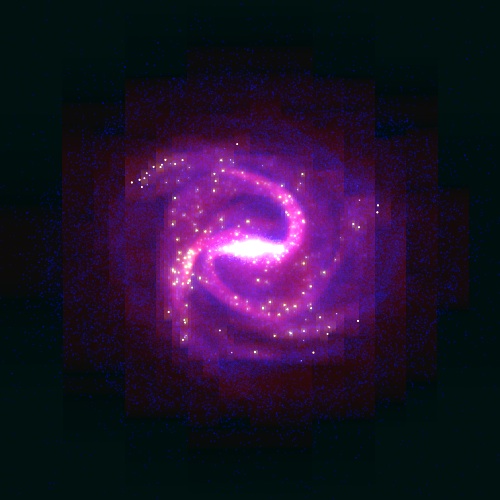}}
\hspace{0.02cm}
\fbox{\subfloat[Camera 3]{\includegraphics[width=0.27\columnwidth]{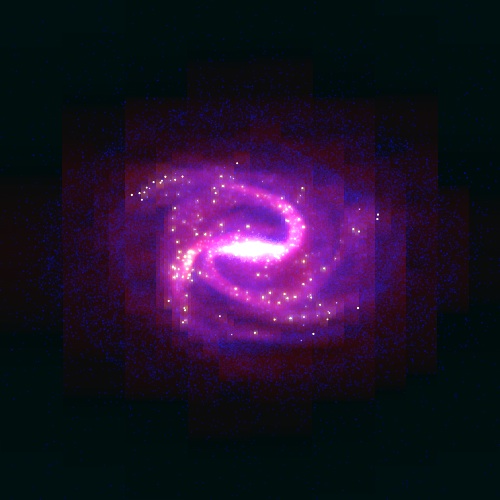}}}
\subfloat[Camera 4]{\includegraphics[width=0.27\columnwidth]{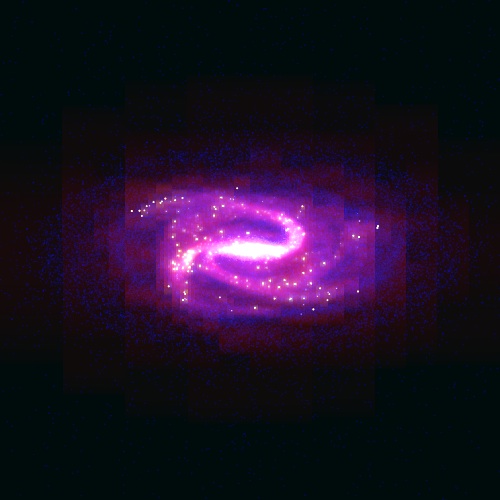}}
\hspace{0.02cm}
\fbox{\subfloat[Camera 5]{\includegraphics[width=0.27\columnwidth]{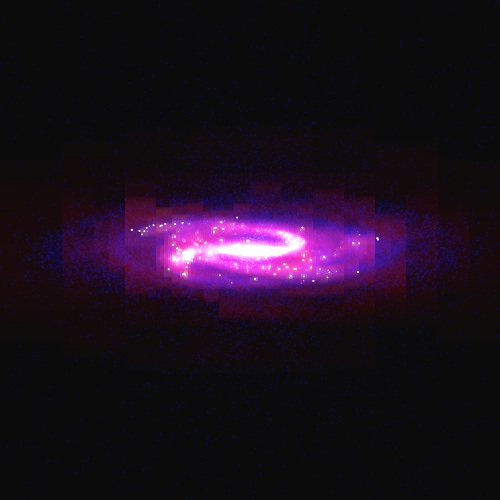}}}
\subfloat[Camera 6]{\includegraphics[width=0.27\columnwidth]{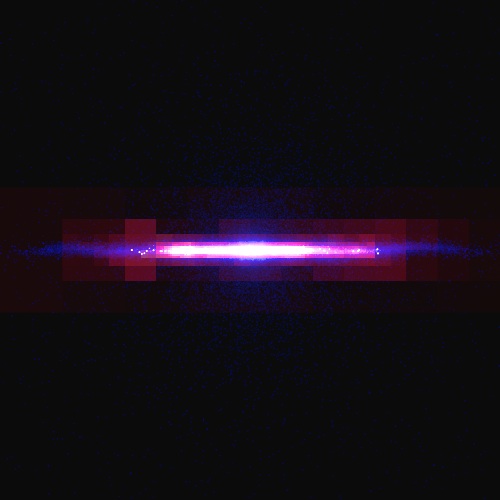}}
\caption{Composite images of the isolated disc simulation studied here, viewed from seven different viewing angles with particular wavelengths of interest mapped to different colours; $g^\prime$ band is shown as blue, while 24\,$\mu$m is green and 250\,$\mu$m is red. The views which we have used for the \magphys\ analysis (cameras 0, 3 \&\ 5) are outlined.}
\label{fig:simulations}
\end{figure*}

We evolve the simulated galaxy for 1 Gyr. As evident in \fref{fig:simulations}, at this time, the galaxy is a thin disc with spiral arms and a notable bar;
the latter two features develop self-consistently from an initially axisymmetric smooth disc.
The global morphology of the galaxy is `reasonable', although the ISM is unresolved on scales $\la 100$ pc.

Using the outputs of the hydrodynamical simulation to specify the 3-D distributions of sources of emission (star particles and the BH) and dust,
we perform dust radiative transfer to forward-model spatially resolved spectral energy distributions (SEDs) spanning the UV--mm for 7 viewing
angles, ranging from face-on to edge-on (see \fref{fig:simulations}). We employ the 3-D Monte Carlo radiative transfer code \sunrise \citep{pj06,
pj10}. For star particles with ages $>10$ Myr, \sunrise assigns single-age stellar population SEDs appropriate for their age and metallicity
using {\sc starburst99} \citep{leitherer99}. For younger star particles, it employs the models of \citet{groves08}, which account for the effects of
H{\sc ii} regions and photodissociation regions (PDRs). However, for reasons detailed below, we assume a PDR covering fraction of 0. We assume an initial mass function from \citet{kroupa01}. 
The BH is assigned an SED using the luminosity-dependent templates of \citet{hopkins07}, which are based on observations of un-reddened quasars.
The gas-phase metal distribution in the hydrodynamical simulation is used to specify the dust distribution, assuming a constant dust-to-metal density ratio of
0.4 \citep{dwek98,james02}. To calculate the effects of dust absorption and scattering, Monte Carlo radiative transfer is performed using $10^7$
photon packets, which we have ensured is sufficient to maintain Monte Carlo noise (in the resulting integrated SED) of a few per cent or
less.\footnote{Of course, there will always be some surface brightness below which the results are too noisy; we address this below.}
The intrinsic dust grain properties are specified assuming the Milky Way $R = 3.1$ dust model of \citet{draine07}. Dust temperatures are calculated
assuming thermal equilibrium, and the resulting IR emission is propagated through the simulated ISM. Dust self-absorption is accounted
for by iterating the photon packet propagation and dust temperature calculation steps until the dust temperatures are sufficiently converged,
but given the modest optical depths in this simulated galaxy, this step is unimportant for this particular study.

Neglecting the physical uncertainties inherent in the hydrodynamical simulation, the most significant uncertainty in the radiative transfer calculations
is the treatment of the sub-resolution ISM structure. We typically employ two extreme approaches to estimate the level of uncertainty: (1) that the cold
clouds implicit in the \citet{sh03} model have zero volume-filling factor and thus do not participate in the radiative transfer (`multiphase-on')
and (2) that the dust is uniformly distributed on sub-resolution scales (i.e. both the diffuse ISM and cold clouds contribute; we refer to this as
`multiphase-off'). When employing the latter assumption, we assume a PDR covering fraction of zero to avoid `double-counting' dust.
This issue has been discussed extensively in many previous works \citep[e.g.][]{H11,Snyder13,Lanz14,HS15,safarzadeh17}, so we will not
discuss it in detail here. We simply note that we employ the `multiphase-off' assumption because (1) owing to the higher optical depths,
recovery of galaxy properties is more challenging and (2) it is in principle possible to recover the total true dust mass because all dust implicit
in the simulation emits thermal radiation, unlike in the `multiphase-on' case. To check the robustness of our results to the sub-resolution ISM
structure, we re-ran the \sunrise calculation with the `multiphase-on' assumption and fit the $100 \times 100$ camera 0 image. Although the
results quantitatively differed from the multiphase-off case, as they should (e.g. the true attenuation is less in the multiphase-on case, and this
is also true for the recovered values), the conclusions were unchanged.

As noted above, the results of our calculations consist of spatially resolved UV--mm SEDs for the simulated galaxy viewed from seven different
viewing angles (although we only analyze three of them in detail in this work). We set the field-of-view to 50 kpc and used $500 \times 500$
pixels, yielding an intrinsic image resolution of 100 pc (which is approximately the smallest physical scale resolved in the \gadgetthree
simulations).

The SEDs predicted using simulations such as that studied here have been demonstrated to be consistent with e.g. local `normal' galaxies
\citep{pj10}, local (ultra-)luminous IR galaxies \citep{younger09,Lanz14}, high-redshift ULIRGs \citep[e.g.][]{H11,H12,roebuck16},
and obscured AGNs \citep{Snyder13,roebuck16}, post-starburst galaxies \citep{snyder11}, and compact quiescent galaxies
\citep{Wuyts10}. Thus, despite the simulations not being perfectly representative of real galaxies, it is reasonable to employ them for
`controlled experiments' such as this study (see also e.g. \citealt{Michalowski14,
H14,HS15,SH15}) in which we attempt to recover physical quantities from (synthetic) observables and compare the results with the
`ground truth' values, which is not possible when studying real galaxies.

\subsection{The \magphys\ model}

\magphys\ is described in detail in \citet{dacunha08}, however we will provide a brief overview here. \magphys\ uses an empirical but physically-motivated model to produce consistent SEDs of galaxies from the rest-frame ultra-violet to the far-infrared. It does this by using a simple two-component dust model from \citet{charlot00} to absorb photons in the stellar birth clouds and diffuse interstellar medium, and re-radiate this energy in the far-infrared. \magphys\ includes a library of optical to near-infrared stellar spectra based on 50,000 exponentially declining star formation histories (with random bursts superposed) derived using the latest version of the \citet{bc03} stellar population libraries (the ``CB07" library, unpublished), assuming an initial mass function from \citet{chabrier03}, and a range of metallicities between $0.0 < Z < 2.0$ (in solar units). This library is consistently linked to a dust library containing 50,000 dust emission SED models (each consisting of multiple continuum components associated with dust grains with a range of sizes and temperatures, as well as emission lines from polycyclic aromatic hydrocarbons) according to an energy balance criterion. The energy balance criterion specifies that the energy absorbed from starlight by dust primarily at UV\slash optical wavelengths is re-radiated in the far-infrared, and this ensures that the model SEDs are physically consistent across the full wavelength range. 
While this criterion is reasonable for modelling whole galaxies, it may not be the case for regions within galaxies, since emission from neighbouring areas can dominate the heating of dust in a given region. It is therefore essential to test how well \magphys\ works in the spatially resolved regime. 

\magphys\ produces model photometry based on passing the emergent SED from each combination of the libraries satisfying the energy balance criterion through the filter curves appropriate for the available observations. These model fluxes are then compared to the empirical values and uncertainties using the $\chi^2$ algorithm. By assuming that $P^\prime \propto \exp \left(-\chi^2/2 \right)$, the relative likelihood for each star formation history (or dust emission SED model) in the library is then recorded and used to produce marginalised probability distribution functions (PDFs) for each parameter in the model. 

In \citet{SH15}, we described modifications to \magphys\ to allow these probabilities to be output for each star formation history in the stellar library, enabling e.g. average star formation histories to be determined. Outputting these data also allows PDFs to be derived in post-processing without having to further modify and re-run the \magphys\ fitting (which is reasonably CPU-intensive, taking $\sim$8 minutes per galaxy on a single 2.5\,GHz processor), by taking the input parameters from the \magphys\ library of stellar SEDs (which also include e.g. metallicity, mass-weighted age, and $A_V$\footnote{We note that the values of $A_V$ used in this analysis are calculated by comparing the fluxes transmitted by the intrinsic and transmitted spectra through the $V$ band filter curve, both for the \magphys\ libraries, and for the model galaxy data cubes.}), along with the marginalised probability distribution for each. In this way, we can produce best-fit as well as median-likelihood parameter estimates, along with uncertainties estimated using the 16$^\mathrm{th}$ and 84$^\mathrm{th}$ percentiles of the PDF, in exactly the same way as the default version of the code does, but for a wider range of parameters, since the supplied version of \magphys\ treats metallicity and mass-weighted age as nuisance parameters, and does not include $A_V$ values (though they can be trivially engineered from the spectral libraries). This can all be done with minimal modification of the fortran source code, requiring only the straightforward modifications to output a list of the probabilities for each SFH. 

In this way, we can use \magphys\ to estimate stellar metallicities and ages based on the full range of multi-wavelength photometry, a step forward relative to recent works based on optical photometry alone (e.g. \citealt{smail01,lee07,carter09,sanders13,acquaviva16}, though see also \citealt{amblard17}). It is generally recognised that the best metallicity and age estimates can only be obtained through deep spectroscopy \citep[e.g.][]{poggianti01}, but it is potentially very useful if it is possible to obtain reliable metallicities from photometry alone. Though the benefit of including far-infrared data for estimating stellar metallicities and ages may not be initially clear, far-infrared data can potentially play a key role in breaking the age-metallicity degeneracy \citep[e.g.][]{worthey94,carter09}, since galaxies dominated by old stellar populations are in general far less luminous at far-infrared wavelengths than their younger counterparts. In this work we therefore also perform exploratory tests to uncover the extent of the \magphys\ formalism's ability to recover these non-standard outputs.

\subsection{SED fitting using \magphys}

The simulation datacubes (which are analogous to integral field spectrograph data, i.e. they have two spatial dimensions that correspond to the 2D projected surface of the galaxy for a given camera angle and a third dimension that corresponds to wavelength) have an intrinsic spatial resolution of $500 \times 500$ pixels, each 0.1\,kpc in size. We rebinned the emergent photometry by resampling the spatial dimensions of the datacubes into lower-resolution versions 250, 100, 50, 20, 10, 5, 2 \&\ 1 pixels on a side (corresponding to spatial resolutions of 0.2, 0.5, 1.0, 2.5, 5.0, 10, 25 and 50\,kpc), and fit each individual element of the resulting arrays of photometry independently using \magphys. This means that for each viewing angle we performed a total of 75,530 individual SED fits requiring approximately $10^4$ CPU-hours for each. In order to see how our results vary depending on viewing angle, we applied the model to cameras 0, 3 and 5, which are oriented such that $(\theta, \phi) = (0^{\circ},0^{\circ})$, $(45^{\circ}, 0^{\circ})$ and $(75^{\circ}, 0^{\circ})$ 
for the three cameras respectively,\footnote{We adopt the notation that $\theta$ and $\phi$ represent the polar and azimuthal angles, respectively.} with the galaxy's disc being oriented along the $xy$ plane. Broadly speaking, camera 0 views the disc face-on, while camera 5's view is close to edge-on, with camera 3 in between.  

In this analysis, we arbitrarily choose to use 21 bands of photometry, generated by convolving the emergent spectra at each location with the relevant filter curves. The bands we include are the following: FUV and NUV data from {\it GALEX} \citep{martin05}, $ugriz$ bands from the Sloan Digital Sky Survey \citep{york00}, $J$, $H$ and $K$ band near-infrared filters from the UK Infrared Deep Sky Survey \citep[UKIDSS;][]{lawrence07}, as well as 3.6, 4.5, 5.8, 8.0, 24 and 70\,$\mu$m filters from the IRAC \citep{fazio04} and MIPS \citep{rieke04} instruments on the {\it Spitzer Space Telescope} \citep{werner04}, plus 100, 160, 250, 350 and 500\,$\mu$m transmission functions for the PACS \citep{poglitsch10} and SPIRE \citep{griffin10} instruments on the {\it Herschel Space Observatory} \citep{pilbratt10}. The synthetic images that are produced by the simulations do not contain photometric uncertainties; however, in order to be able to fit the emergent photometry with \magphys, we follow \citet{SH15} and assume a signal-to-noise ratio of five in every photometric band.

\section{Results} \label{S:results}

In this section we will show the results of our \magphys\ fitting in a range of different ways, in a bid to critically gauge how well \magphys\ can perform in this challenging scenario. In the main body of the paper we include the results for the face-on simulation (camera 0), while in the appendices we include selected analogous figures derived based on the model photometry for the other viewing angles. We will discuss how well \magphys\ recovers each of seven parameters (stellar mass, SFR, specific SFR, dust mass, stellar metallicity, $A_V$, and mass-weighted age) in turn, and as a function of resolution in the next subsections. First, we begin by quantifying what fraction of the time \magphys\ is able to produce good fits to the data. 

\subsection{Can \magphys\ produce good fits to spatially-resolved data?}
\label{sec:goodfits}

\magphys\ is designed using an empirical, but ``physically motivated'' model to describe the integrated properties of galaxies; this idealised model may therefore not be applicable to the spatially-resolved case. However, the \magphys\ model is not strictly correct in the integrated case either, though we have shown \citep{HS15,SH15} that this does not preclude it from producing reliable results. One key test that we must make before we can even begin to address whether the results it produces are sensible, is to determine whether the \magphys\ model can produce statistically acceptable fits to the data. We do this using the $\chi^2$ criterion discussed in \citet{smith12}, which provides a means of determining a $\chi^2$ threshold that depends only on the number of bands used in the fitting, above which there is a chance of less than 1 per cent that the data are consistent with the model. Since we include 21 photometric bands in our fitting, we identify ``good fits'' as those which have $\chi^2 < 30.6$. 

\begin{figure}
\centering 
\includegraphics[width=\columnwidth]{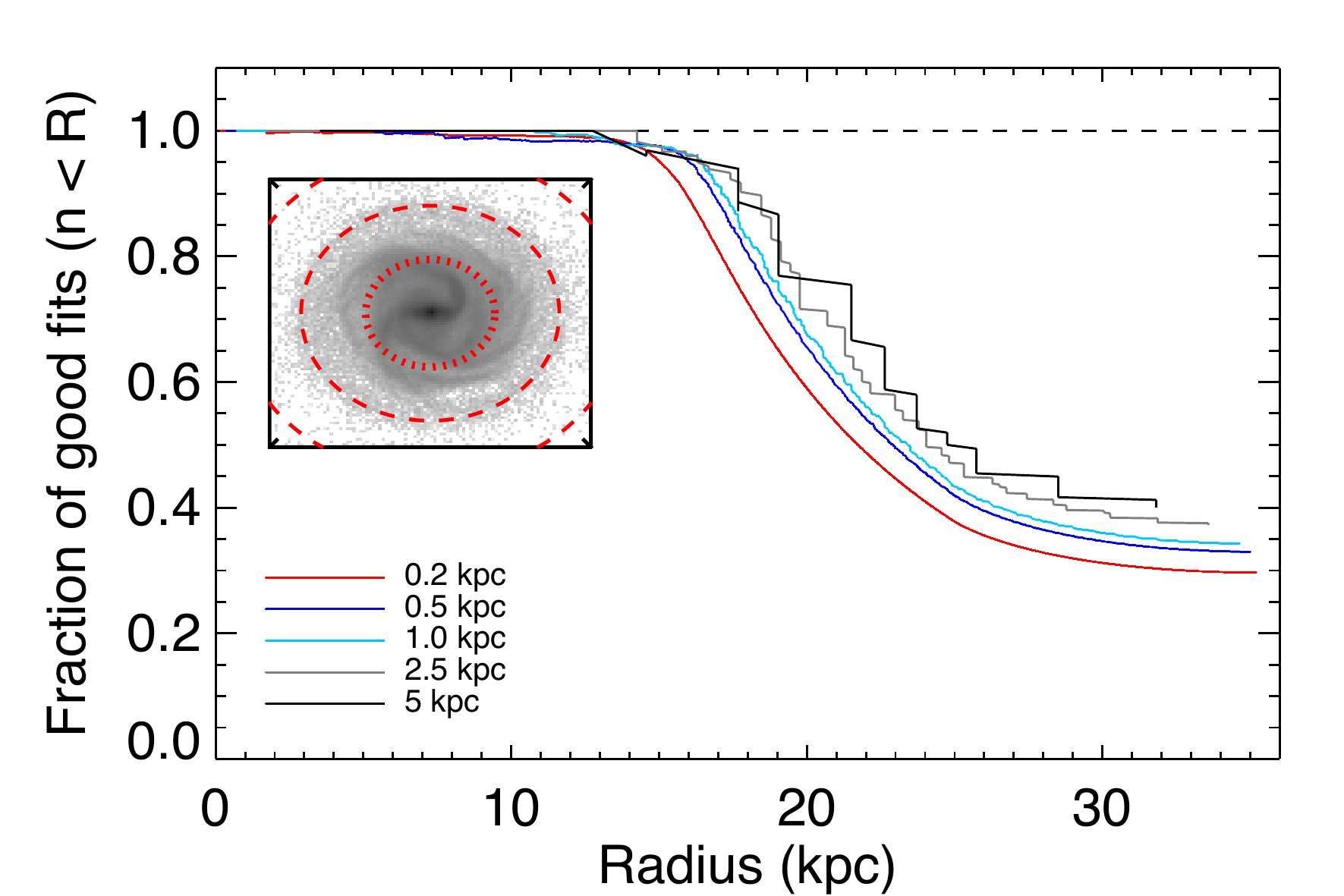}
\caption{Fraction of good fits within each radius, shown for spatial resolutions 250, 100, 50, 20 and 10 pixels on a side (corresponding to 0.2, 0.5, 1.0, 2.5 and 5.0\,kpc, as detailed in the legend). To give some idea of the physical scale, the inset greyscale image shows the true stellar mass distribution and is overlaid with circles with radii of 10, 20 and 30 kpc, shown as the red lines (dotted, dashed and dot-dashed, respectively). The $r$-band effective radius is 5.7~kpc. The pixels for which we do not obtain good fits contain a maximum of 2.3 and 1.0 per cent of the total stellar mass and SFR, respectively.}
\label{fig:fgood}
\end{figure}

Figure \ref{fig:fgood} shows the fraction of good fits as a function of radial separation from the centre of the galaxy for the five highest resolution runs (ranging from $250\times 250$ to $10\times 10$, with colours as shown in the legend to the lower left). The fraction of good fits is a strong function of radius, and we are able to fit the vast majority of pixels ($\ge 99$ per cent, in all runs) within 10 kpc (indicated by the dotted red line on the inset image of the true stellar mass distribution), while that figure drops to between 59 and 77 per cent of the pixels within 20\,kpc (dashed red line in the inset), and the fraction of good fits is lowest for the higher resolution runs at large separations. To put these values in further context, the $r$-band effective radius of the galaxy is 5.7\,kpc, and 99\,per cent of the $r$-band light is emitted within 19.2\,kpc. The pixels for which we do not obtain good fits contain a maximum of 2.3 per cent of the total stellar mass and 1.0 per cent of the total SFR. 

\begin{figure}
\centering 
\includegraphics[height=0.58\columnwidth,trim=0cm 0cm 0cm 0cm,clip]{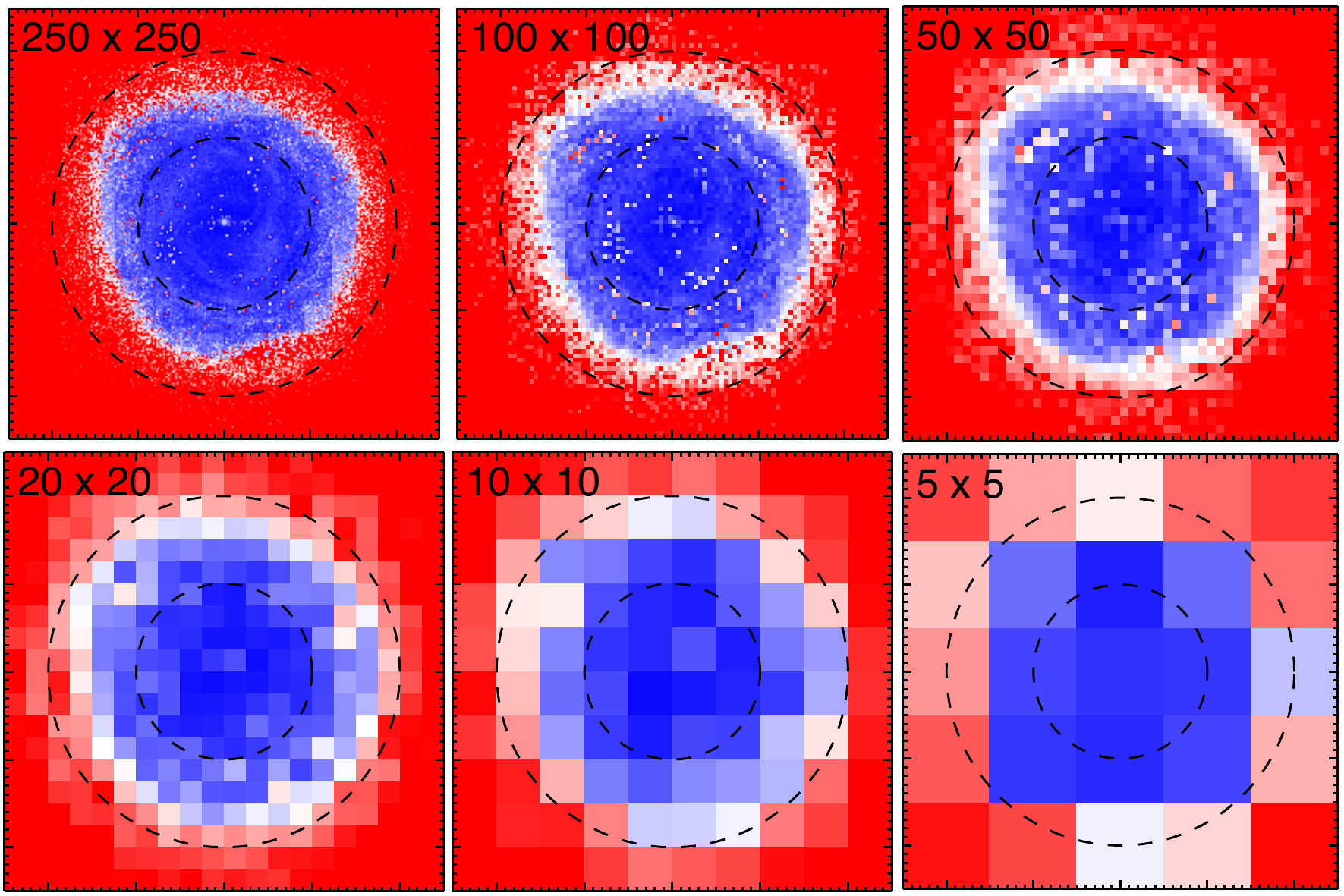}
\includegraphics[height=0.58\columnwidth,trim=2.5cm 0cm 1cm 0cm,clip=true]{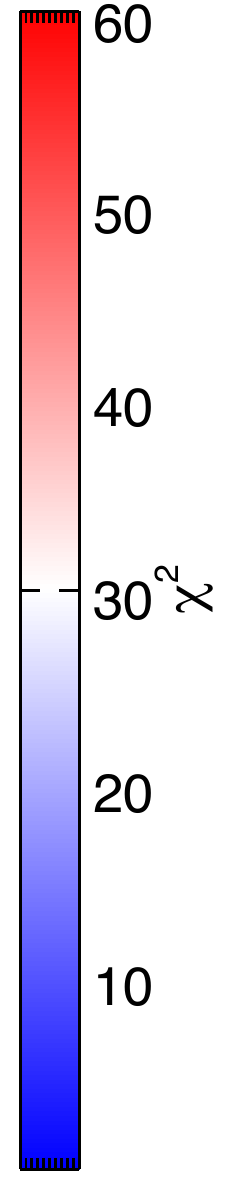}
\caption{Maps showing the best-fit values of $\chi^2$ as a function of position for six of the different resolution runs, ranging from the $250\times 250$ pixel (0.2\,kpc resolution) to the $5\times 5$ (10\,kpc resolution) run. The  bar to the right indicates the colour scale, which is common to each panel, and which is centred on $\chi^2 = 30.6$, the cut-off value that delineates acceptable and unacceptable fits. In this scheme, the pixels for which a good fit was obtained are shown in blue (the bluer, the lower the $\chi^2$), while the rejected fits are shown in red, and those close to the threshold are white. Dashed circles indicate the regions within 10 and 20\,kpc from the centre of the map, which is 50\,kpc on a side in every case.}
\label{fig:chi2maps}
\end{figure}

In figure \ref{fig:chi2maps} we look at this in more detail for the six highest resolution runs, showing the best-fit $\chi^2$ value as a function of position on the map, colour coded such that good fits are in blue, and fits that were rejected by our $\chi^2$ criterion are shown in red, with increasingly strong colour indicating how far away from the threshold they are. The lack of spiral structure in the blue parts of these figures offers encouragement \magphys\ is working as expected across a broad range of structure present in the galaxy. As noted above, the vast majority of the bad fits are in the faint outer regions of the disk, where the simulations begin to encounter Monte Carlo effects due to the quantization of the photon packets (see discussion of $A_V$ in section \ref{subsec:age_recovery}, below). Examining the results of the \magphys\ fitting of these unacceptable pixels does not reveal any one particularly discrepant wavelength regime providing the dominant source of the $\chi^2$, rather a steady progression towards worse $\chi^2$ with increasing distance from the centre. 

In the coming sections, we will show parameter maps and scatter plots for each parameter that we attempt to recover; in all of these we will only show those pixels for which we are able to obtain acceptable fits to the data, and as we shall see, we are able to recover a range of total\slash effective properties very well. Although our work is based on the \magphys\ libraries built using the CB07 stellar templates, we re-ran a subset of the fits using the \citet{bc03} libraries currently recommended on the \magphys\ website, finding no significant differences in our results which follow.

\subsection{\magphys\ recovery of stellar mass, $M_\mathrm{stars}$}

\begin{figure*}
\centering
\subfloat[]{
\begin{minipage}{0.4\textwidth}
\includegraphics[width=\textwidth,trim=0cm 0cm 0cm 0cm,clip]{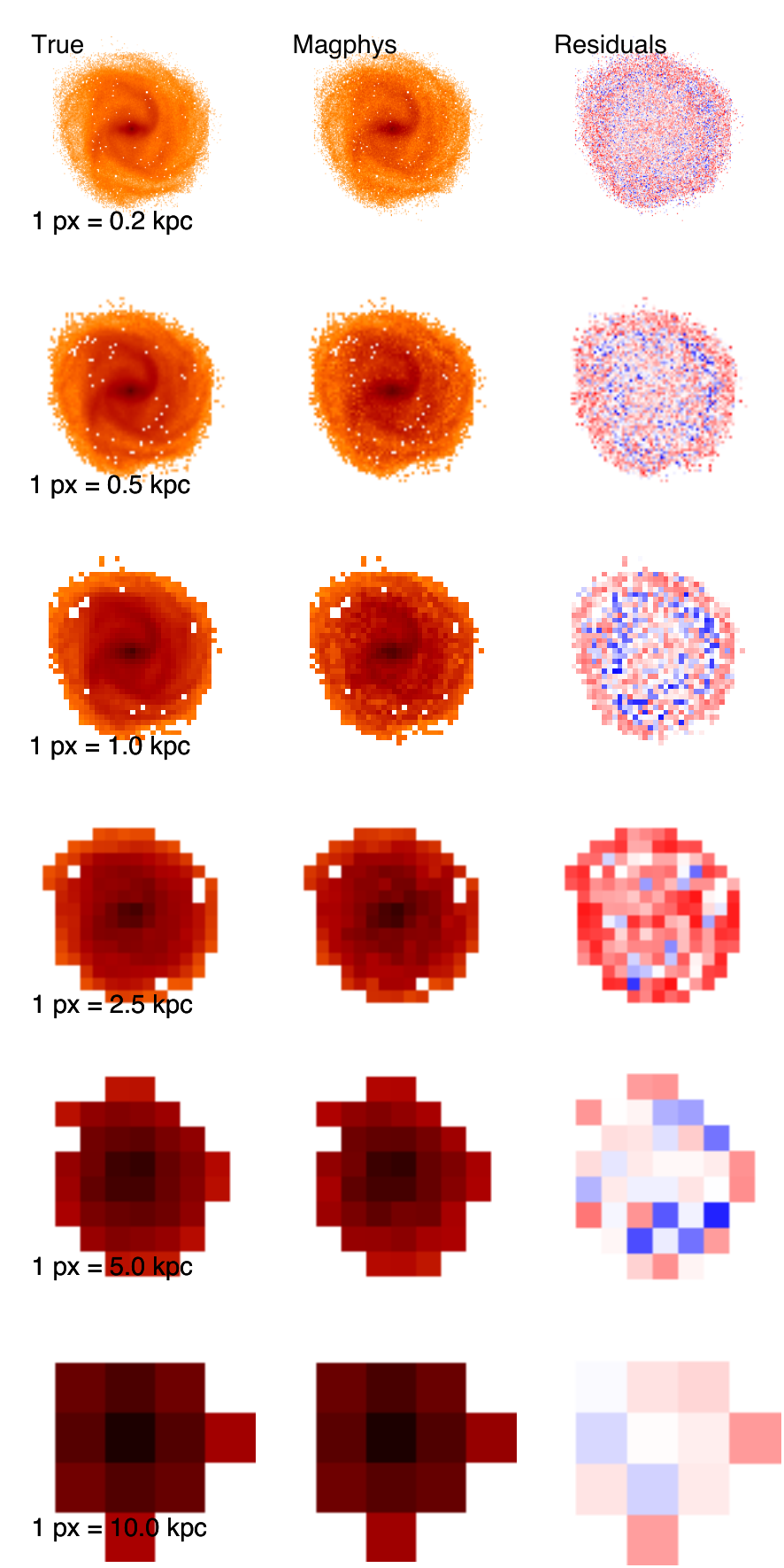} \\\includegraphics[width=\textwidth,trim=0.0cm 1.2cm 0.0cm 2.4cm,clip=true]{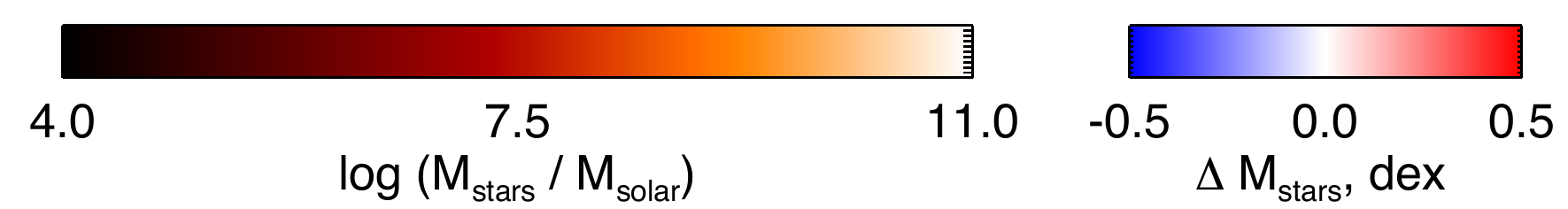}
\end{minipage}
}
\begin{minipage}{0.6\textwidth}
\subfloat[]{\includegraphics[width=\columnwidth,trim=0cm 0cm 0cm 0.85cm,clip]{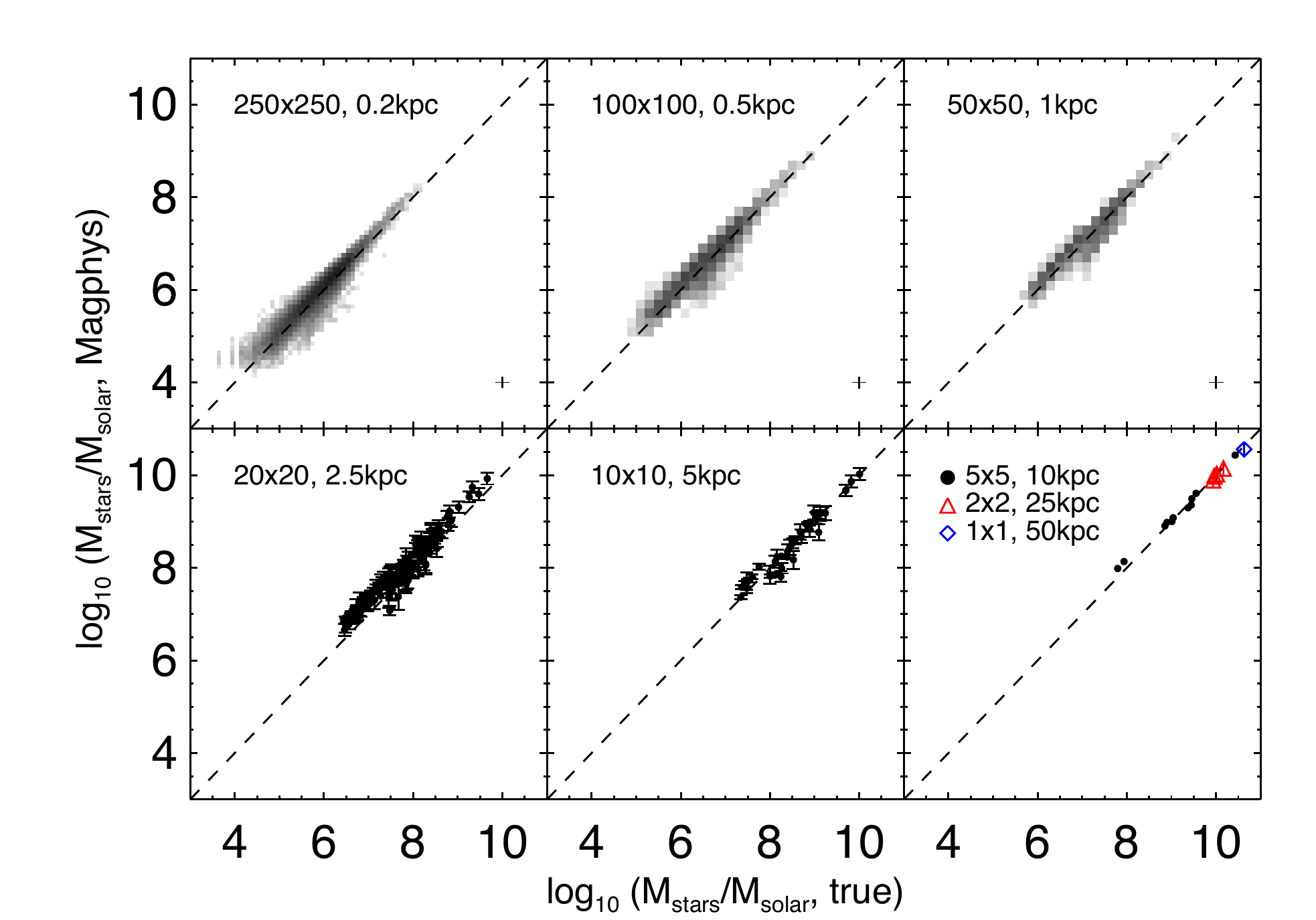}}\\
\vspace{1.0cm}\\
\subfloat[]{\includegraphics[width=\columnwidth,trim=0cm 0cm 0cm 0.8cm,clip]{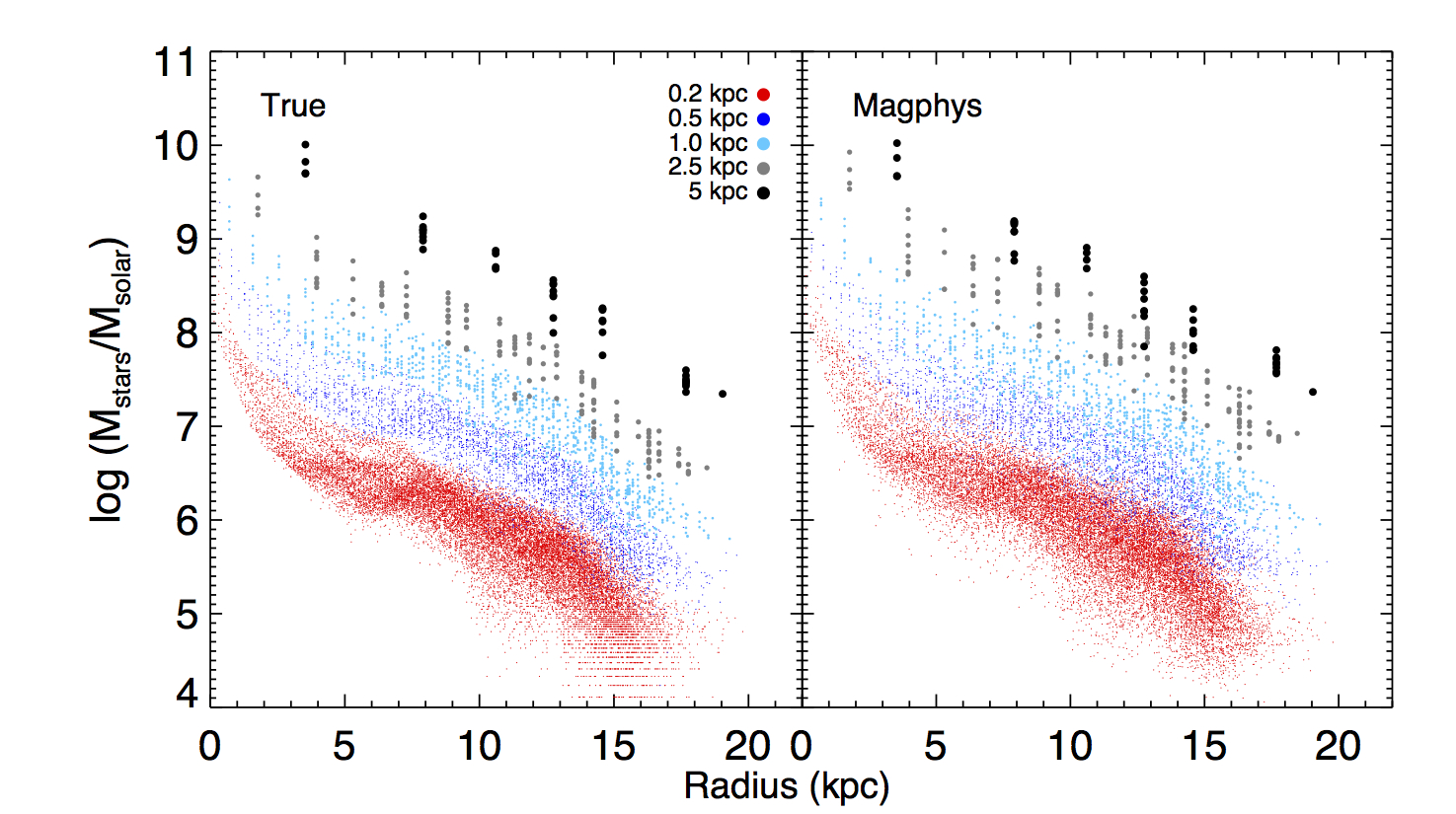}}
\end{minipage}
\caption{Recovery of stellar mass for camera 0. In this and similar figures below, only those pixels for which \magphys\ obtains good fits according to the $\chi^2$ criterion defined in \citet{smith12} are shown. In the left-hand {\bf panel (a)}, we show three columns of images, detailing the true stellar mass distribution (left), and the \magphys\ estimates (center) with spatial resolution decreasing from top (0.2\,kpc pixels) to bottom (10\,kpc pixels), as labelled in bottom left corner of the left column. The colour bar (shown at the bottom) has been chosen to represent values in the range $4 < \log (M_\mathrm{stars}/\mathrm{M}_\odot) < 11$. The right-hand column shows the difference between the \magphys\ estimates and the true values, in dex, with the colour bar at the bottom indicating the scale. 
In {\bf panel (b)}, located on the upper right hand side, we show the recovered stellar masses as a function of the true values taken from the simulations, with different resolutions shown in each panel, as labelled in the legend. The diagonal dashed line in each panel indicates the 1:1 relation. The top three panels are shown as a 2D histogram with a colour scale linear in the logarithm of the bin occupancy, ranging from $0 < log N < N_{\mathrm{max}}$ and a typical error bar shown in the lower-right corner, while in the lower panels the comparatively lower number of pixels allows us to also overlay the uncertainties returned by \magphys, shown by the vertical error bars. In {\bf panel (c)} we show the variation of the stellar mass as a function of the radial distance from the centre of the galaxy in kpc, with the different resolutions shown as different colours detailed in the legend. The true values are shown in the left panel, while the corresponding \magphys values are shown to the right. At all resolutions, the stellar masses of individual pixels are recovered extremely well, and consequently, so are the projected stellar mass maps and radial distributions.}
\label{fig:mass_recovery}
\end{figure*}

Figure \ref{fig:mass_recovery} shows the fidelity of the stellar mass recovery in three different ways. Figure \ref{fig:mass_recovery}\,(a) shows images of the true and recovered stellar mass in the left and centre columns, respectively, with the difference between the two (\magphys\ - True) indicated in the right-hand column. The spatial resolution decreases from 0.2\,kpc in the upper row to 10\,kpc in the bottom row. Figure \ref{fig:mass_recovery}\,(b) shows the true and \magphys\ values plotted directly against one another for the full range of resolutions; in the top row (corresponding to the highest spatial resolutions) we show only a 2D histogram of the best-fit values (with the colour bar linear in the logarithm of the number of pixels in each bin, between $0 < \log N < \log N_\mathrm{max}$, and with a typical error bar on an individual data point shown in the bottom right), while for the lower row we show the best fit values with uncertainties on the \magphys\ values based on the 16th and 84th percentiles of the stellar mass PDF. Figure \ref{fig:mass_recovery}\,(c) shows the radial dependence of the stellar mass (true values on the left, recovered \magphys\ values on the right). For the $250\times 250$, 0.2\,kpc resolution scenario, the median $M_\mathrm{stars}^\mathrm{true} - M_\mathrm{stars}^\mathrm{magphys}$ = -0.08\,dex with 68 per cent of the data within 0.16\,dex across the whole mass range, equivalent to the $1\sigma$ interval in the limit of Gaussian statistics.

It is clear that \magphys\ produces excellent estimates of stellar mass, both in the unresolved case \citep[as demonstrated in][]{HS15}, and now shown here in the resolved case, at resolutions of up to 0.2\,kpc. \magphys\ is able to reliably recover best-fit stellar mass values as low as $10^4$\,M$_\odot$, and produces good fits at the vast majority of locations across the disc (as discussed in section \ref{sec:goodfits}). The spiral arm structure is apparent in both the true and \magphys\ mass maps in Figure \ref{fig:mass_recovery}\,(a), while the \magphys\ values are located along the 1:1 line in each panel of Figure \ref{fig:mass_recovery}\,(b), and the recovery of the radial dependence shown in Figure \ref{fig:mass_recovery} (c) is also excellent. 

\begin{figure*}
	\centering 
	\subfloat[Camera 0]{\includegraphics[width=1.02\textwidth, trim=0cm 0.90cm 0cm 0.43cm, clip=true]{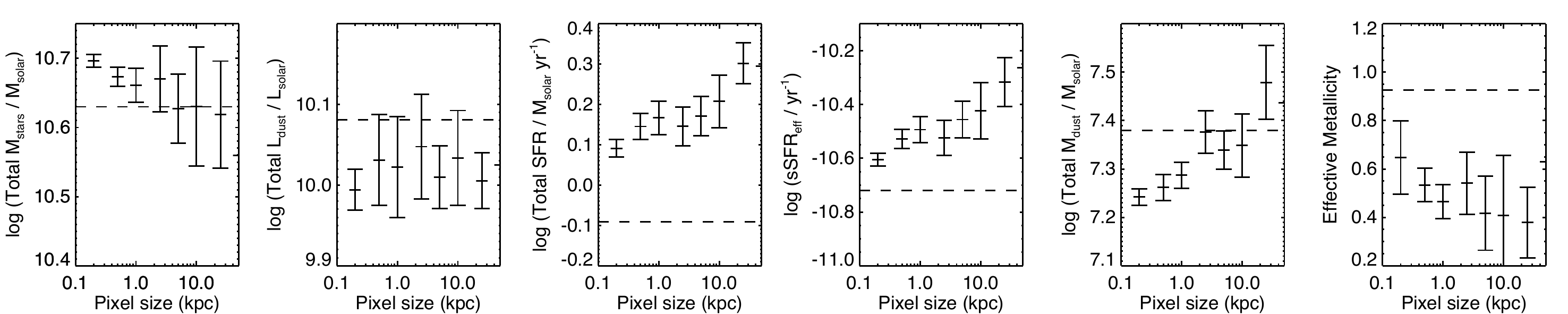}}\\
	\vspace{-0.3cm}
	\subfloat[Camera 3]{\includegraphics[width=1.02\textwidth, trim=0cm 0.90cm 0cm 0.43cm, clip=true]{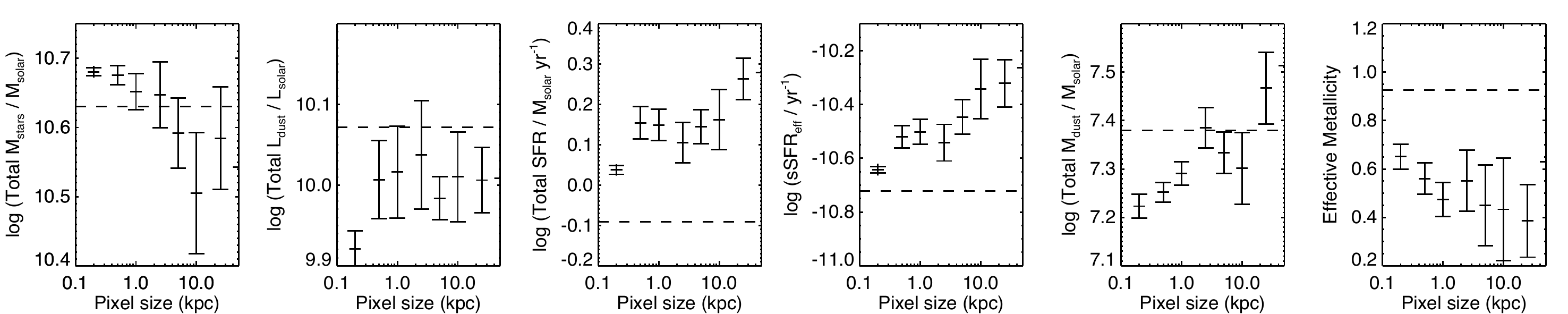}}\\
	\vspace{-0.3cm}
	\subfloat[Camera 5]{\includegraphics[width=1.02\textwidth, trim=0cm 0cm 0cm 0.43cm, clip=true]{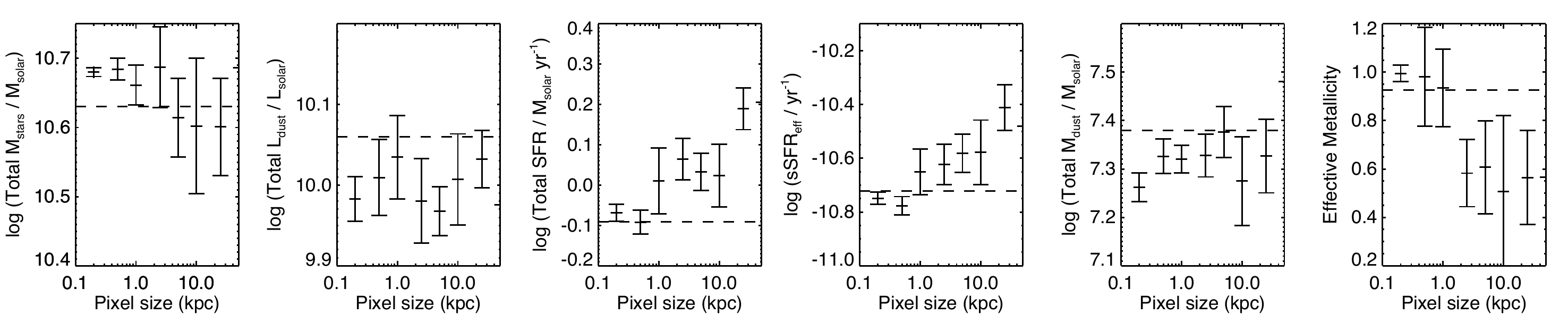}}
	\caption{Plots showing (from the leftmost to rightmost column) the total stellar mass, dust luminosity, SFR, effective sSFR, total dust mass and effective stellar metallicity, obtained by summing the values recovered by \magphys\ for each pixel,
	 as a function of the pixel size. The true values taken from the simulation are shown as the horizontal dashed lines. Stellar mass and dust luminosity are recovered superbly at all resolutions. The SFR and sSFR values are overestimated somewhat,
	 increasingly so at lower resolution, with maximum overestimates of $\sim 0.4$ dex. The dust mass is underestimated by $\la 0.2$ dex at the highest resolutions and recovered to within 0.1 dex for pixel sizes greater than a
	 few kpc. The metallicity recovery is poor except for the highest-resolution, edge-on photometry. 
	 }
	\label{fig:totals_vs_resolution}
\end{figure*}

As mentioned in section \ref{S:intro}, the work by \citet{sorba15} suggested that the stellar masses estimated for galaxies with only integrated photometry may have been underestimated relative to the sum of the individual components. In the left column of Figure \ref{fig:totals_vs_resolution}, which the three rows corresponding to the three viewing angles chosen for this study, we test for the influence of resolution on the total stellar mass obtained using \magphys, by calculating the sum of the individual masses at each resolution (including propagating the error bars), and comparing with the true stellar mass (shown as the horizontal dashed line). Figure \ref{fig:totals_vs_resolution} shows that there is no strong evidence for a resolution-dependent bias in the stellar mass estimates; if there is a discrepancy it is located at the spatially resolved end \citep[i.e. the opposite of what was found by][]{sorba15}, where the error bars are inconsistent with the true stellar mass (though as discussed above, the ``discrepancy" is less then 0.1\,dex) and this trend is consistent for each of the three viewing angles. We will return to this figure in the coming sections, with reference to the other panels.

\subsection{\magphys\ recovery of dust luminosity}

\begin{figure*}
\centering
\subfloat[]{
\begin{minipage}{0.4\textwidth}
\includegraphics[width=\textwidth,trim=0cm 0cm 0cm 0cm,clip]{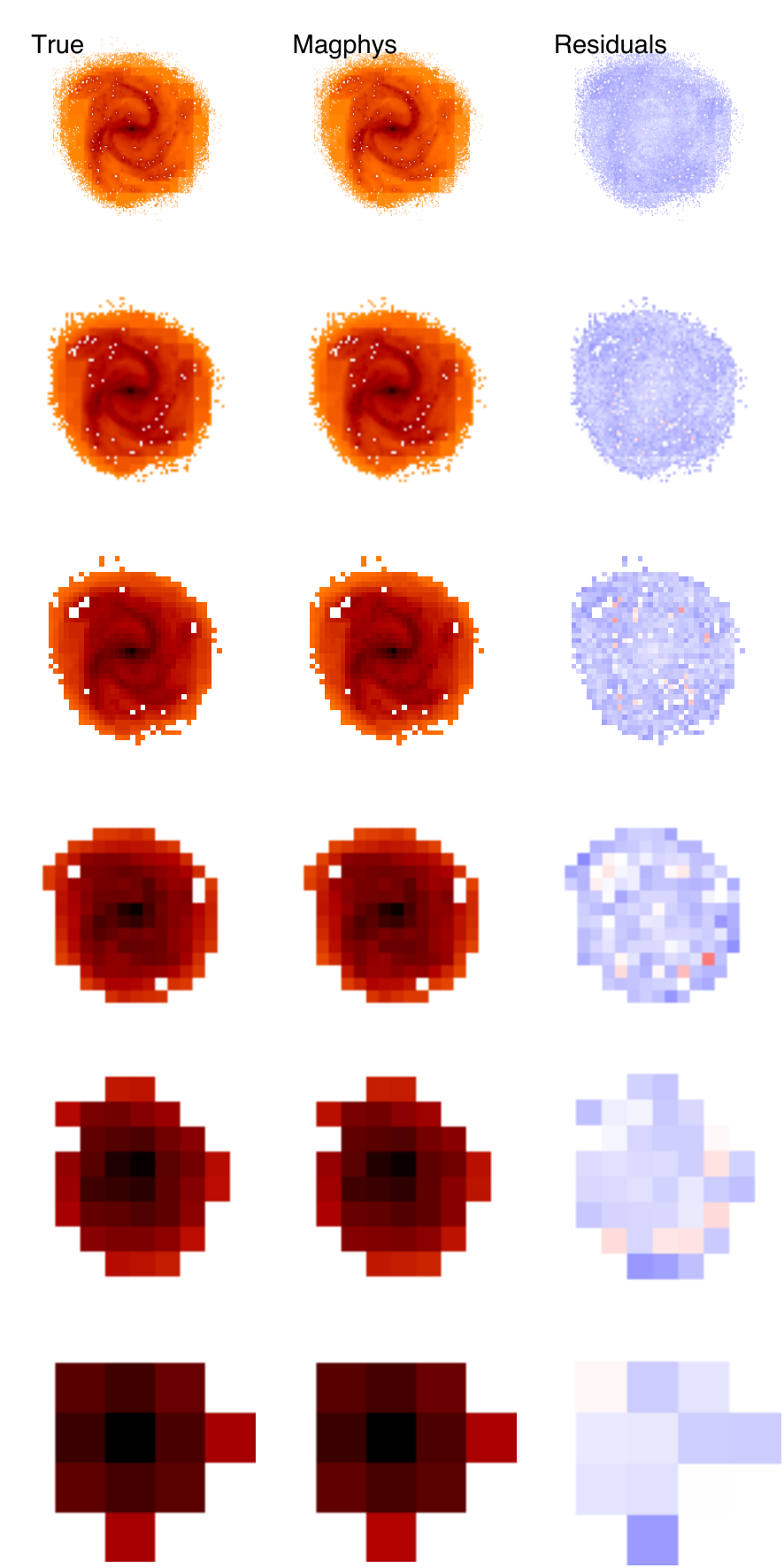} \\
\includegraphics[width=\textwidth,trim=0cm 1.2cm 0cm 2.4cm]{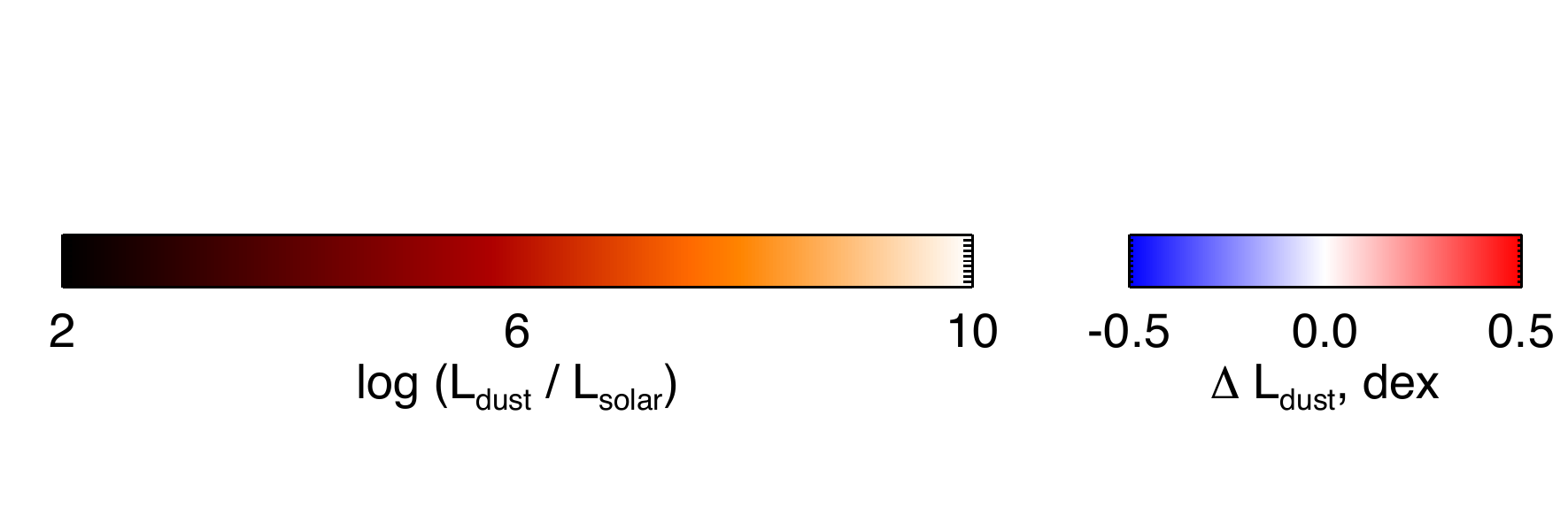}
\end{minipage}
}
\begin{minipage}{0.6\textwidth}
\subfloat[]{\includegraphics[width=\columnwidth,trim=0cm 0cm 0cm 0.85cm,clip]{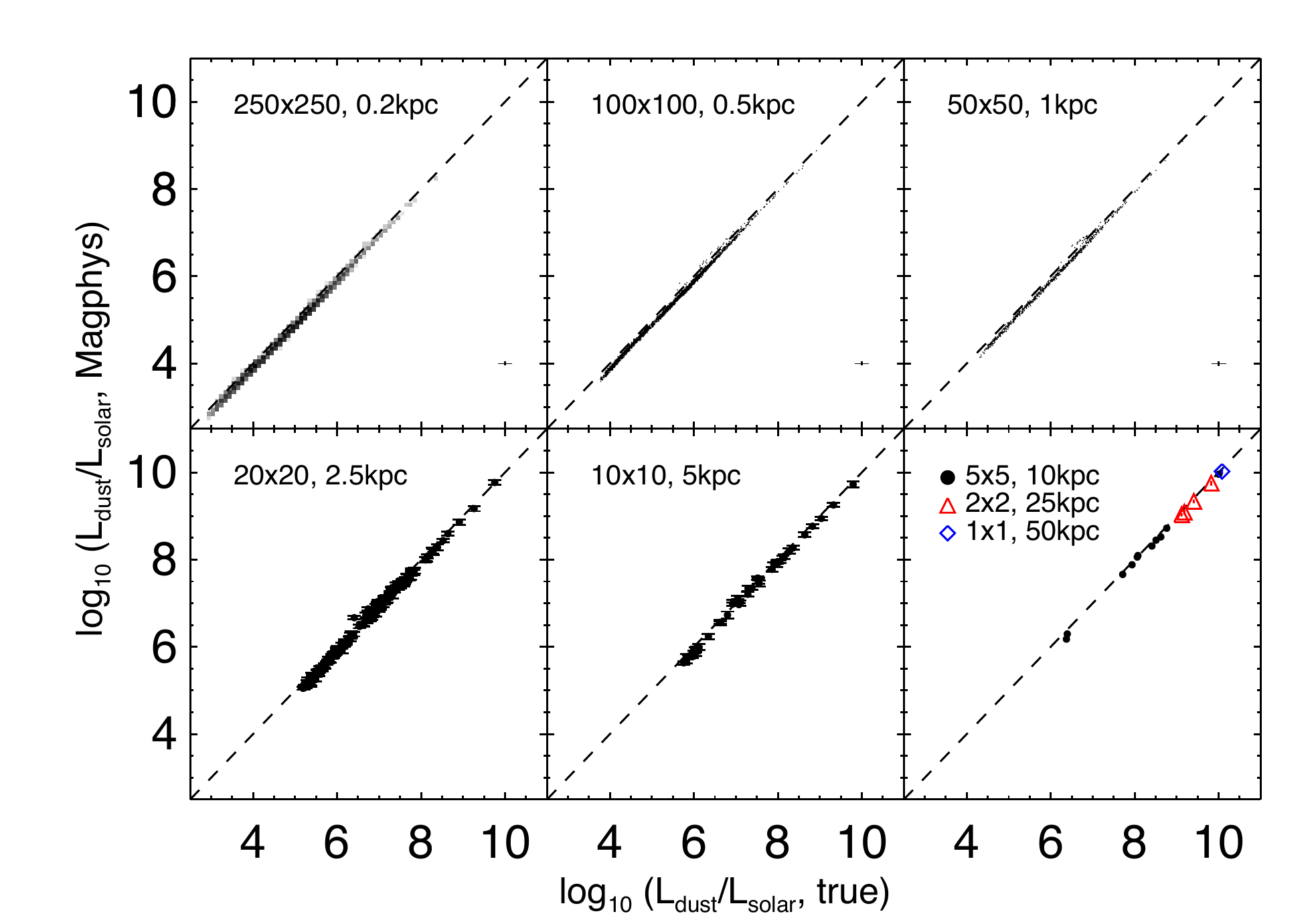}}\\
\vspace{1.0cm}\\
\subfloat[]{\includegraphics[width=\columnwidth,trim=0cm 0cm 0cm 0.8cm,clip]{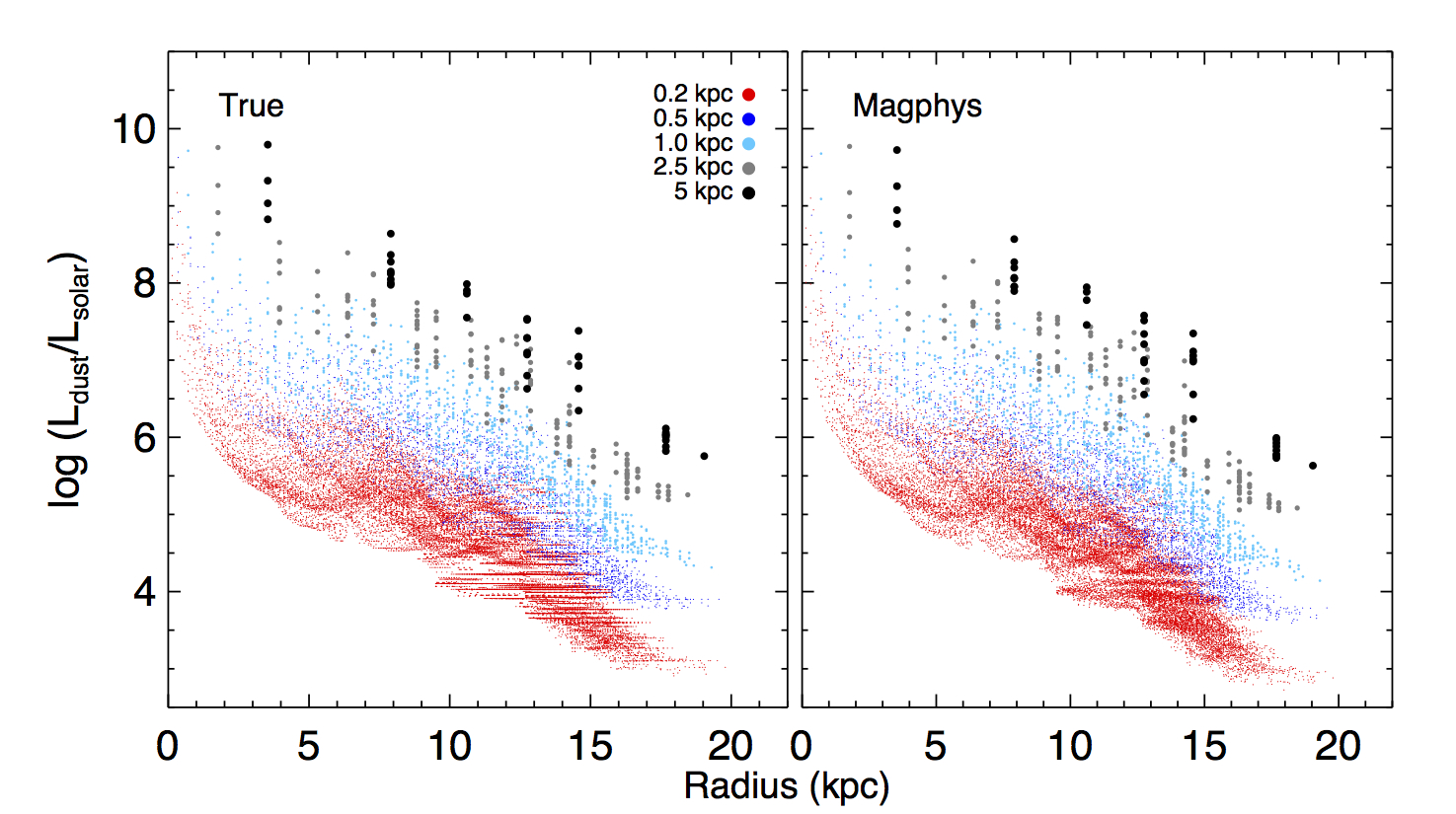}}
\end{minipage}
\caption{Recovery of dust luminosity for camera 0. The colour scales for the images showing the true and \magphys-estimated luminosity distributions are given by the colour bars at the bottom of each column of panel (a). For more details, see the caption of Figure \ref{fig:mass_recovery}. The details of the dust luminosity maps (e.g. spiral arms) are well recovered by \magphys, as are the radial profiles. At all resolutions, the dust luminosity contained in individual pixels are recovered very well.}
\label{fig:ldust_recovery}
\end{figure*}

The panels of Figure \ref{fig:ldust_recovery} are similar to Figure \ref{fig:mass_recovery}, however they now detail the extent of our ability to recovery the dust luminosity across the disk. It is unsurprising, given the 21-bands of SNR=5 photometry that we have assumed in our numerical experiment, that our tests show that \magphys\ produces very high fidelity dust luminosities across the whole disk, and even down to luminosities as low as $10^3\,L_\odot$. We are able to clearly make out the spiral structure in the recovered parameter maps (shown in the centre column of figure \ref{fig:ldust_recovery}\,a) and in the obvious radial structures present in the true data cube and in the \magphys\ recovered values (figure \ref{fig:ldust_recovery}\,c).

For the $250\times 250$ test, the median $L_\mathrm{dust}^\mathrm{true} - L_\mathrm{dust}^\mathrm{magphys}$ = 0.12\,dex, reflected by the light blue colour of the majority of the pixels in the right-hand column of figure \ref{fig:ldust_recovery} (a). This value is very similar to the one found for the integrated case discussed in \citet{HS15}, however we now show virtually identical behaviour in the spatially resolved case. In the second columns of figure \ref{fig:totals_vs_resolution}, we investigate the possibility of resolution bias in estimates of the total far-infrared luminosity, by summing the dust luminosity in each of the individual pixels and propagating the error bars at each individual resolutions (as for the stellar mass analysis discussed above), finding no evidence for such an effect. 

\subsection{\magphys\ recovery of SFR}

\begin{figure*}
\centering
\subfloat[]{
\begin{minipage}{0.4\textwidth}
\includegraphics[width=\textwidth,trim=0cm 0cm 0cm 0cm,clip]{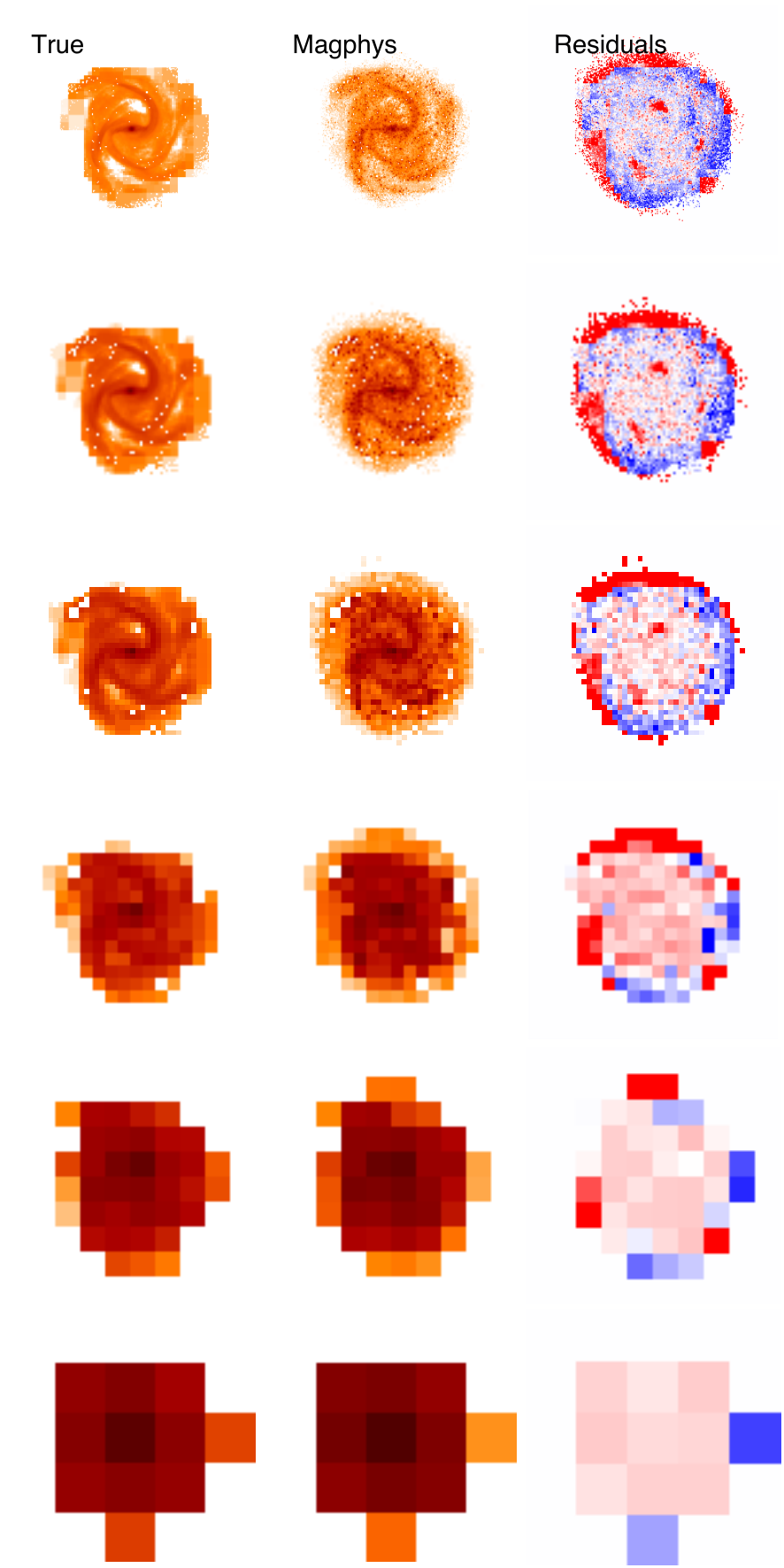} \\
\includegraphics[width=\textwidth,trim=0cm 1.2cm 0cm 2.4cm]{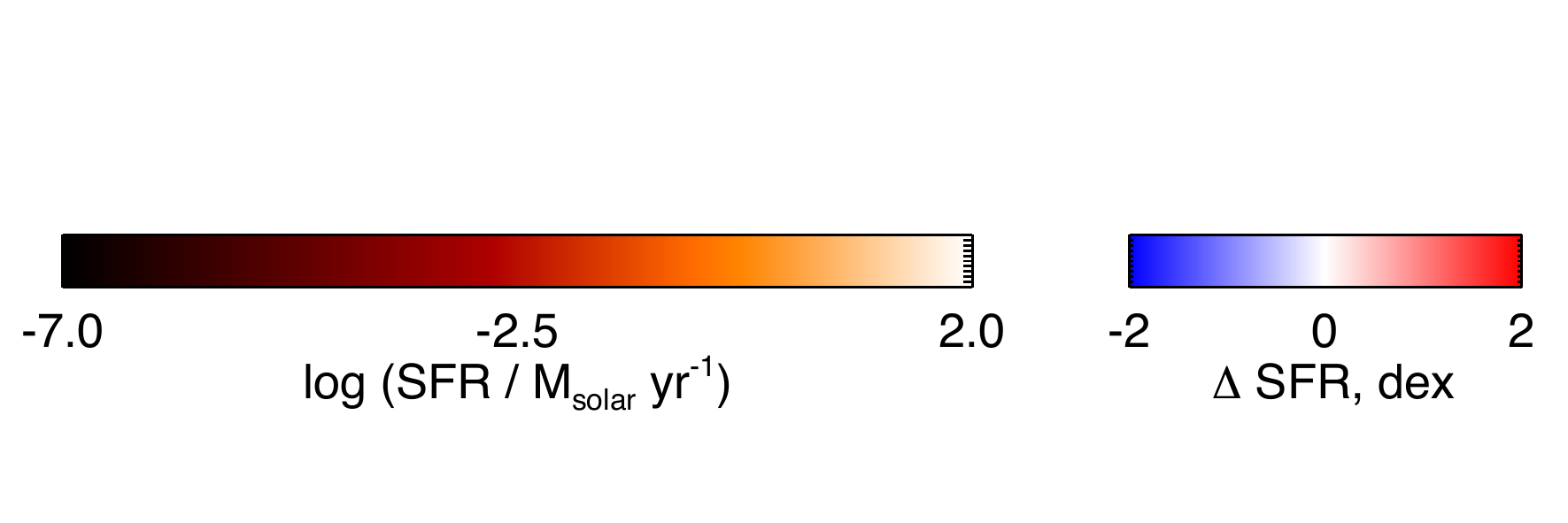}
\end{minipage}
}
\begin{minipage}{0.6\textwidth}
\subfloat[]{\includegraphics[width=\columnwidth,trim=0cm 0cm 0cm 0.85cm,clip]{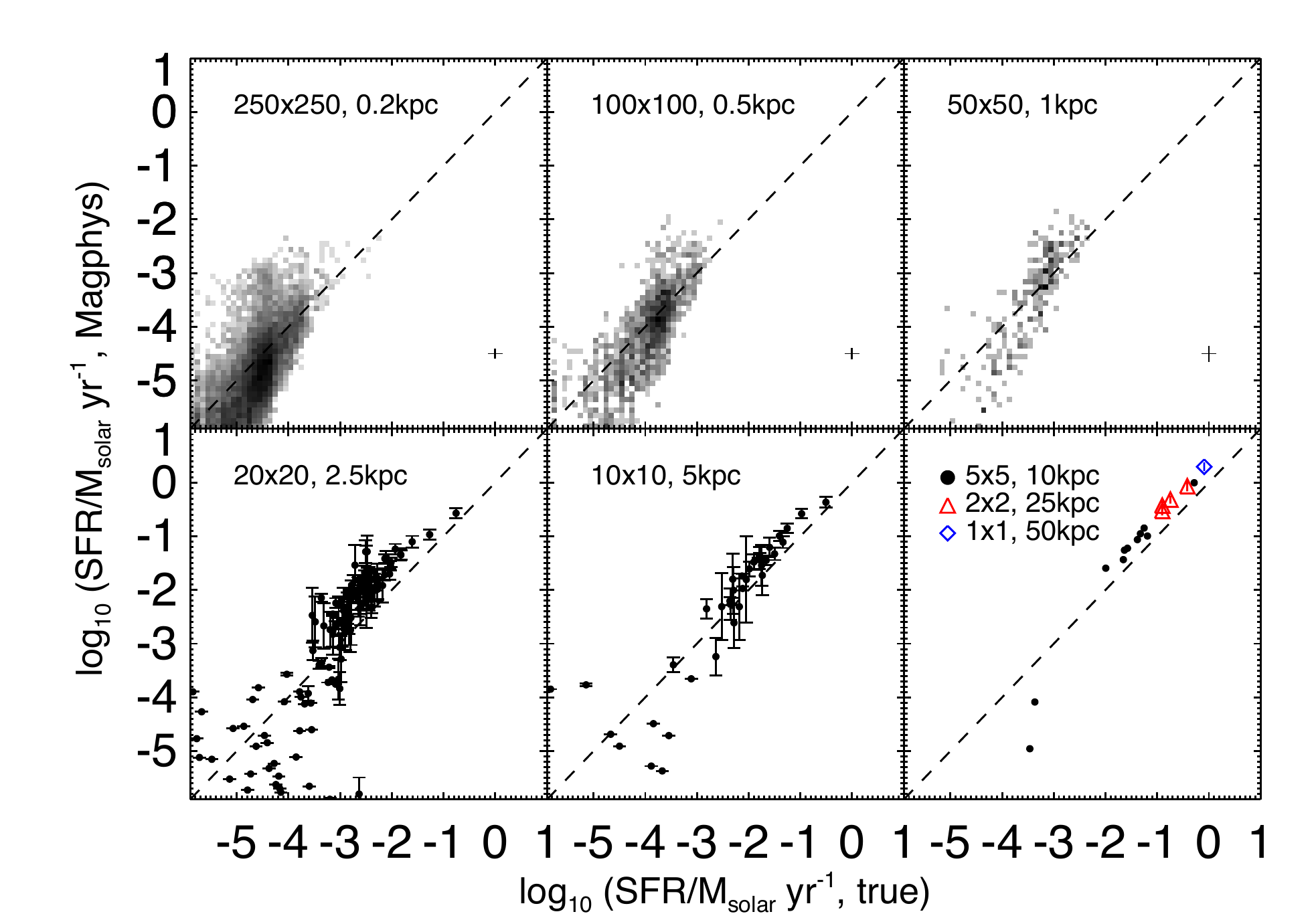}}\\
\vspace{1.0cm}\\
\subfloat[]{\includegraphics[width=\columnwidth,trim=0cm 0cm 0cm 0.8cm,clip]{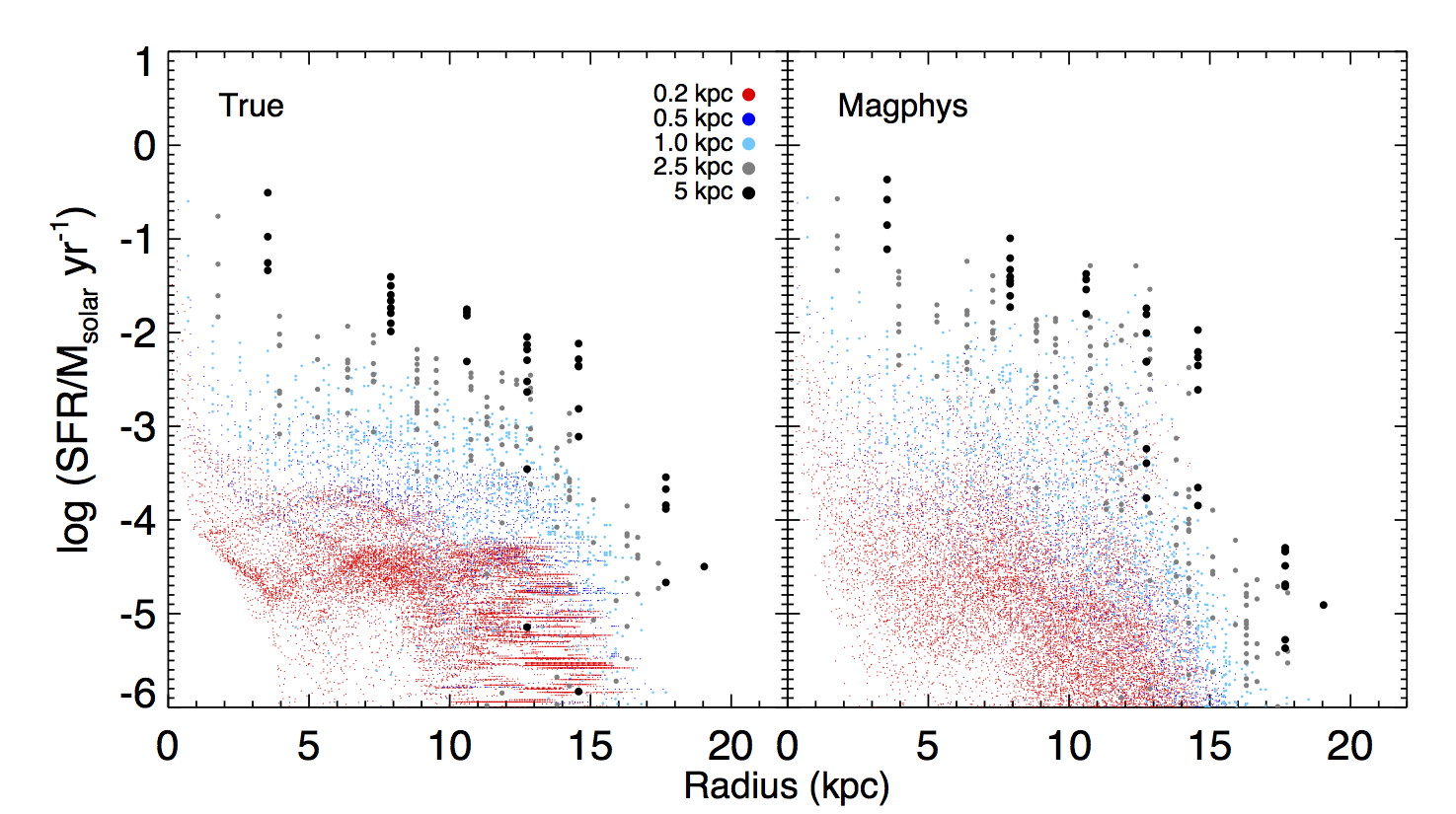}}
\end{minipage}
\caption{Recovery of SFR for camera 0. In panel (a) the colour scales used for the SFR and residual images are as detailed in the bars at the bottom of the corresponding column. For more details, see the caption of Figure \ref{fig:mass_recovery}. The SFRs of individual pixels are typically recovered well for pixel scales $\ga 1$ kpc, although the true values are slightly systematically overestimated, resulting in the systematic overestimate of the total SFR evident in Figure \ref{fig:totals_vs_resolution}. For smaller scales, there is considerable scatter around the one-to-one relation (although the recovery of the total SFR is better at these resolutions; see Figure \ref{fig:totals_vs_resolution}). Moreover, \magphys\ overestimates the SFR in the inter-arm regions. Although the radial dependence of the SFR is approximately recovered, at the highest resolutions, detailed features are lost.}
\label{fig:sfr_recovery}
\end{figure*}

In figure \ref{fig:sfr_recovery} we now compare the \magphys\ SFR estimates with the true values. Figure \ref{fig:sfr_recovery} again reveals that, despite each individual pixel being fit independently of the neighbouring pixels, the spiral arm structure is again quite clear in the recovered SFR map, despite the increased scatter between the \magphys\ and true SFRs shown in the right column of Figure \ref{fig:sfr_recovery}\,(a) and Figure \ref{fig:sfr_recovery}\,(b) relative to the situation for the stellar masses. The spiral arm structures visible in Figure \ref{fig:sfr_recovery}\,(a) are also visible as structures in the true radial SFR values -- shown in the left panel of Figure \ref{fig:sfr_recovery}\,(c) -- though they are washed out in the \magphys\ values. However, the trend of decreasing SFR with increasing radius is clearly apparent in the \magphys\ results.

Though the scatter in individual SFRs is considerable (especially fitting to pixels scales $< 0.5$\,kpc), the total SFR recovered at each resolution compares reasonably well with the true total SFR (0.81 M$_\odot$\,yr$^{-1}$). These values are shown in the second column of plots in Figure \ref{fig:totals_vs_resolution}, where we find a modest offset of 0.1-0.3\,dex between the true values and the \magphys\ estimates \citep[consistent with the values found for a broader range of simulation timesteps in][]{HS15}. We also find evidence for a resolution dependence, with the low spatial resolution values having the largest overestimates of the true total SFR, though the offset is driven primarily by the highest-resolution data point, which we have already identified as potentially problematic in Figure \ref{fig:sfr_recovery}\,(b). This trend of increasing recovered SFR with increasing pixel size remains apparent for each viewing angle to the disc. 

\begin{figure*}
\centering 
\subfloat{\includegraphics[height=0.85\columnwidth, trim=1.9cm 0.6cm 1.3cm 1.3cm, clip=true]{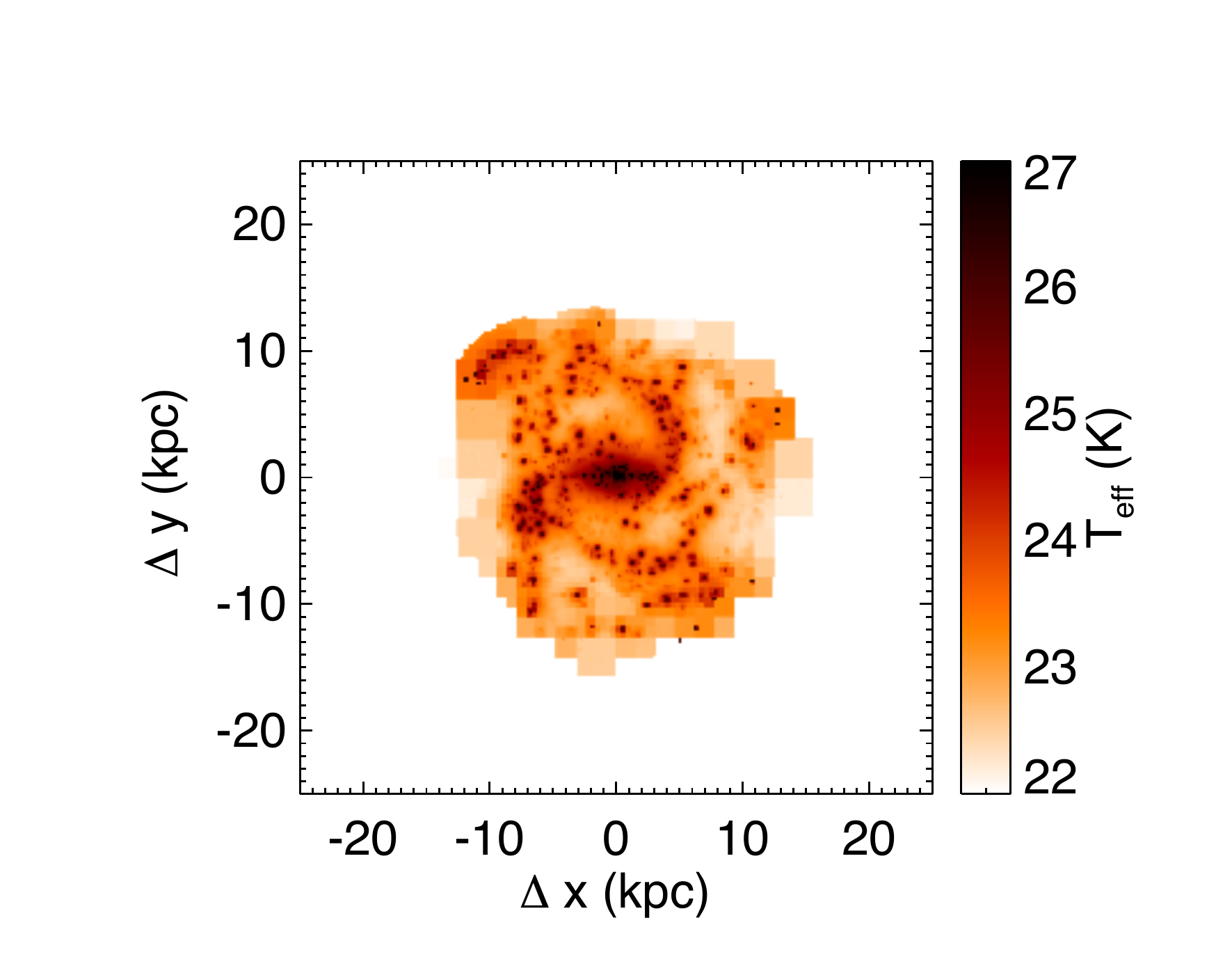}}
\hspace{0.8cm}
\subfloat{\includegraphics[height=0.85\columnwidth, trim=1.9cm 0.6cm 4.3cm 1.3cm, clip=true]{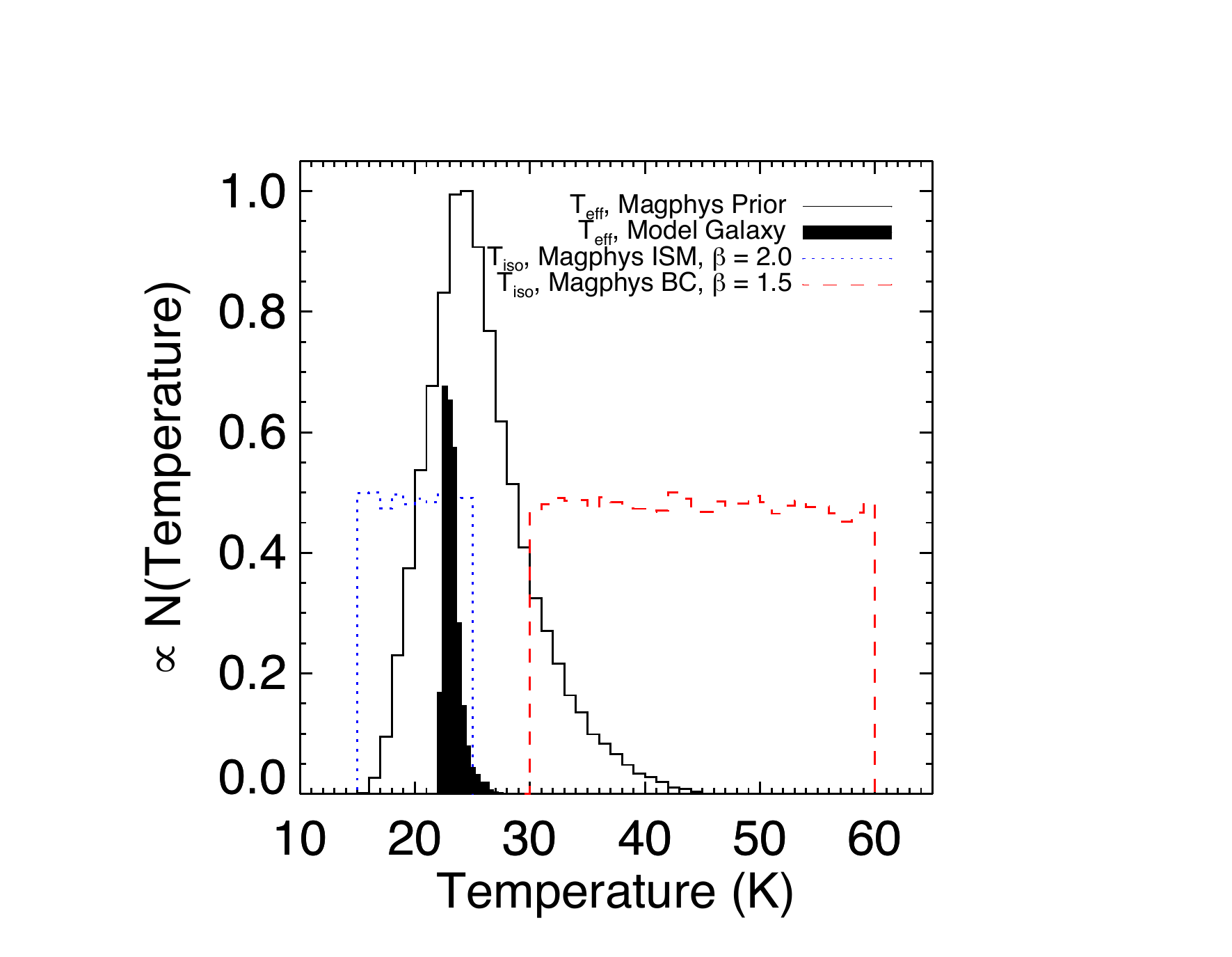}}
\caption{{\bf Left:} Map of the effective temperature of the far-IR dust SED, estimated from the ratio between the MIPS 70 and PACS 160\,$\mu$m model fluxes and assuming an emissivity index of 1.82 from \citet{smith13}. The temperature scaling is indicated by the colour bar to the right, with values over the range $22 < T_{\mathrm{eff}} < 27$\,K, spanning the full range of effective dust temperatures in the disc. The dust around young star particles is hottest, whereas that in inter-arm regions, which is primarily heated by the diffuse interstellar radiation field, is coldest. {\bf Right:} A comparison between histograms showing the range of {\it effective} dust temperatures spanned by the default \magphys\ far-infrared library (black solid line), and the effective temperatures of the SEDs of individual pixels in our highest-resolution simulation (i.e. the values shown in the left panel, shown as the filled black hisogram). Also overlaid are the {\it isothermal} temperature distributions of the hot and cold components of the \magphys\ library (red dashed and blue dotted histograms, respectively). This comparison demonstrates that while the individual components may have a discontinuous prior distribution, the prior of the overall effective dust temperature in \magphys\ is sufficiently broad.}
\label{fig:tmap}
\end{figure*}

Two main regions of discrepancy are apparent. The first is in the outer regions of the spiral arms, where the \magphys\ results suggest a halo of low-level star formation that is not seen in the true values (this feature is particularly visible as an annular excess around the edge of the galaxy in the 3rd and 4th rows of figure \ref{fig:sfr_recovery}\,(a), corresponding to pixel sizes of 1kpc and 2.5kpc, but it is also apparent in the other panels). 
The second is in the inter-arm regions, where \magphys\ returns considerably higher values than the ground-truth. The left-hand panel of Figure \ref{fig:tmap} highlights that these regions are predominantly the parts of the galaxy surface with the lowest effective dust temperatures. To ensure that these discrepancies are not due to the influence of the choice of dust temperature priors in the default \magphys\ library, and in an attempt to marginalise over the multi-component dust model \magphys\ assumes to build the dust SEDs, in the right-hand panel of Figure \ref{fig:tmap}, we show a comparison between the range of {\it effective} dust temperatures of the FIR SEDs of individual pixels of the model galaxy (filled histogram) and those of the individual \magphys\ IR SED templates (outlined histogram), with the priors on the ISM and birthcloud temperatures also overlaid (as blue dotted and red dashed histograms, respectively). The effective temperatures are estimated based on remapping the 70/160\,$\mu$m flux ratio on a pixel-by-pixel basis, assuming an isothermal model with an emissivity index $\beta = 1.82$ following \citet{smith13}, and the values reveal that the prior is sufficiently broad for our purposes.\footnote{This may seem at odds with there being a gap in the distributions of {\it isothermal} temperature for the ISM and birthcloud components shown in figure \ref{fig:tmap}, however recall that the two components have different emissivity indices, and that the \magphys\ far-infrared library -- which is described fully in \citet{dacunha08} -- is more complicated than simply a sum of two components with adjustable temperatures shown in the right panel of figure \ref{fig:tmap}.} It is possible that the dust heating in these regions is being misinterpreted by \magphys\ as from young stars within these pixels rather than the diffuse interstellar radiation field, which is dominated by neighbouring pixels, leading to the overestimated SFRs in these regions. These regions are prime locations where the assumption of \emph{within-pixel} energy balance may break down, since in the lowest-sSFR pixels, stars lying in neighbouring pixels are likely to contribute significantly to the dust heating. As expected, such discrepancies disappear when sufficiently large pixels are employed because as the pixel size is increased, the minimum values in the true SFR maps (ignoring pixels with zero star formation) increases and thus the within-pixel energy balance assumption becomes increasingly valid.

\subsection{\magphys\ recovery of sSFR}

\begin{figure*}
\centering
\subfloat[]{
\begin{minipage}{0.4\textwidth}
\includegraphics[width=\textwidth,trim=0cm 0cm 0cm 0cm,clip]{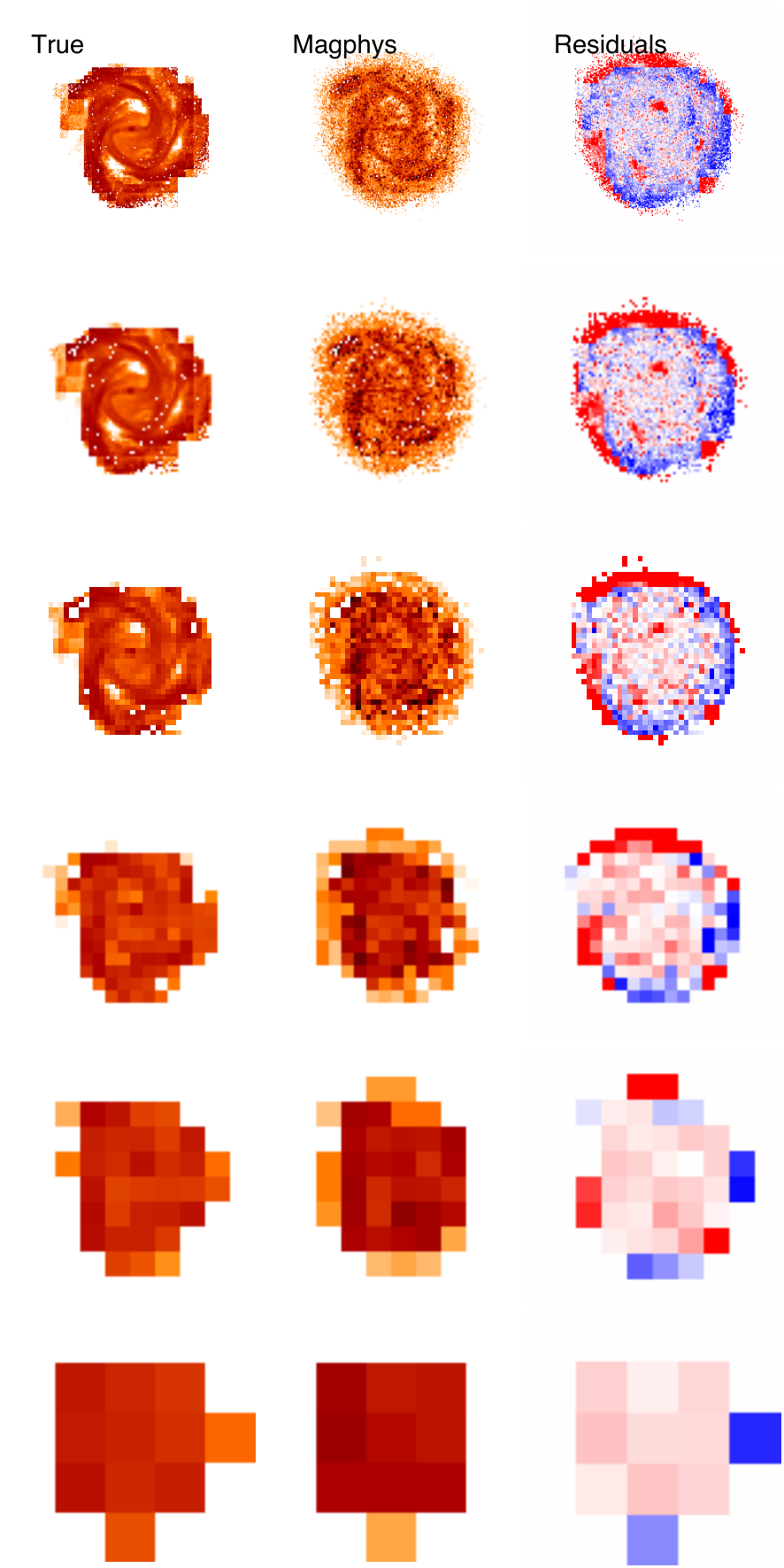} \\
\includegraphics[width=\textwidth,trim=0cm 1.2cm 0cm 2.4cm]{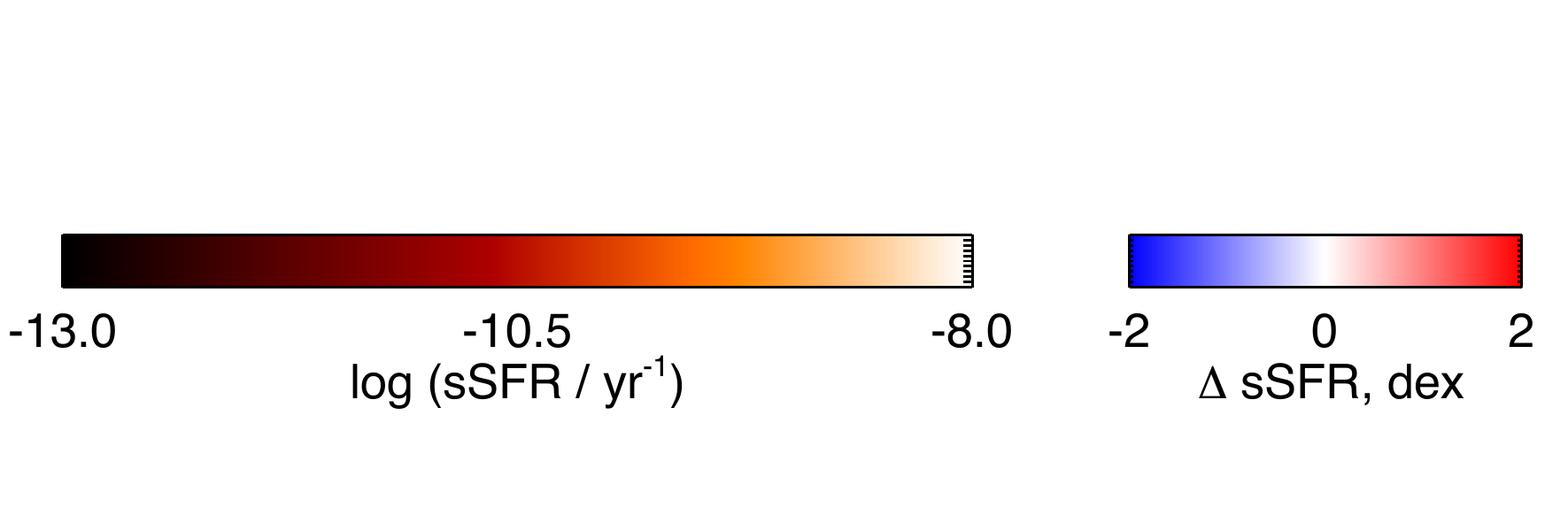}
\end{minipage}
}
\begin{minipage}{0.6\textwidth}
\subfloat[]{\includegraphics[width=\columnwidth,trim=0cm 0cm 0cm 0.85cm,clip]{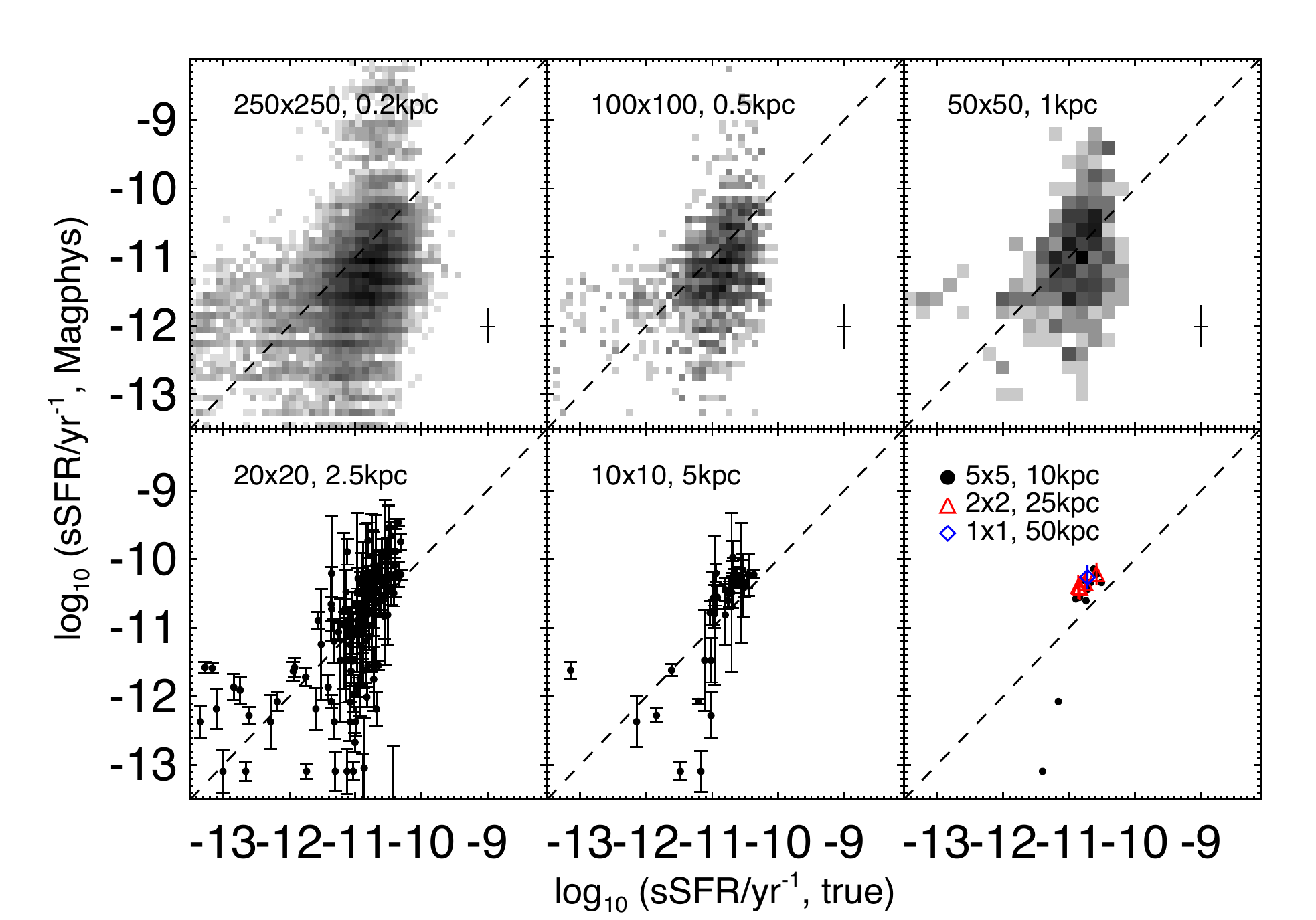}}\\
\vspace{1.0cm}\\
\subfloat[]{\includegraphics[width=\columnwidth,trim=0cm 0cm 0cm 0.8cm,clip]{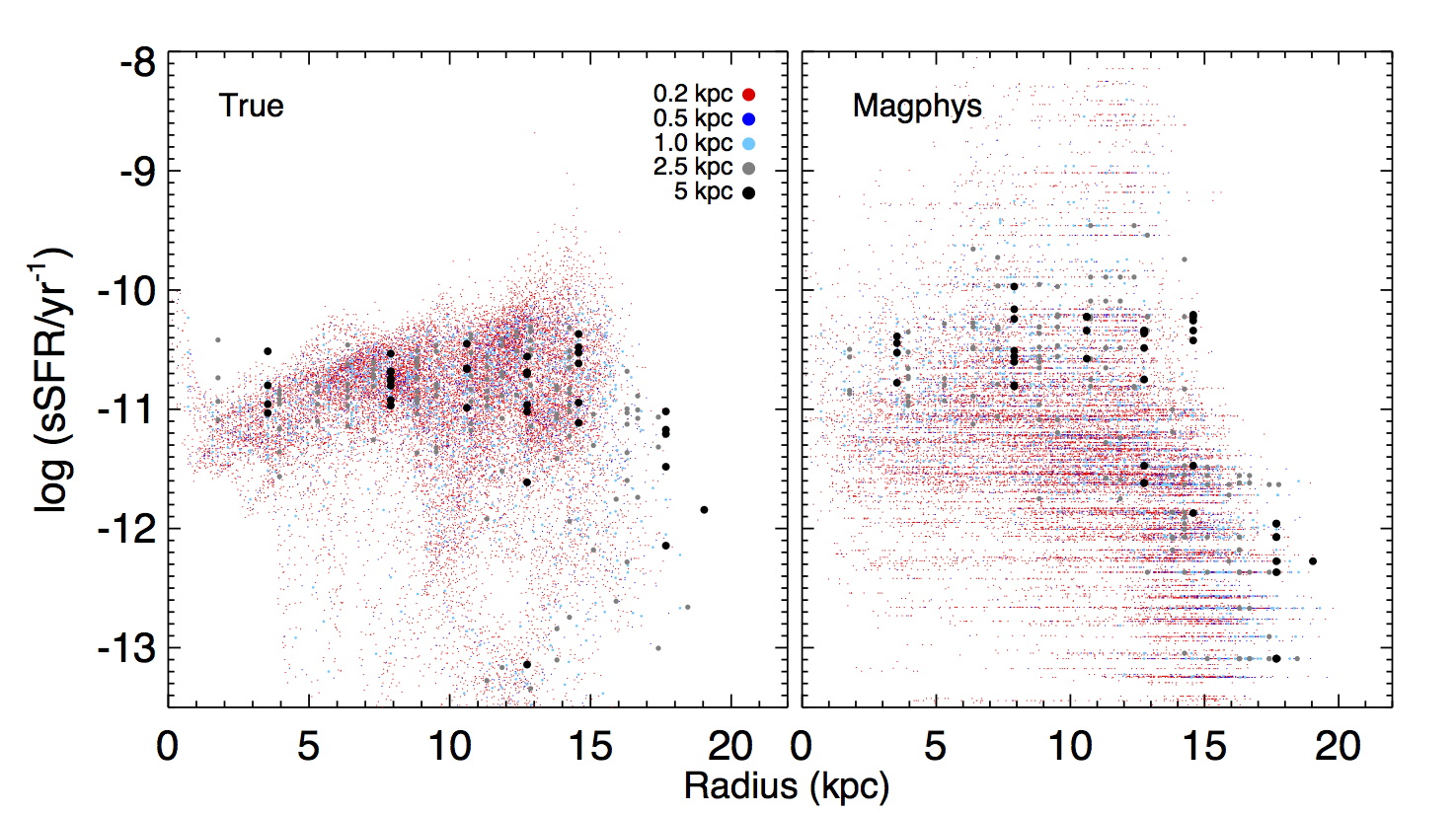}}
\end{minipage}
\caption{Recovery of sSFR for camera 0. The image colour scales for the images are as detailed in the bars underneath the corresponding columns, corresponding to $-13 < \log($sSFR/yr$^{-1}) < -8$ and $-2 < \Delta$ sSFR\ $< 2$\,dex. For more details, see the caption of Figure \ref{fig:mass_recovery}. For pixel sizes less than a few kpc, the scatter in the sSFR values recovered by \magphys\ at fixed true sSFR is large. At lower resolutions, the sSFR is recovered reasonably well, although there is a systematic overestimate of the sSFR, which is caused by the overestimate of the SFR discussed above. The relatively flat sSFR profile is qualitatively recovered by \magphys\, although there is larger scatter at fixed radius in the recovered values compared with the true ones.}
\label{fig:ssfr_recovery}
\end{figure*}

Having investigated the recovery of the \magphys\ SFR and stellar mass in the previous sections, we now investigate the recovery of the specific star formation rates. The spiral arm structures are visible in the maps of the \magphys\ values by virtue of their elevated sSFR relative to the nuclear and inter-arm regions in Figure \ref{fig:ssfr_recovery}\,(a), while Figures \ref{fig:ssfr_recovery}\,(b) and (c) highlight the very large spread in the recovered values about the 1:1 relation, though this is most apparent in the highest-resolution tests with pixels $< 1$\,kpc in size. The values determined for the low-spatial resolution cases (pixels $\ge 1$\,kpc) are more consistent with the true values taken from the simulations.
 
The inter-arm and especially the outer regions of the disc that were also problematic for the SFR recovery represent some of the most challenging areas. These are the regions where the effective dust temperature is lowest, and where the energy balance criterion is perhaps more of a hindrance due to the importance of neighbouring regions for the dust energy budget.

The plots in the centre column of Figure \ref{fig:totals_vs_resolution} show the resolution dependence of the effective sSFR (obtained by dividing the total SFR by the total stellar mass, with both estimated by summing over all of the good fits at each of the resolutions). As expected based on the stellar mass and SFR values \citep[and based on the results of][]{HS15} we find evidence for a modest offset of up to $\sim$0.3\,dex, and possible evidence for a resolution-dependent bias, however -- as for the SFR -- this is primarily driven by the highest-resolution data points.

\subsection{\magphys\ recovery of dust mass, $M_\mathrm{dust}$}

\begin{figure*}
\centering
\subfloat[]{
\begin{minipage}{0.4\textwidth}
\includegraphics[width=\textwidth,trim=0cm 0cm 0cm 0cm,clip]{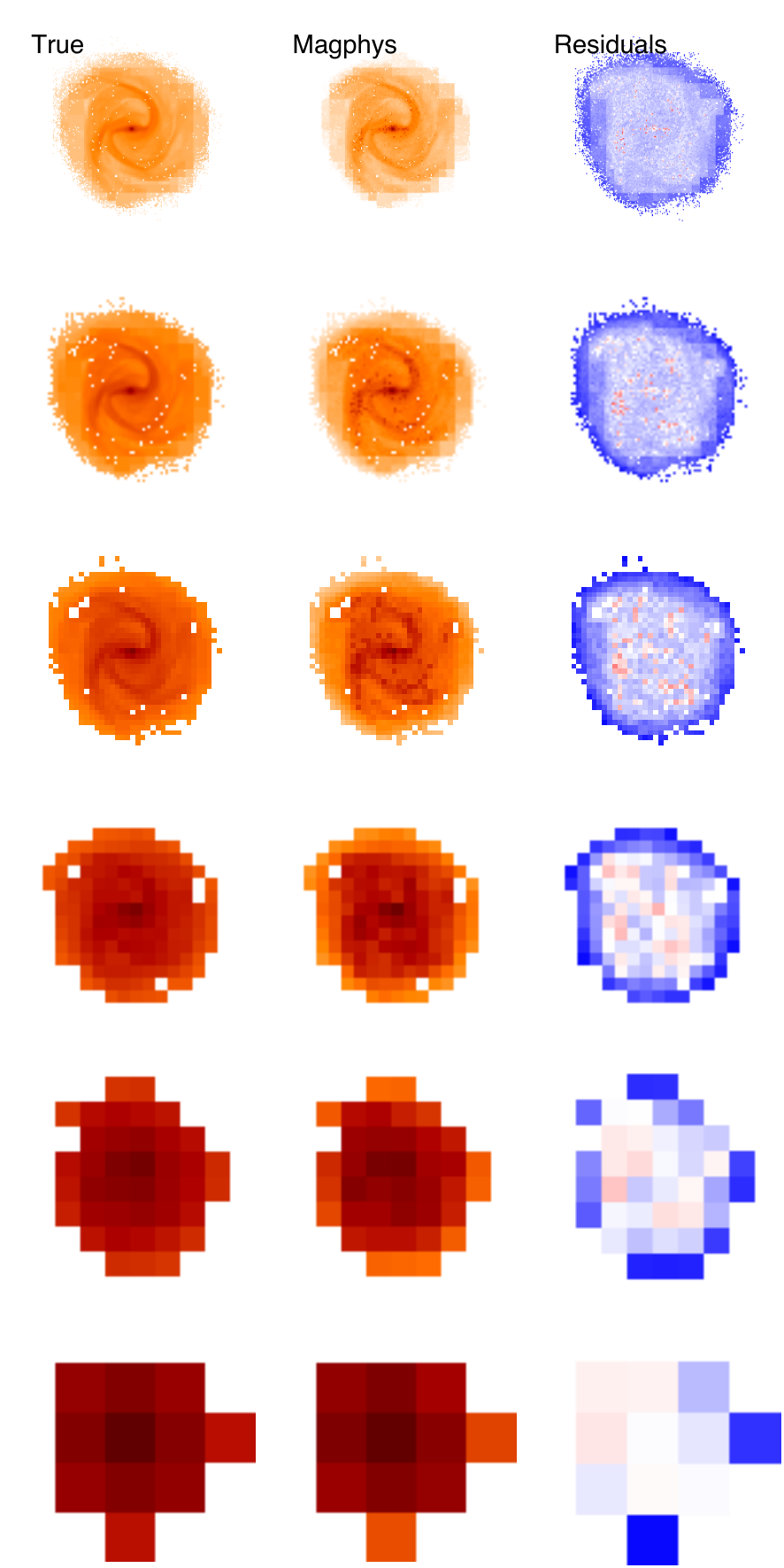} \\
\includegraphics[width=\textwidth,trim=0cm 1.2cm 0cm 2.4cm]{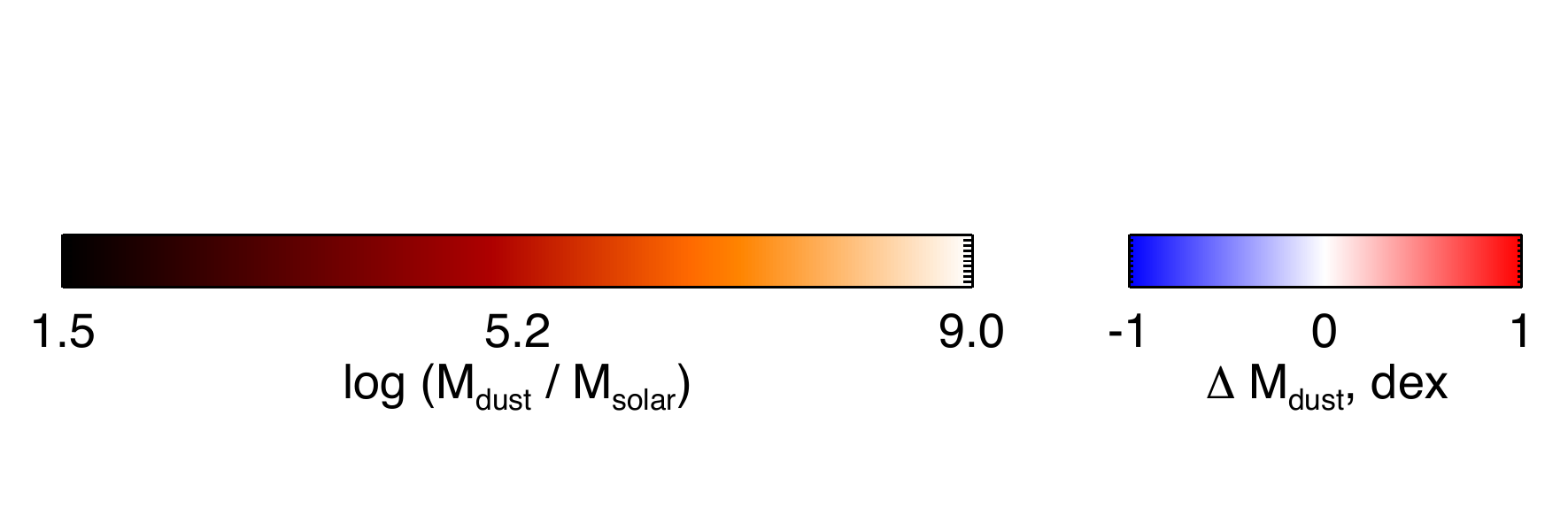}
\end{minipage}
}
\begin{minipage}{0.6\textwidth}
\subfloat[]{\includegraphics[width=\columnwidth,trim=0cm 0cm 0cm 0.85cm,clip]{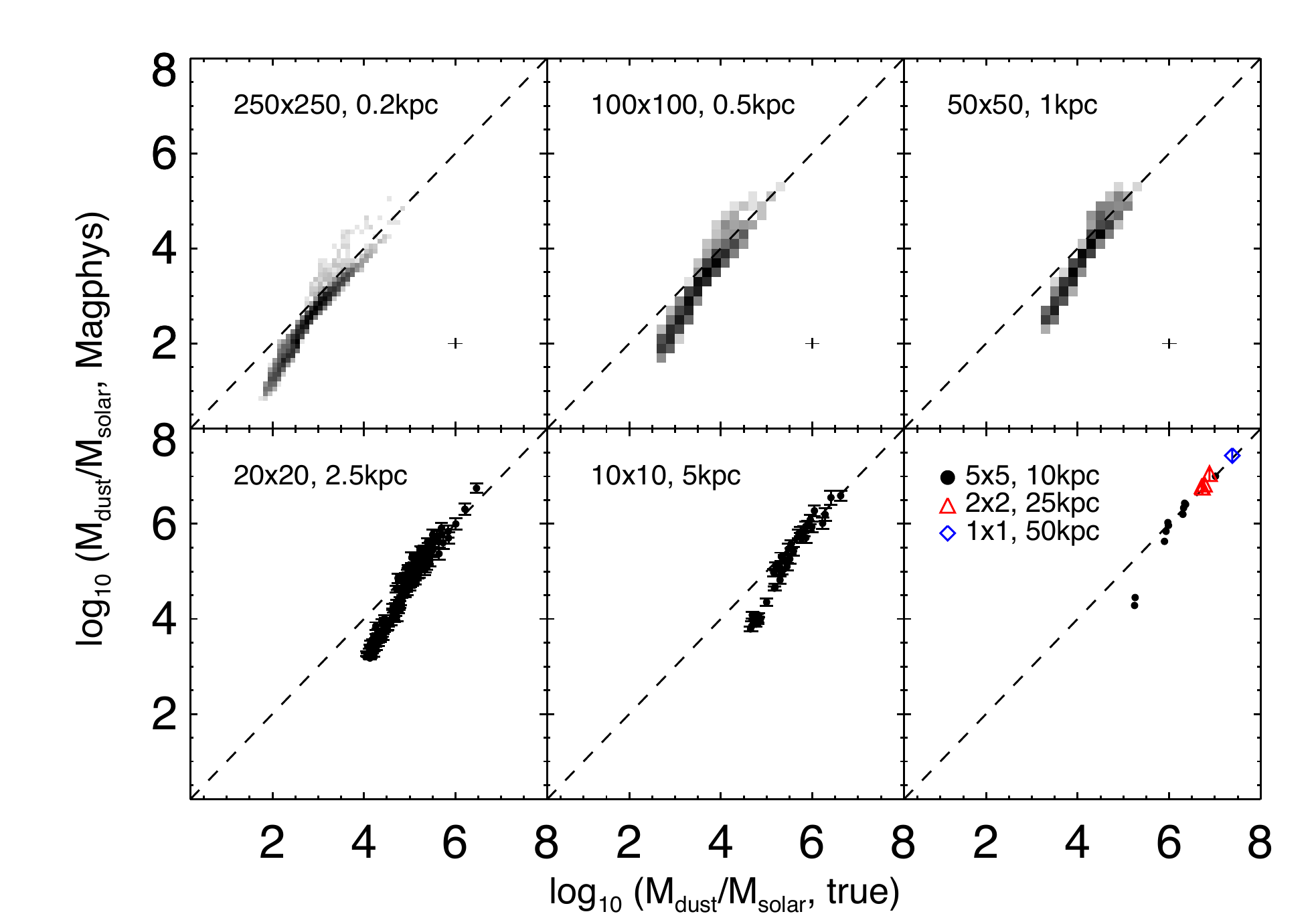}}\\
\vspace{1.0cm}\\
\subfloat[]{\includegraphics[width=\columnwidth,trim=0cm 0cm 0cm 0.8cm,clip]{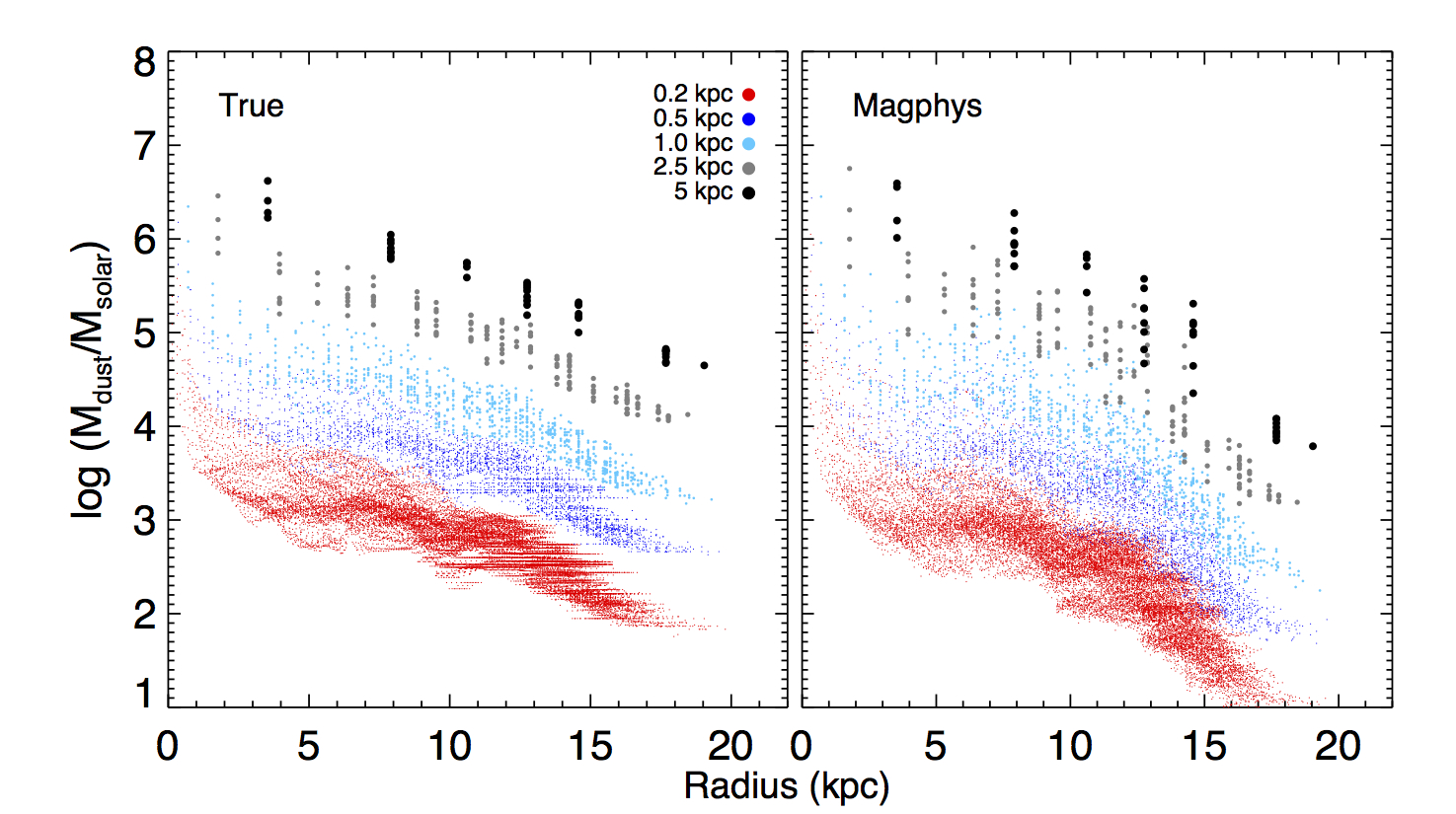}}
\end{minipage}
\caption{Recovery of dust mass for camera 0. The colour scales for the images showing the true and \magphys-estimated mass distributions are given by the colour bars at the bottom of each column of panel (a). For more details, see the caption of Figure \ref{fig:mass_recovery}. The details of the dust mass maps (e.g. spiral arms) are well recovered by \magphys, as are the radial profiles. At all resolutions, the dust masses contained in individual pixels are recovered very well, except for the lowest-dust-mass pixels, for which the dust mass is systematically overestimated.}
\label{fig:mdust_recovery}
\end{figure*}

We now consider how well \magphys\ is able to recover the spatial distribution of the dust mass\footnote{As in \citet{HS15}, the dust emissivities assumed by the two codes differ; in MAGPHYS, the emissivity is normalized by $\kappa_{850\mu m} = 0.77$\,g$^{-1}$\,cm$^2$ following \citet{dunne00}, whereas the MW dust model used in the simulations has $\kappa_{850\mu m} = 0.38$\,g$^{-1}$\,cm$^2$. Consequently, we multiply the dust masses output by MAGPHYS by 2 to account for this difference}. Figure \ref{fig:mdust_recovery}\,(a) reveals that \magphys\ performs very well at this task, with the spiral structures visible in the true dust mass maps apparent and well-defined in the \magphys\ estimates. \magphys\ is able to estimate dust masses as low as 10\,M$_\odot$ in the outer regions of the highest-resolution maps. Although there is little or no evidence for bias in the overall dust mass estimates, or in the more massive pixels (irrespective of their size), Figures \ref{fig:mdust_recovery}\,(a) and \ref{fig:mdust_recovery}\,(b) highlight that there is a mass-dependent bias, such that the lowest dust masses are underestimated by $\sim 1$\,dex. Figure \ref{fig:mdust_recovery}\,(a) shows that the regions where this occurs are primarily in the outer regions of the galaxy, which are bright blue in the residual maps; these pixels are located at radial distances $> 10$\,kpc, where Figure \ref{fig:tmap} indicates that the effective dust temperature is lowest. This discrepancy is not due to the \magphys\ dust temperature priors, which include models for the ambient ISM as cold as 15\,K,  considerably colder than the coldest regions of the isolated disc simulation that we use in this work (after accounting for the difference in $\beta = 1.82$ used by \magphys and the intrinsic $\beta = 2.0$ of the dust grain model assumed in the radiative transfer calculations). Remarkably, the spiral structures visible in the true radial distributions of dust mass within $10\,$kpc at the highest spatial resolution (shown as the red points in Figure \ref{fig:mdust_recovery}\,c) are also apparent in the \magphys estimates.

The total dust mass recovered estimated as a function of spatial resolution is shown in Figure \ref{fig:totals_vs_resolution}, revealing that at spatial scales $\gtsim 1$\,kpc, the total dust mass is recovered well, while at higher resolution (i.e. smaller pixel sizes) the total dust mass is slightly underestimated by around 0.15\,dex, presumably due to the large number of low-dust-mass pixels at large distances from the nucleus.

\subsection{\magphys\ recovery of non-standard parameters}

In this section we analyse the extent of \magphys' ability to recover the additional parameters -- metallicity, $A_V$ and mass-weighted age -- for which modification of the default version of the code was required. Sections \ref{subsec:metallicity_recovery} -- \ref{subsec:age_recovery} represent exploratory tests, using the same methodology as the previous sections, of how well the \magphys\ formalism can obtain reliable estimates of these parameters. 

\subsubsection{\magphys\ recovery of stellar metallicity}
\label{subsec:metallicity_recovery}

\begin{figure*}
\centering
\subfloat[]{
\begin{minipage}{0.4\textwidth}
\includegraphics[width=\textwidth,trim=0cm 0cm 0cm 0cm,clip]{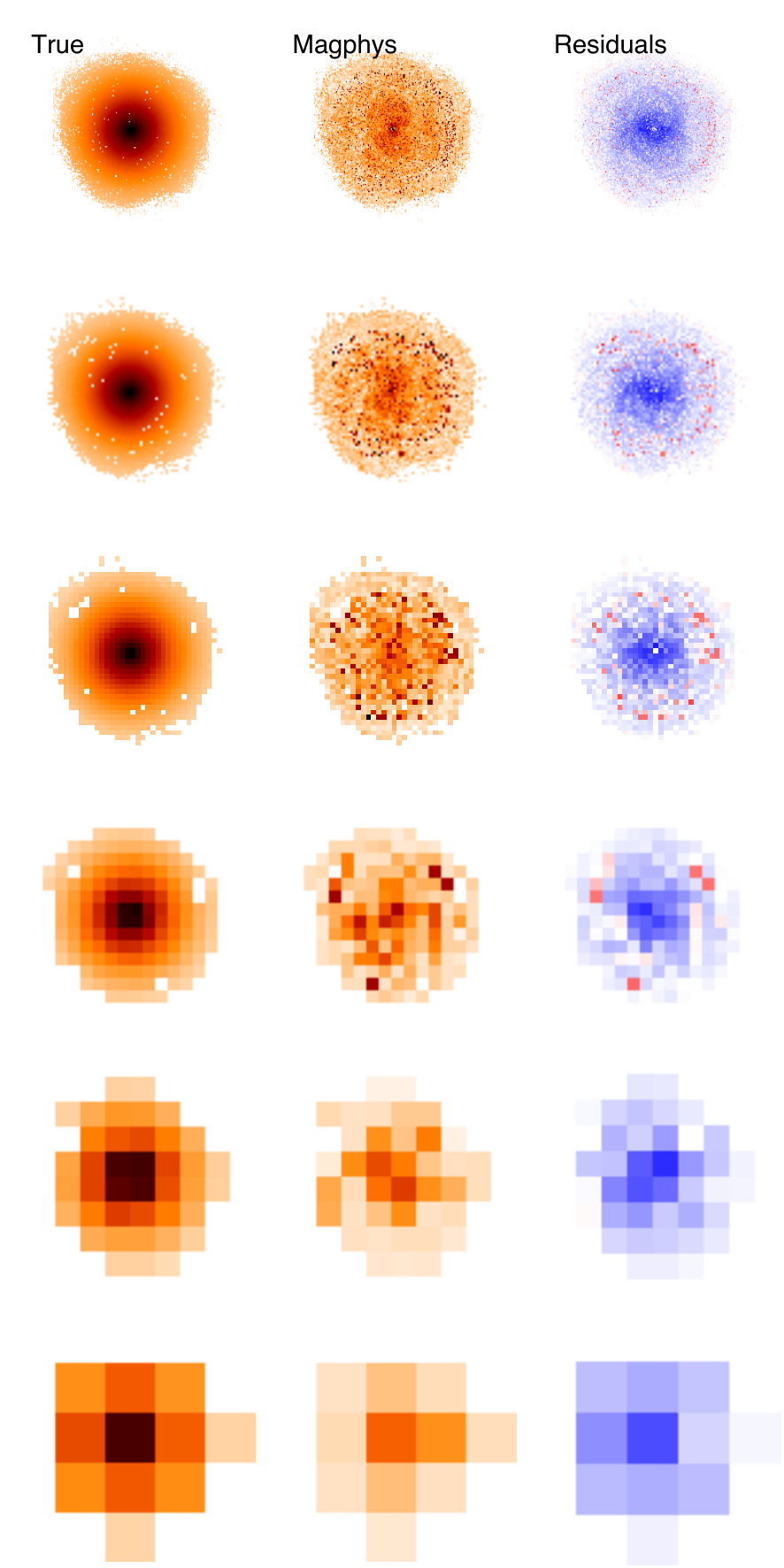} \\
\includegraphics[width=\textwidth,trim=0cm 1.2cm 0cm 2.4cm]{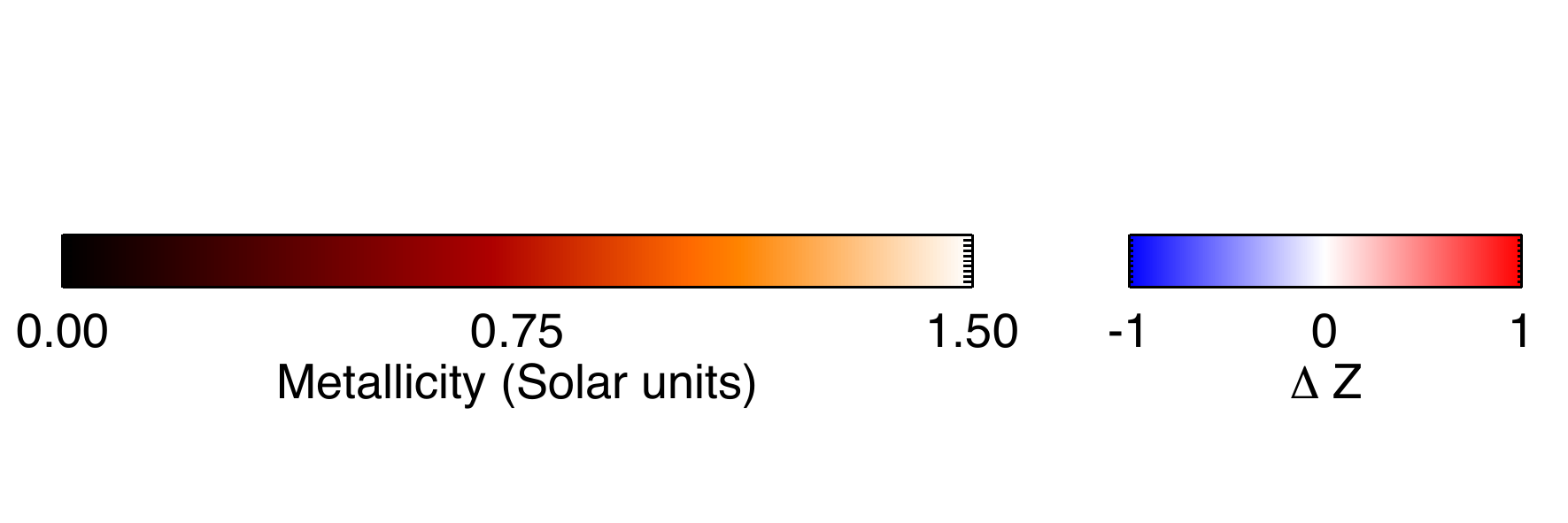}
\end{minipage}
}
\begin{minipage}{0.6\textwidth}
\subfloat[]{\includegraphics[width=\columnwidth,trim=0cm 0cm 0cm 0.85cm,clip]{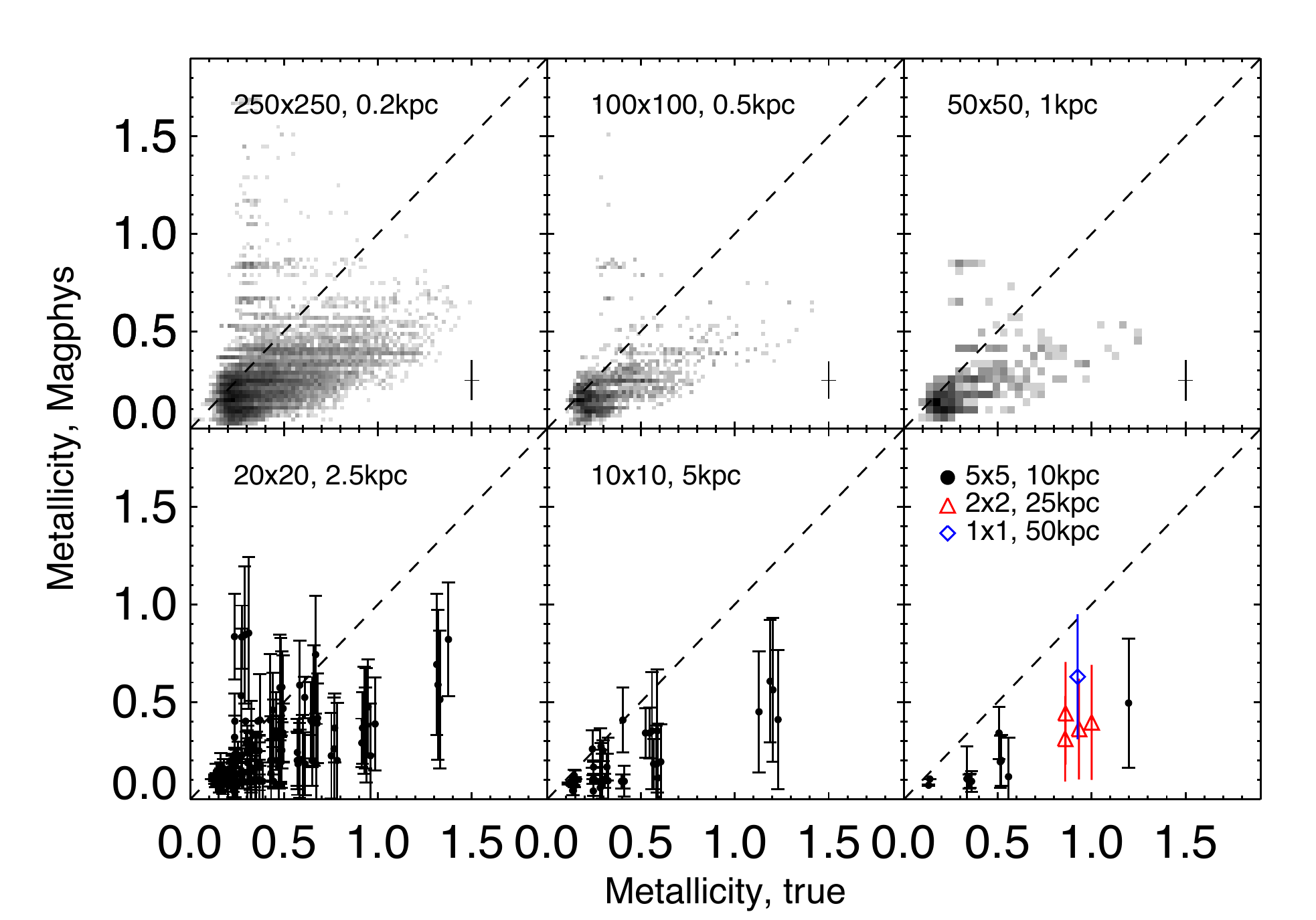}}\\
\vspace{1.0cm}\\
\subfloat[]{\includegraphics[width=\columnwidth,trim=0cm 0cm 0cm 0.8cm,clip]{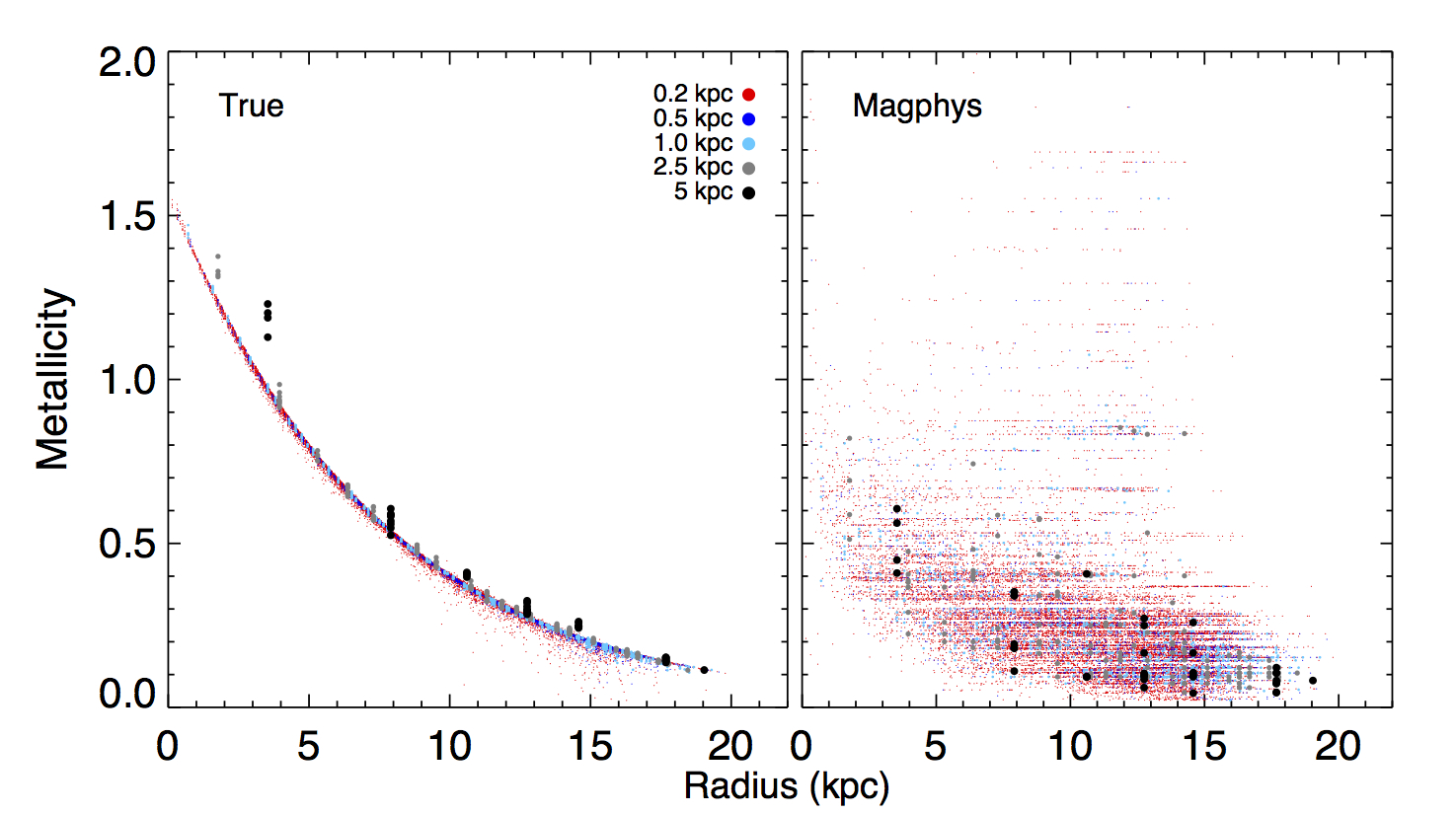}}
\end{minipage}
\caption{Recovery of metallicity for camera 0. The colour scales for each of the metallicity maps in panel (a) are given by the corresponding colour bars at the bottom of each column. For more details, see the caption of Figure \ref{fig:mass_recovery}. The effective metallicity is systematically underestimated by \magphys\ at all spatial resolutions, and there is considerable scatter in the recovered metallicity at fixed true metallicity (as also evident from the noisiness of the metallicity maps). Although \magphys\ correctly infers that the metallicity decreases with radius, the detailed shape is not well recovered. Overall, these results demonstrate that recovering metallicity from even noiseless photometry that densely samples the UV through FIR is very challenging.}
\label{fig:z_recovery}
\end{figure*}

Figure \ref{fig:z_recovery}\,(a) shows a comparison between metallicity maps taken from the simulations and those obtained from the \magphys\ estimates. It is immediately clear that the \magphys\ estimates (centre column) are noisier than the true values (left column). Although the general radial dependence is apparent in both the images and in Figure \ref{fig:z_recovery}\,(c), the recovered values at fixed radius exhibit larger scatter than the true approximately exponential profile, with the largest discrepancy in the central regions (see the right column of panel a).

Figure \ref{fig:z_recovery}\,(b) shows the relationship between the true and recovered metallicities for each of the different spatial resolutions; the \magphys\ values are nearly always underestimates (reflected by the predominantly blue residual maps in figure \ref{fig:z_recovery}\,a), and they exhibit considerable scatter at fixed true metallicity. This is at odds with the results of \citet[see figure 9 therein]{dye08}, who were able to recover the metallicity of a suite of mock composite stellar populations using only optical through near-infrared photometry. The discrepancy between our results is perhaps due to the far greater degree of realism in the current work (for example, the metallicity and degree of attenuation vary across the simulated disc employed in this work, and we employ full radiative transfer to produce the emergent SEDs). The difficulty in recovering metallicities using optical photometry alone is well demonstrated by \citet[their figure 11]{pacifici12}, who employed a suite of model galaxies from a semi-analytic model, which exhibit complex star formation histories, and \citet{mentuch12} further highlighted the difficulty of this task even using panchromatic data (albeit without energy balance). The \magphys\ method of recovering metallicities is therefore potentially of great interest, since as mentioned in section \ref{S:intro}, the inclusion of far-infrared data can potentially break the well known degeneracy between age and metallicity that affects photometry at optical\slash near-infrared wavelengths. However, this method is clearly not without its shortcomings, even though we are able to recover the general trend of decreasing metallicity with increasing distance from the nucleus (but not the detailed radial profile). We intend to further investigate the ability of \magphys\ to estimate metallicities from photometry using spectroscopic samples in a forthcoming investigation. 

Figure \ref{fig:totals_vs_resolution} shows the resolution dependence of the effective metallicity, calculated by summing the product of the metallicity and the stellar mass in each resolution element, and dividing the total by the total stellar mass (and propagating the uncertainties). The metallicity estimates show little evidence for variation as a function of resolution, with the values being broadly consistent within the error bars for the most face-on viewing angles. The one exception is the edge-on view, which is suggestive of a slight tendency to underestimate the metallicity in the unresolved case, though it is hard to find any conclusive evidence given the size of the error bars on the total values.

\subsubsection{\magphys\ recovery of $V$-band attenuation, $A_V$}
\label{subsec:av}

\begin{figure*}
\centering
\subfloat[]{
\begin{minipage}{0.4\textwidth}
\includegraphics[width=\textwidth,trim=0cm 0cm 0cm 0cm,clip]{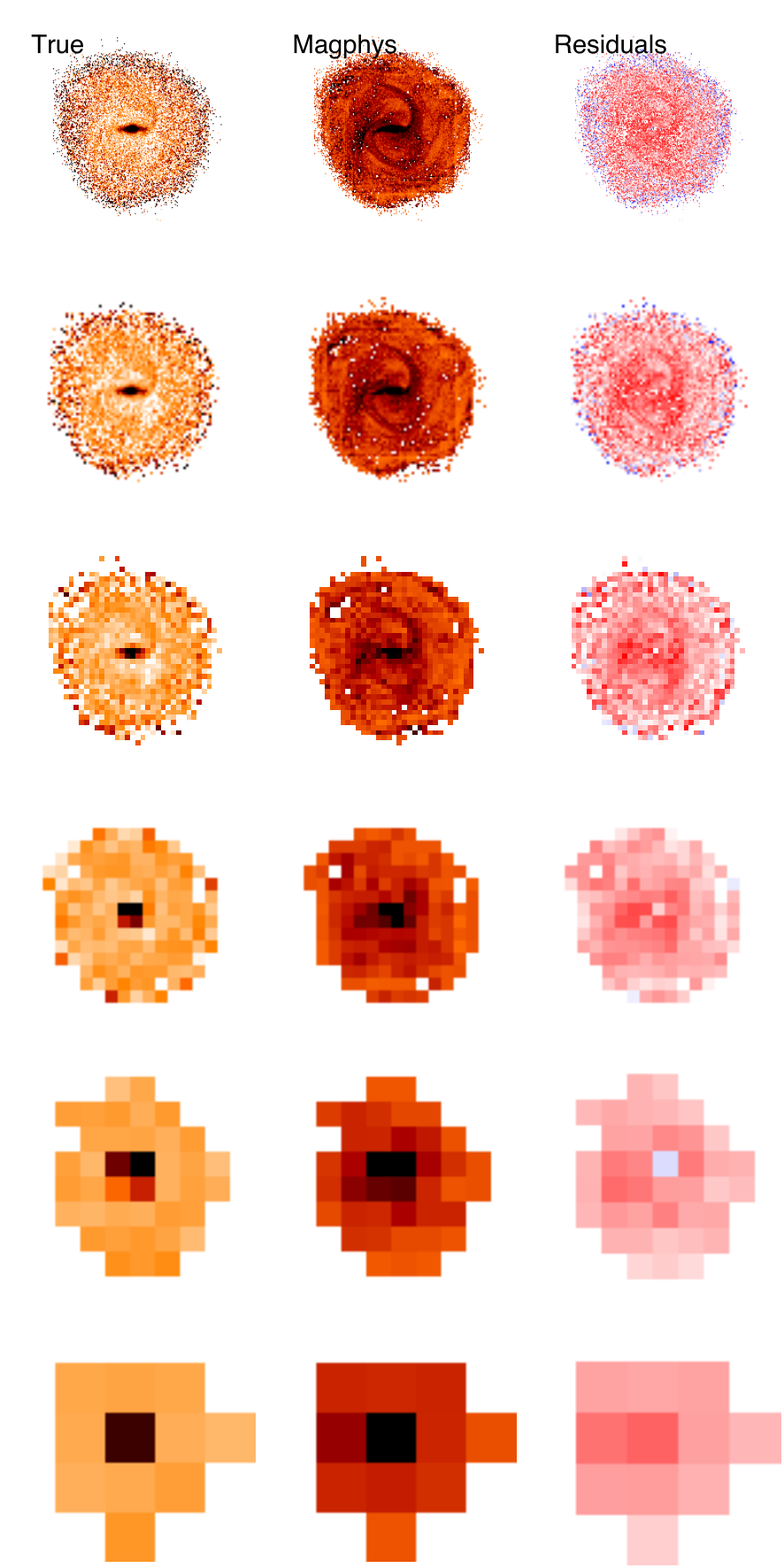} \\
\includegraphics[width=\textwidth,trim=0cm 1.2cm 0cm 2.4cm]{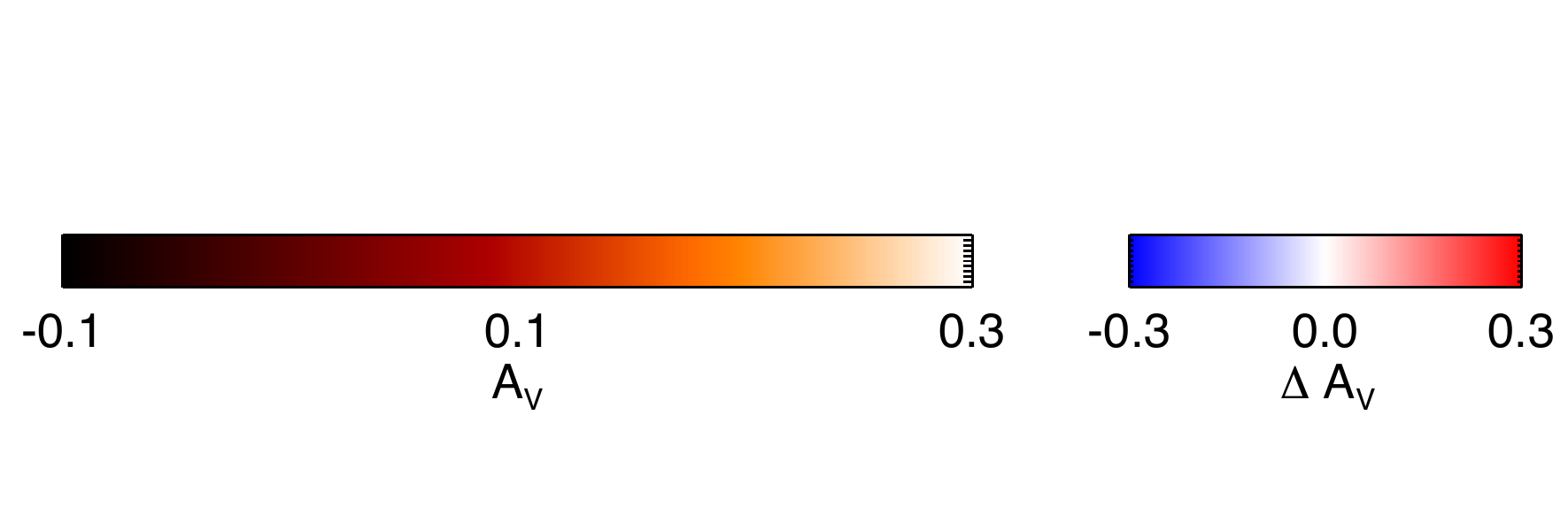}
\end{minipage}
}
\begin{minipage}{0.6\textwidth}
\subfloat[]{\includegraphics[width=\columnwidth,trim=0cm 0cm 0cm 0.85cm,clip]{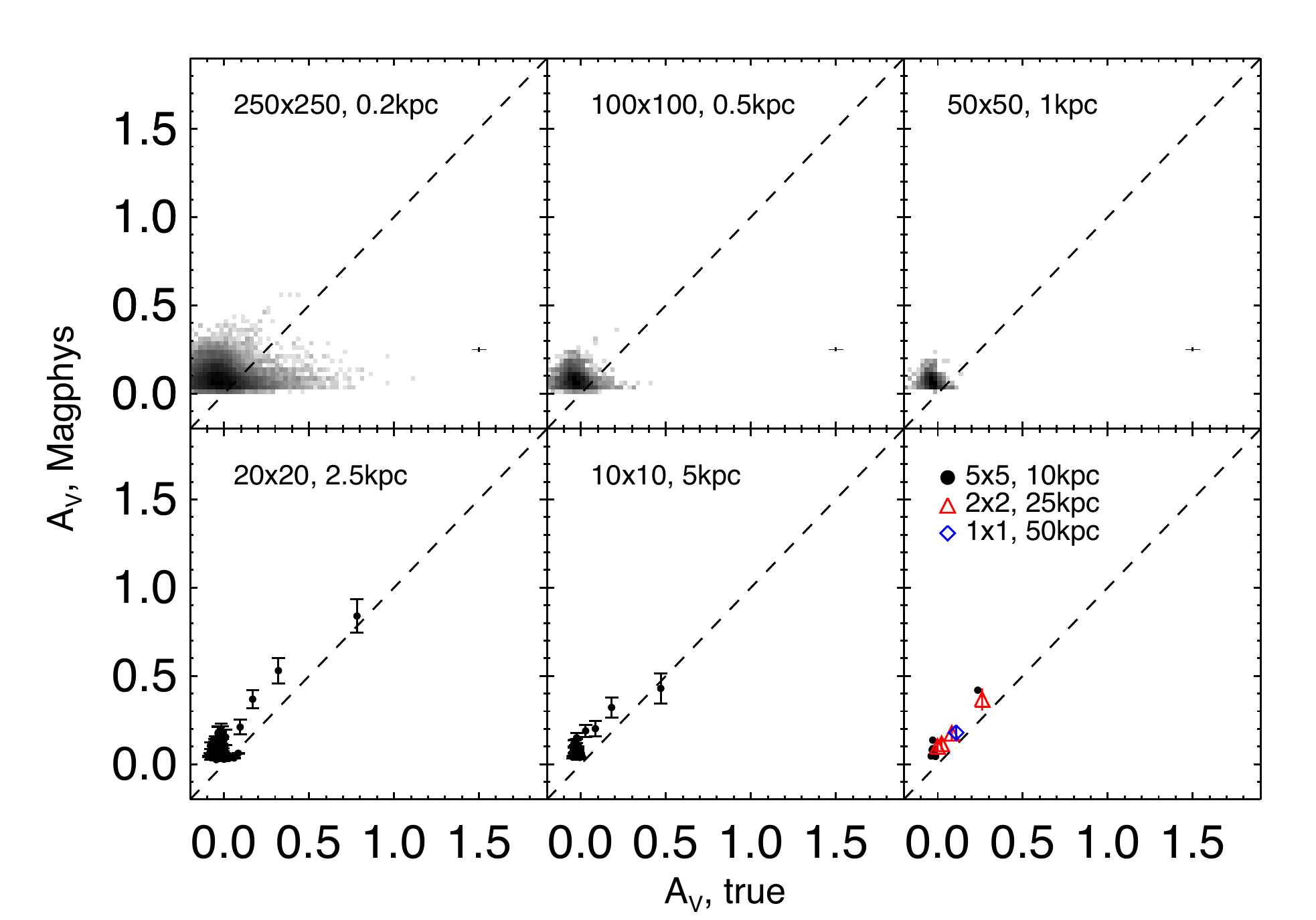}}\\
\vspace{1.0cm}\\
\subfloat[]{\includegraphics[width=\columnwidth,trim=0cm 0cm 0cm 0.8cm,clip]{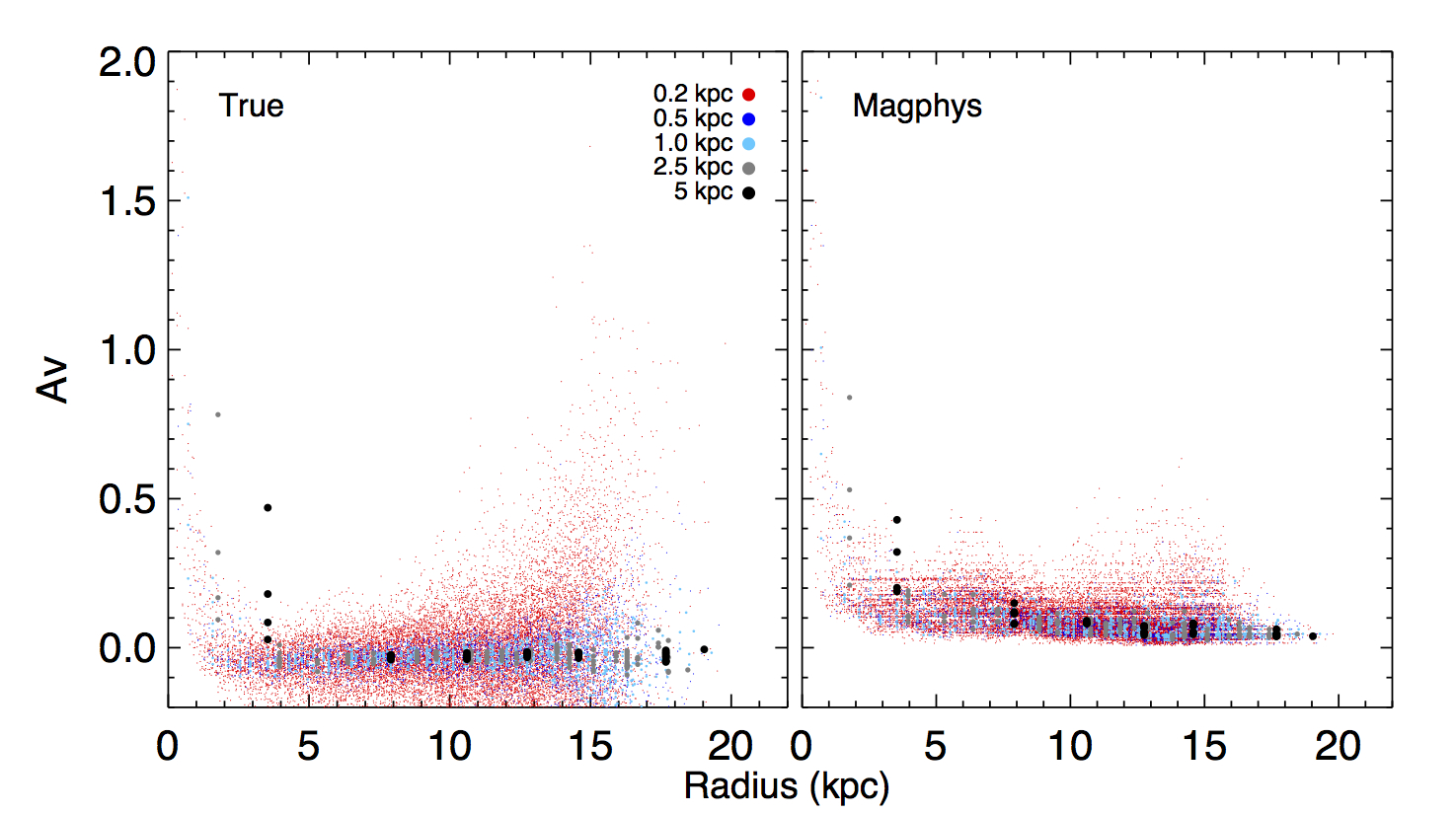}}
\end{minipage}
\caption{Recovery of $A_V$ for camera 0. As in previous figures, the colour bars showing the scale for each map of $A_V$ (true, \magphys and difference, from left to right in panel a) are located at the bottom of the corresponding columns. For more details see the caption of Figure \ref{fig:mass_recovery}. The fact the $A_V$ is highest in the nucleus is
recovered by \magphys, but the contrast in $A_V$ between the spiral arms and inter-arm regions is overestimated. For pixel sizes greater than a few kpc, $A_V$ is reasonably well recovered, albeit slightly
overestimated (by $\sim 0.1$ mag). The negative $A_V$ values present in the simulations (which are due to scattered light and, in the low surface-brightess outskirts, can also be an artifact of Monte Carlo noise in
the radiative transfer calculations), cannot be recovered by \magphys\ owing to its assumed prior for $A_V$.}
\label{fig:av_recovery}
\end{figure*}

In the three sub-plots of Figure \ref{fig:av_recovery} we show the extent of our ability to estimate the visual extinction as a function of position across the disc. From examining Figure \ref{fig:av_recovery} it is clear that \magphys\ systematically overestimates the true $A_V$, however in many cases it is clear that this is because the true values are negative. This feature of the simulations comes about for two primary reasons. Firstly, there is a physical effect (in both the simulations and in the real Universe): scattering of photons means that the emergent spectrum can sometimes be brighter than the intrinsic spectrum at a given position \citep{WiseSilva96,BaesDejonghe02}. Secondly, in low-surface brightness regions of the simulation, Monte Carlo noise can cause the emergent luminosity to be greater than the emitted luminosity (in the limit in which the luminosity is less than that carried by a single photon packet). Figure \ref{fig:av_recovery}\,(a) shows once more that \magphys\ is able to broadly recover the trend of $A_V$ being largest in the nuclear region, and this feature is also visible in Figure \ref{fig:av_recovery}\,(c). While the high-resolution panels of Figure \ref{fig:av_recovery}\,(b) show that \magphys\ is not especially effective at recovering low $A_V$, the situation is much improved at pixel sizes $\gtsim 2.5$\,kpc, where we are able to obtain excellent $A_V$ estimates. At these lower resolutions, the scattering and small number Monte-Carlo effects are much less noticeable since the individual pixels contain more flux, and it is therefore not surprising that our $A_V$ estimates are much more effective in this regime. The aforementioned effects may also explain the erroneous appearance of enhanced spiral arm structure in the high-resolution images of A$_{V}$ values in Figure \ref{fig:av_recovery}\,(a). 
That the pixels with negative $A_V$ are comparatively faint (they all have less than 1 per cent of the peak $V$ band flux density), offers further support to our hypothesis.

We therefore conclude that \magphys\ is reasonably effective at recovering trends in visual extinction within the simulated galaxies, albeit with a residual bias of up to 0.2 magnitudes, primarily dominated by regions with $A_V^\mathrm{true} \ltsim 0$. It is, however, able to recover the increased obscuration in the nuclear region of this isolated disc galaxy irrespective of the spatial scales sampled.

\subsubsection{\magphys\ recovery of mass-weighted age, Age$_M$}
\label{subsec:age_recovery}

\begin{figure*}
\centering
\subfloat[]{
\begin{minipage}{0.4\textwidth}
\includegraphics[width=\textwidth,trim=0cm 0cm 0cm 0cm,clip]{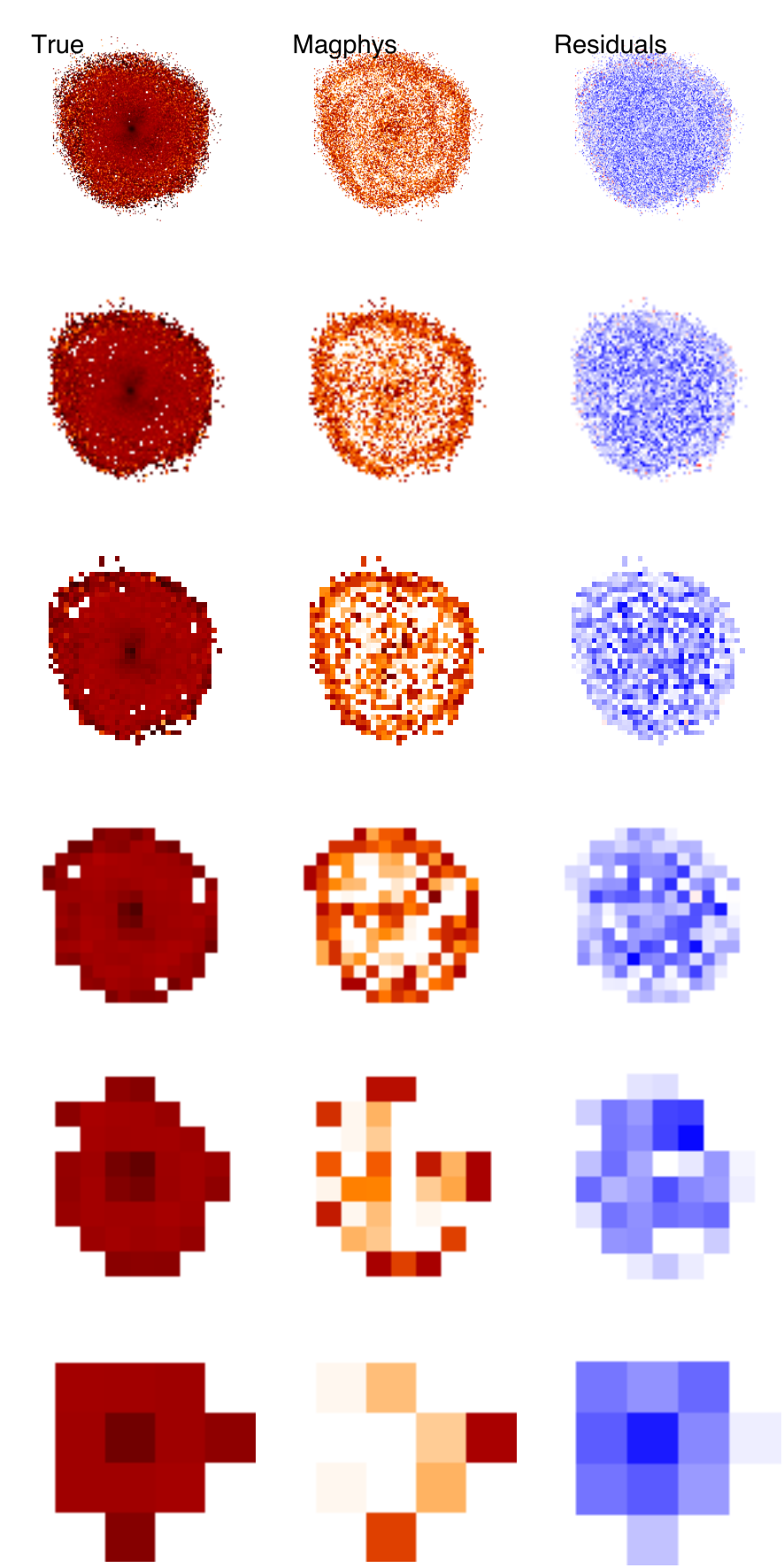} \\
\includegraphics[width=\textwidth,trim=0cm 1.2cm 0cm 2.4cm]{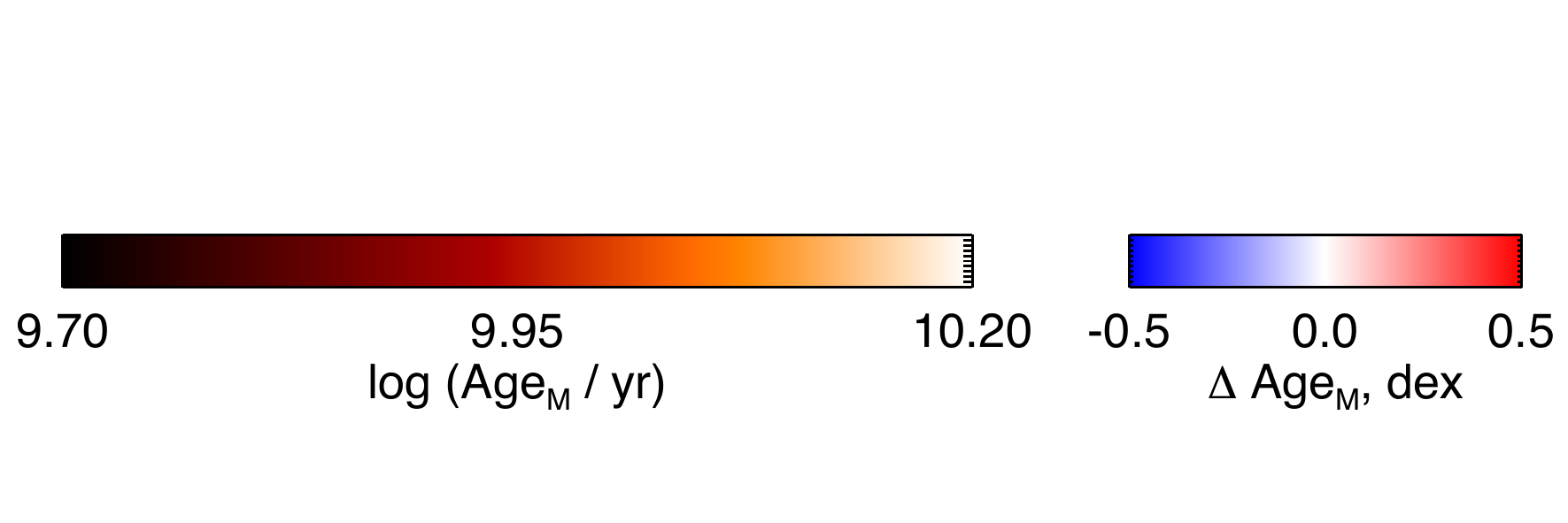}
\end{minipage}
}
\begin{minipage}{0.6\textwidth}
\subfloat[]{\includegraphics[width=\columnwidth,trim=0cm 0cm 0cm 0.85cm,clip]{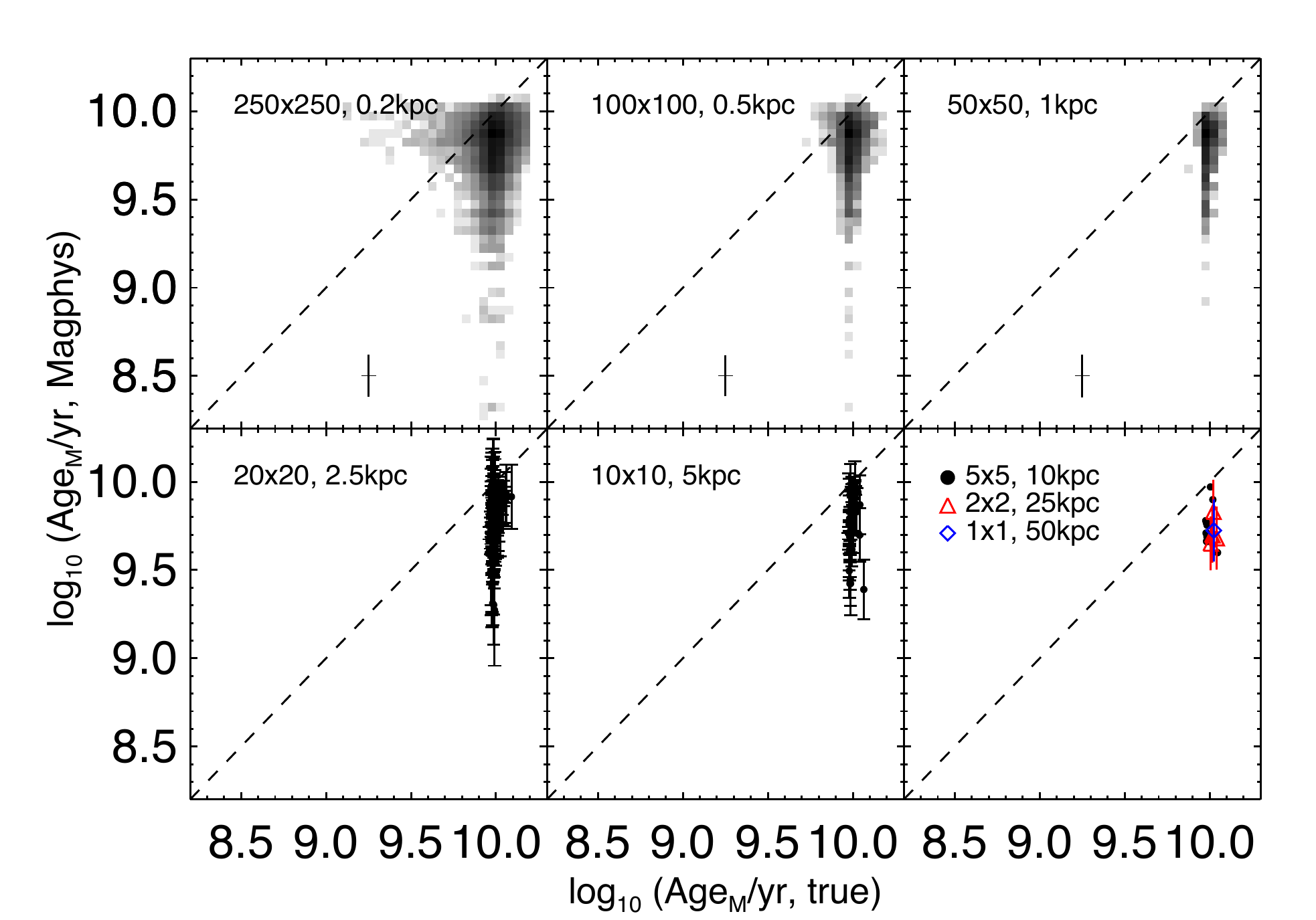}}\\
\vspace{1.0cm}\\
\subfloat[]{\includegraphics[width=\columnwidth,trim=0cm 0cm 0cm 0.8cm,clip]{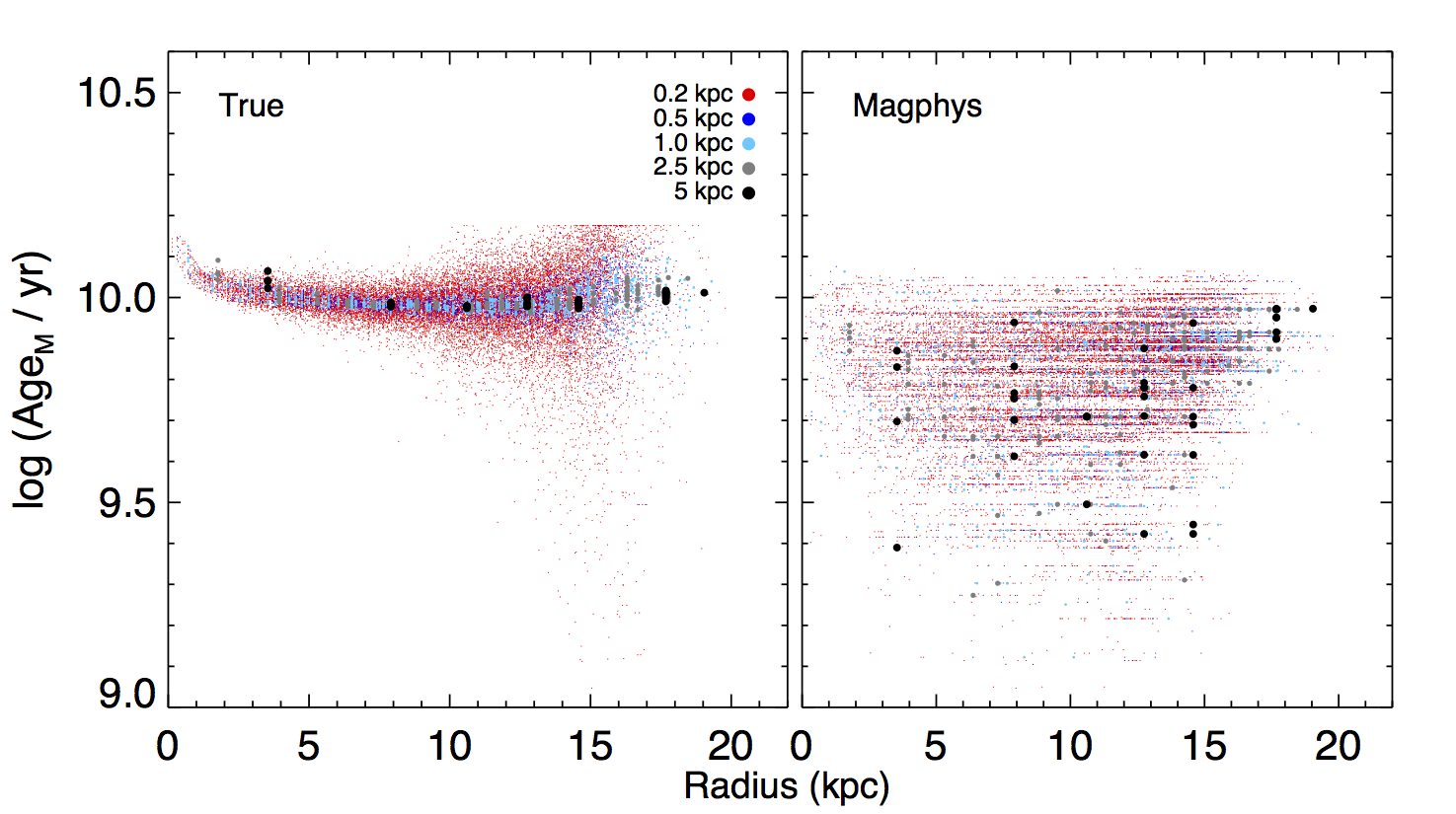}}
\end{minipage}
\caption{Recovery of mass-weighted age for camera 0. For more details, see the caption of Figure \ref{fig:mass_recovery}. \magphys\ tends to underestimate the mass-weighted age, likely because the bulk of the pixels are dominated by $\sim 10$ Gyr-old stellar populations, and integrated photometry does not depend strongly on age for stellar populations with ages $\gg 1$ Gyr.}
\label{fig:age_recovery}
\end{figure*}

Finally, we show the extent of our ability to recover the mass-weighted age using \magphys\ in Figure \ref{fig:age_recovery}.  The scatter in the \magphys\ estimates is large at all spatial scales, though the situation is best for spatial pixels $> 10$\,kpc in size, where the average age underestimate is around 0.3\,dex, slightly larger than the \magphys\ error bars. While the radial dependence of the true stellar population age occupies a tight locus in the left panel of Figure \ref{fig:age_recovery}\,(c), the same can not be said of the \magphys\ estimates, for which the scatter is also around 0.2-0.3\,dex. Perhaps the best way to describe the \magphys\ estimates of mass-weighted age is that they are of limited use since the offset scatter is so large, however it will be of particular interest to see whether this trend persists for a larger sample of simulated galaxies. 

\section{Discussion} \label{S:discussion}

\subsection{Acceptable fits}

In this work we have performed a controlled experiment on the efficacy of \magphys\ for recovering the spatially-resolved properties of galaxies, a task for which it was not designed. \magphys\ is able to recover statistically acceptable fits to $> 99$\,per cent of the pixels enclosed within the $r$-band effective radius of the simulated galaxy, and between 61 and 77 per cent of the pixels within the region emitting 99 per cent of the total $r$-band light (at 200\,pc and 5\,kpc resolution, respectively). With \magphys, we are able to produce a larger fraction of acceptable fits out to larger radii at lower spatial resolution, consistent with larger regions being better approximated by the simple galaxy model used by \magphys. These results are in general agreement with the \magphys\ study of M31 by \citet{viaene14}, however since we are basing our study on a simulation, we are also able to determine whether or not the parameters that \magphys\ derives are reasonable, and we will discuss this in the coming sub-sections.

\subsection{The influence of spatial resolution on integrated parameters}

As well as producing statistically acceptable fits, we are able to recover realistic integrated galaxy properties, that in general show little strong evidence for dependence on spatial resolution once the uncertainties on the individual parameter estimates (e.g. the \magphys\ stellar mass estimates in each pixel) are taken into account. \magphys\ produces the best integrated parameter estimates on the almost edge-on simulation, consistent with the idea that along these sightlines we are observing through a larger volume of the galaxy, for which the assumed energy balance criterion is more reasonable than e.g. for the face-on case (in which the photometry comes from integrating through a geometrically thin disc). 

The lack of resolution dependence in the stellar mass estimation is apparently at odds with \citet{sorba15}, who found that spatially-resolved photometry underestimates the true mass of galaxies, especially those with higher specific star formation rates. While our best-fit stellar mass estimates are smaller than the sum of the resolved values, they are consistent with each other within the uncertainties. In addition, since our simulation-based approach enables us to know the true stellar mass, we are able to see that it is the high-resolution stellar mass estimate which is biased relative to the truth, by $\sim 0.05-0.07$\,dex. Furthermore, according to the \citet{sorba15} relation for the bias in stellar mass estimates, the size of the expected bias at the true specific SFR of the galaxy snapshot used in this work ($\log_{10}$ sSFR $\approx -10.72$\,yr$^{-1}$) is a factor of 0.95 for this comparatively quiescent source. This factor is smaller than the uncertainties associated with our (and indeed, {\it all}) stellar mass estimates; it is possible therefore that the trend they report will be clearer if we were to study more actively star forming galaxy simulations (which are predicted to have larger biases according to the \citealt{sorba15} formula). \citet{sorba15} also suggested a sharp decrease in the pixel-by-pixel stellar mass estimates at a physical scale of 3kpc, for which we see no evidence in this study.  

While the integrated stellar mass estimates show little evidence for resolution bias, there is more evidence in other parameters. With SFR (and sSFR) and dust mass, this is primarily due to the sub-kpc-resolution datacubes, which are of course those for which the \magphys\ model is least appropriate. The lower-resolution datacubes with $\gtsim$\,kpc scale pixels show little or no evidence for resolution-dependent bias. Interestingly, the effective metallicity does not exhibit a resolution bias for any of the three viewing angles to the simulated galaxy, but the recovery is poor, in general. These potential resolution-dependent biases may be of particular concern if they still hold after we have considered a larger suite of simulations that span a wider range of galaxy properties.

\subsection{Radial dependence, and variation across the projected disc}

\magphys\ produces reasonable results when we attempt to recover the radial and projected distributions of stellar mass, dust luminosity, SFR, sSFR, dust mass, and $A_V$, especially when the pixel scale is $\ga 1$ kpc. The stellar, dust luminosity and dust mass estimates are sufficiently accurate that we can clearly discern spiral structure even in the radial distributions of the recovered parameters, including stellar masses as low as 10$^4$\,M$_\odot$ and dust luminosities and masses as low as $10^3 L_\odot/10-100\,M_\odot$. 

Our results for sSFR highlight the difficulties associated with estimating values for individual pixels; the derived sSFRs are highly uncertain, and even with model panchromatic photometry, and in the absence of resolution effects, it is difficult to recover any real radial dependence. Though there is little evidence for overall bias in the sSFR estimates, there is considerable scatter, especially in the higher-resolution simulations for lower surface-brightness pixels. These results should therefore serve as a note of caution for works attempting to detect decreased sSFR near the centres of galaxies, as evidence of an ``inside-out'' quenching mechanism in galaxies based on photometry alone \citep[e.g.][at low- and high-redshift, respectively]{viaene14,jung17}, and similar issues could arise if sSFRs are simply estimated from photometry without fitting SEDs \citep[as in][for example]{tacchella15,tacchella17}. At the same time, it is possible that these uncertainties on sSFR may be reduced by making some assumptions about the prior distribution of e.g. stellar mass, rather than fitting individual pixels independently, and as we have noted, the sSFR estimates are far better when averaging over larger regions. Our \magphys\ results also reveal an anomalous annular region of low-sSFR pixels at the edge of the disc; this region is not present in the real data, and a similar effect is reported in the outermost ring of M31 in the \magphys\ study by \citet{viaene14}.

Among the non-standard outputs that we have extracted from \magphys\ by making simple modifications to the code and using the standard libraries, we found that \magphys\ is effective for making reliable estimates of $A_V$ for pixel sizes $\ga 2.5$\,kpc,. 
Although the radial trend in metallicity is qualitatively recovered, the quantitative success of the recovery is more limited for metallicity than for the other properties considered. The mass-weighted age is not well recovered in detail, likely because the bulk of the pixels are dominated by $\sim 10$ Gyr-old stellar populations, and integrated photometry does not depend strongly on age for stellar populations with ages $\gg 1$ Gyr. That \magphys\ is able to at least qualitatively recover the metallicity gradient in the simulated galaxy is particularly interesting, despite the bias towards lower values and scatter of around $\delta Z \approx 0.2$ in the recovered values. With \citet{amblard17}, for example, finding evidence for radial dependence in the metallicity of NGC\,2768 and IC\,1459, albeit using the {\tt CIGALEMC} code from \citep[][which also has an energy balance criterion at its core]{serra11}, further validation of the ability of energy balance codes to estimate metallicities from photometry is essential. Forthcoming observations with new and forthcoming arcminute-scale integral field spectrographs, as well as comparing integrated photometry with already existing spectroscopic data sets, will be ideal for these purposes.

\subsection{Limitations}

In this work, we have studied `only' a single simulated isolated disc galaxy, observed at a single snapshot in time and assuming a constant signal-to-noise ratio in 21 different bands of photometry, with just three different viewing angles. Moreover, the simulation was designed to be representative (in terms of e.g. gas fraction) of galaxies of similar mass in the local Universe, and its dust obscuration is modest. Thus, our conclusions may not be generalisable to higher-redshift or/and more highly dust-obscured galaxies.

Still, to perform this simple test, we produced a total 75,530 different SED fits for each camera angle (i.e. 226,590 SED fits in total), requiring around $3 \times 10^4$ CPU hours just for the \magphys\ SED fitting. This work therefore represents a very encouraging first look at the issue of how well energy-balance SED fitting can be applied to spatially-resolved photometry.  It will be of particular interest to study how well this technique works for a range of different simulations, spanning a range of different redshifts, in different stages of activity, with widely varying levels of dust content, and with different photometric coverage, alongside comparisons with new and forthcoming spectroscopic data sets to provide important additional cross-examination. We intend to build on our growing body of work in this area to lead the effort required to tackle these important questions.

\section{Conclusions} \label{S:conclusions}

We have performed a numerical experiment, in which we have used a simulation of an isolated disc galaxy to test the ability of the widely-used energy balance spectral energy distribution (SED) fitting code \magphys\ \citep{dacunha08} to estimate physical parameters of spatially-resolved observations of galaxies. We used \magphys\ to fit more than 250,000 individual pixels ranging from 10\,kpc to 0.2\,kpc in size, at three different viewing angles towards the simulated galaxy at a single time snapshot, using 21 bands of multi-wavelength photometry spanning the ultra-violet to the far-infrared. We have used the default version of \magphys, making only trivial modifications similar to those in our previous work \citep[][which focussed on photometric constraints on galaxy star formation histories]{SH15} to allow us to produce estimates of mass-weighted age, metallicity and visual extinction ($A_V$). This task is done in post-processing, and allows us to produce marginalised probability distributions for each parameter, derived using the energy balance criterion. 

\magphys\ is able to produce statistically acceptable fits to $> 99$ per cent of the pixels within the $r$-band effective radius, and between 59 and 77 per cent of the pixels within a 20\,kpc radius (where 99 per cent of the total $r$ band luminosity is emitted).  This is true despite being based on an energy balance criterion, which must break down on at small scales (i.e.~when the dust emission from a given pixel is not predominately powered by stellar emission from within the same pixel) and being supplied with libraries of stellar populations and dust emission that are built with whole galaxies in mind. As well as producing acceptable fits, we use the known true values from the simulations to compare the derived values of stellar mass, dust luminosity, SFR, sSFR, stellar metallicity, $A_V$, dust mass and mass-weighted age. We find that \magphys\ produces reasonable parameter estimates for stellar mass, dust luminosity, SFR, sSFR, $A_V$, and dust mass in these challenging situations for which it was not intended; the recovery of stellar metallicity and mass-weighted age (which are not standard \magphys\ outputs) is less effective. 

While we are unable to reproduce claims in the literature for resolution-dependent bias in stellar mass estimates based on SED fitting, we see tentative evidence for resolution-dependent biases in the SFR, sSFR and dust mass estimates. We intend to investigate this concerning issue more in more detail, using a larger suite of simulations, in a forthcoming work. Despite this, the overall offsets that we see on integrated parameters (produced by summation of the values for individual pixels) are generally consistent with those derived based on the integrated photometry in our previous study \citep{HS15}. 

As well as the integrated parameters, we find that \magphys\ does a generally good job of recovering the radial and projected variation in all of the aforementioned parameters, at least qualitatively, with the possible exception of specific star formation rate, which is difficult to interpret in the highest-resolution simulations, primarily in the inter-arm and outermost regions of the stellar disc simulation (although the uncertainties are large). This may be due to the influence of the energy balance criterion, which does not apply in pixels in which the majority of photons heating the dust originate from nearby regions rather than being emitted by stars within the pixel. These results highlight the need for extreme caution when using photometry to search for evidence of e.g. inside out quenching, indicated by possible radial variation in estimates of sSFR \citep[e.g.][]{viaene14,jung17}.

As mentioned above, we intend to expand on this work in a future publication by analysing a suite of simulated galaxies that span a broader range of galaxy properties. We also intend to further test the efficacy of using \magphys\ to estimate metallicities based on photometry alone, using samples of real galaxies for which panchromatic photometry and spectroscopic metallicity estimates are available.

\acknowledgments
This project made use of the University of Hertfordshire's high-performance computing facility.\footnote{http://stri-cluster.herts.ac.uk/} The authors would like to thank the reviewer, Elisabete da Cunha, as well as S\'ebastien Viaene \&\ Martin Krause for useful comments that improved the quality of this paper. 
DJBS would like the thank the Simons Foundation Center for Computational Astrophysics for their hospitality.
CCH is grateful to the University of Hertfordshire for hospitality. The Flatiron Institute is supported by the Simons Foundation.
This research has made use of NASA's Astrophysics Data System Bibliographic Services.
\\

\bibliography{sfh,method}


\label{lastpage}

\begin{appendix}

\section{Additional figures}

\begin{figure*}
	\centering 	
		\subfloat[Mass]{
		\begin{minipage}{0.25\textwidth}
		\includegraphics[width=\textwidth,trim=0cm 0cm 0cm 0cm,clip]{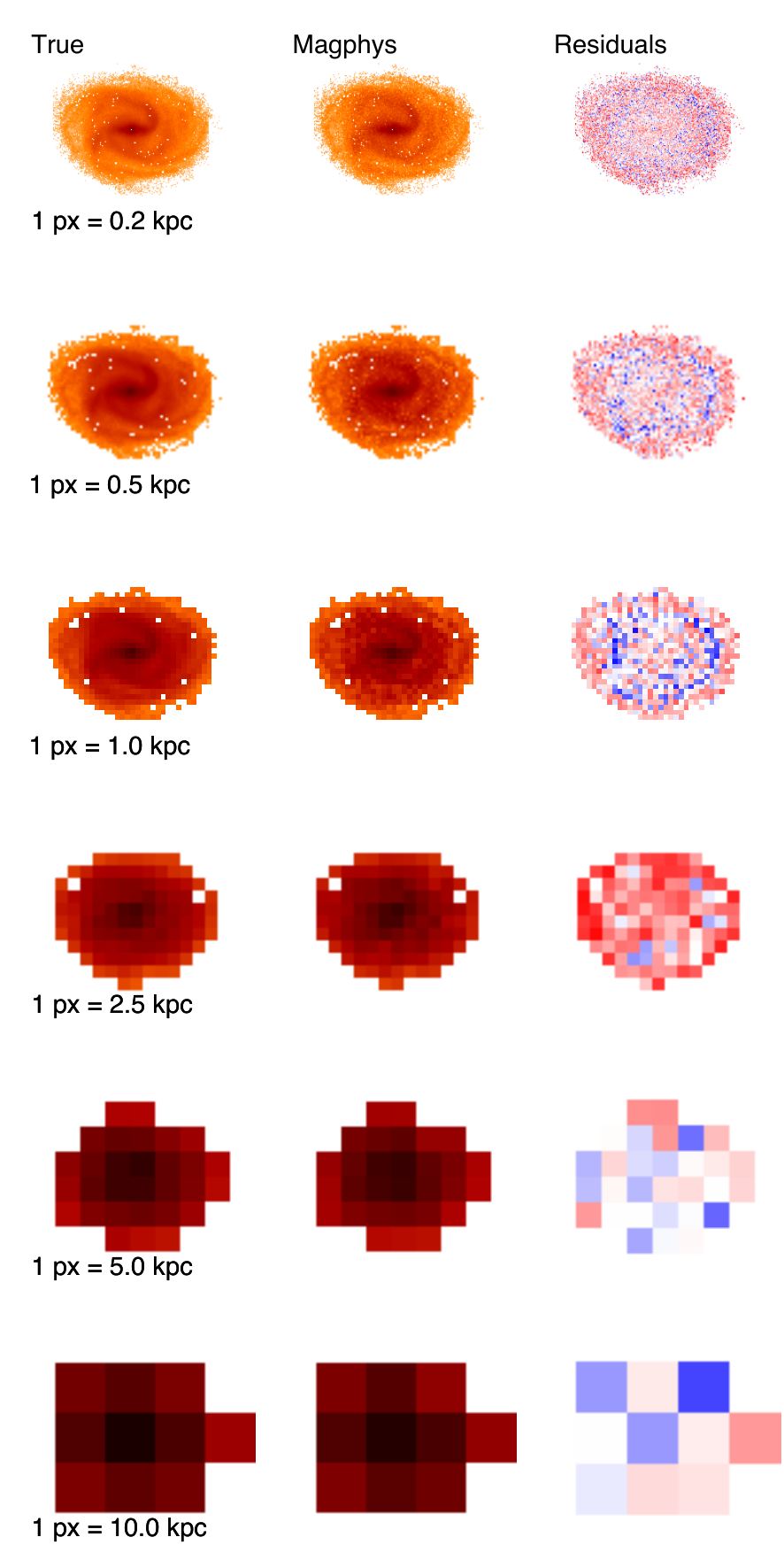} \\
		\includegraphics[width=\textwidth,trim=0.0cm 1.2cm 0.0cm 2.4cm,clip=true]{figures/res_cbar_massmaps.pdf}
		\end{minipage}
		}
		\vline
		\subfloat[Dust luminosity]{
		\begin{minipage}{0.25\textwidth}
		\includegraphics[width=\textwidth,trim=0cm 0cm 0cm 0cm,clip]{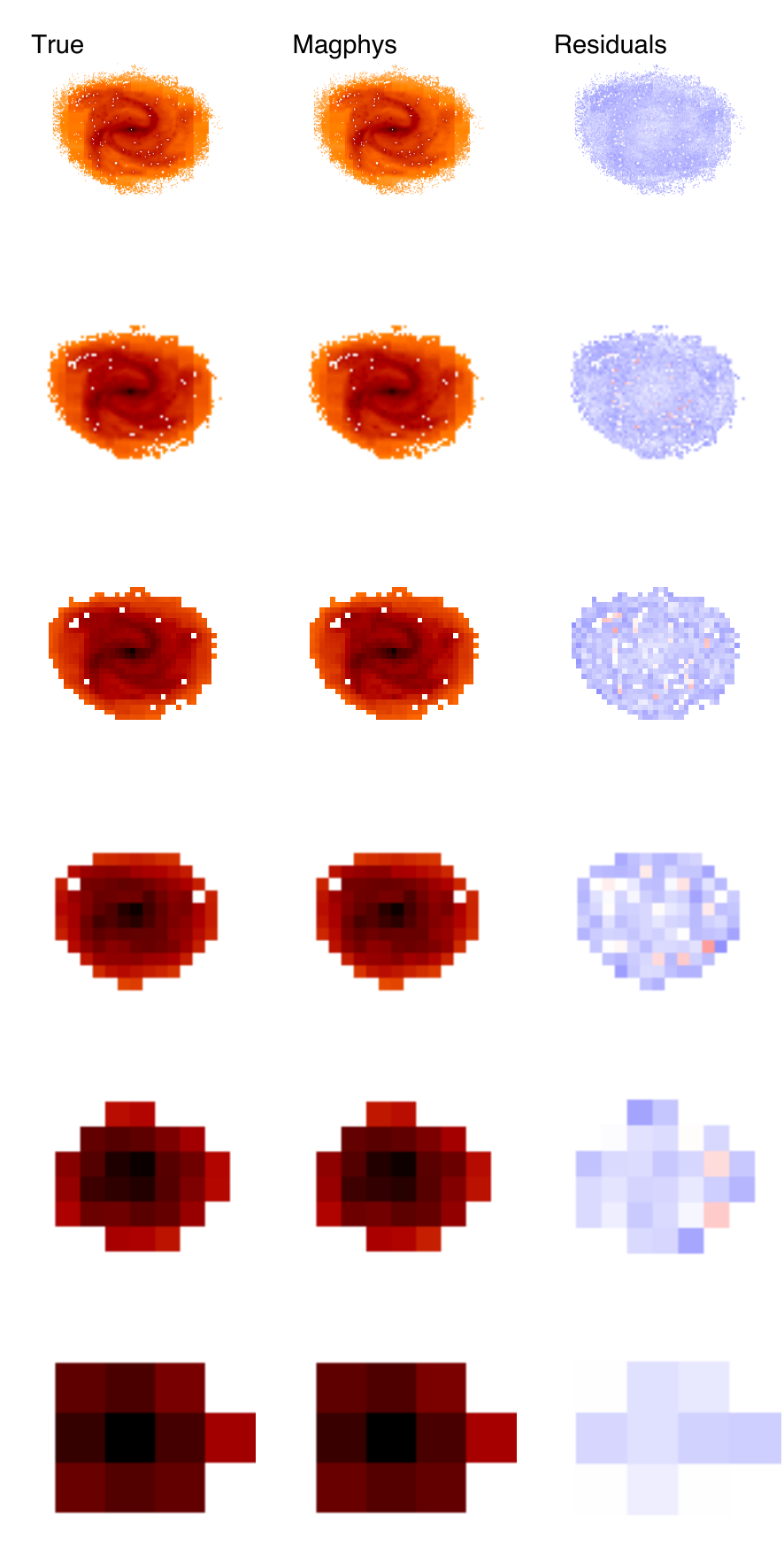}\\
		\includegraphics[width=\textwidth,trim=0cm 1.2cm 0cm 2.4cm]{figures/res_cbar_ldustmaps.pdf}
		\end{minipage}
		}
		\vline
		\subfloat[SFR]{
		\begin{minipage}{0.25\textwidth}
		\includegraphics[width=\textwidth,trim=0cm 0cm 0cm 0cm,clip]{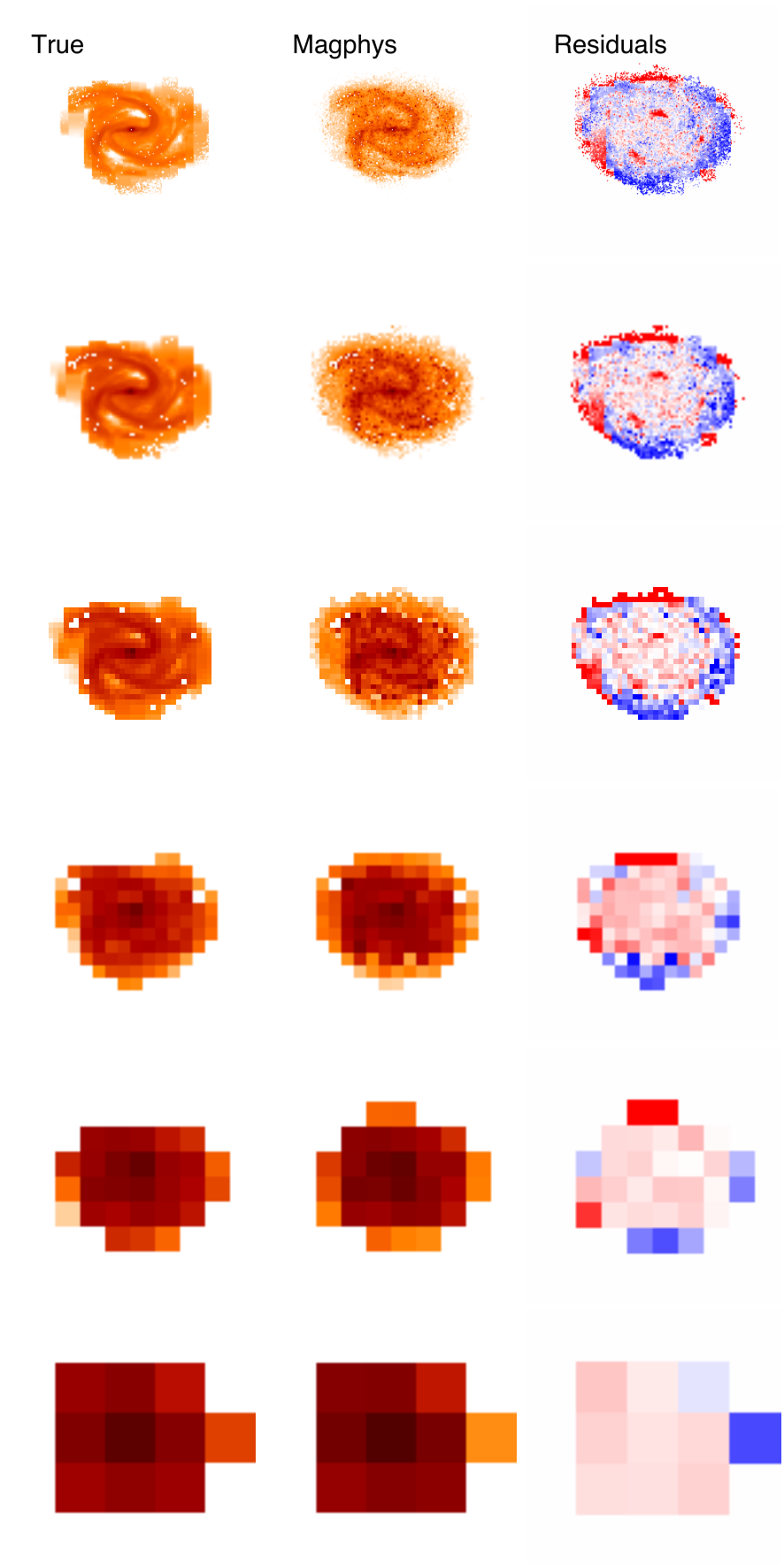} \\
		\includegraphics[width=\textwidth,trim=0cm 1.2cm 0cm 2.4cm]{figures/res_cbar_sfrmaps.pdf}
		\end{minipage}
		}
		\vline
		\subfloat[sSFR]{
		\begin{minipage}{0.25\textwidth}	
		\includegraphics[width=\textwidth,trim=0cm 0cm 0cm 0cm,clip]{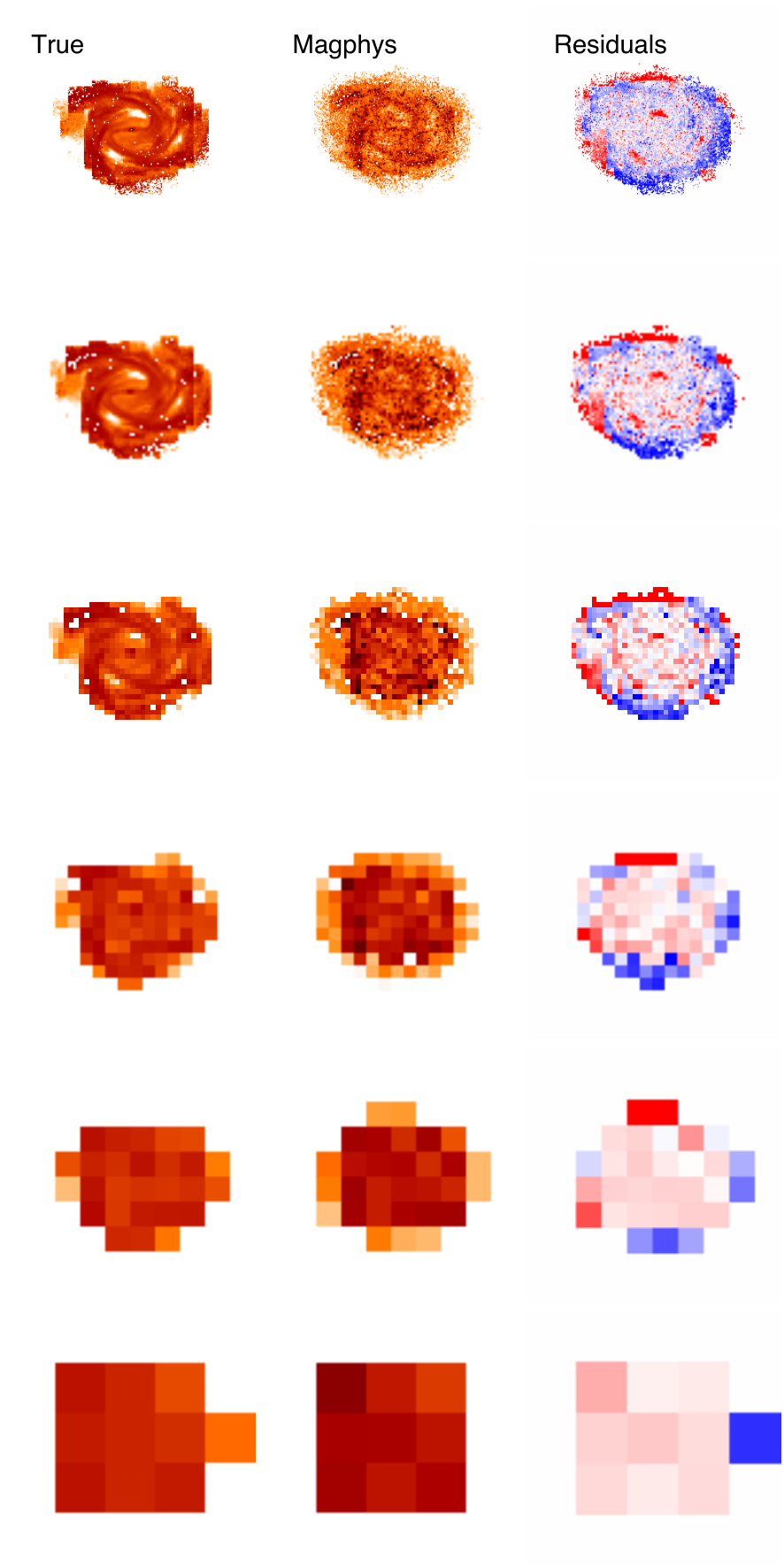} \\
		\includegraphics[width=\textwidth,trim=0cm 1.2cm 0cm 2.4cm]{figures/res_cbar_ssfrmaps.pdf}
		\end{minipage}
		} \\
		\hrule
		\subfloat[Dust mass]{
		\begin{minipage}{0.25\textwidth}
		\includegraphics[width=\textwidth,trim=0cm 0cm 0cm 0cm,clip]{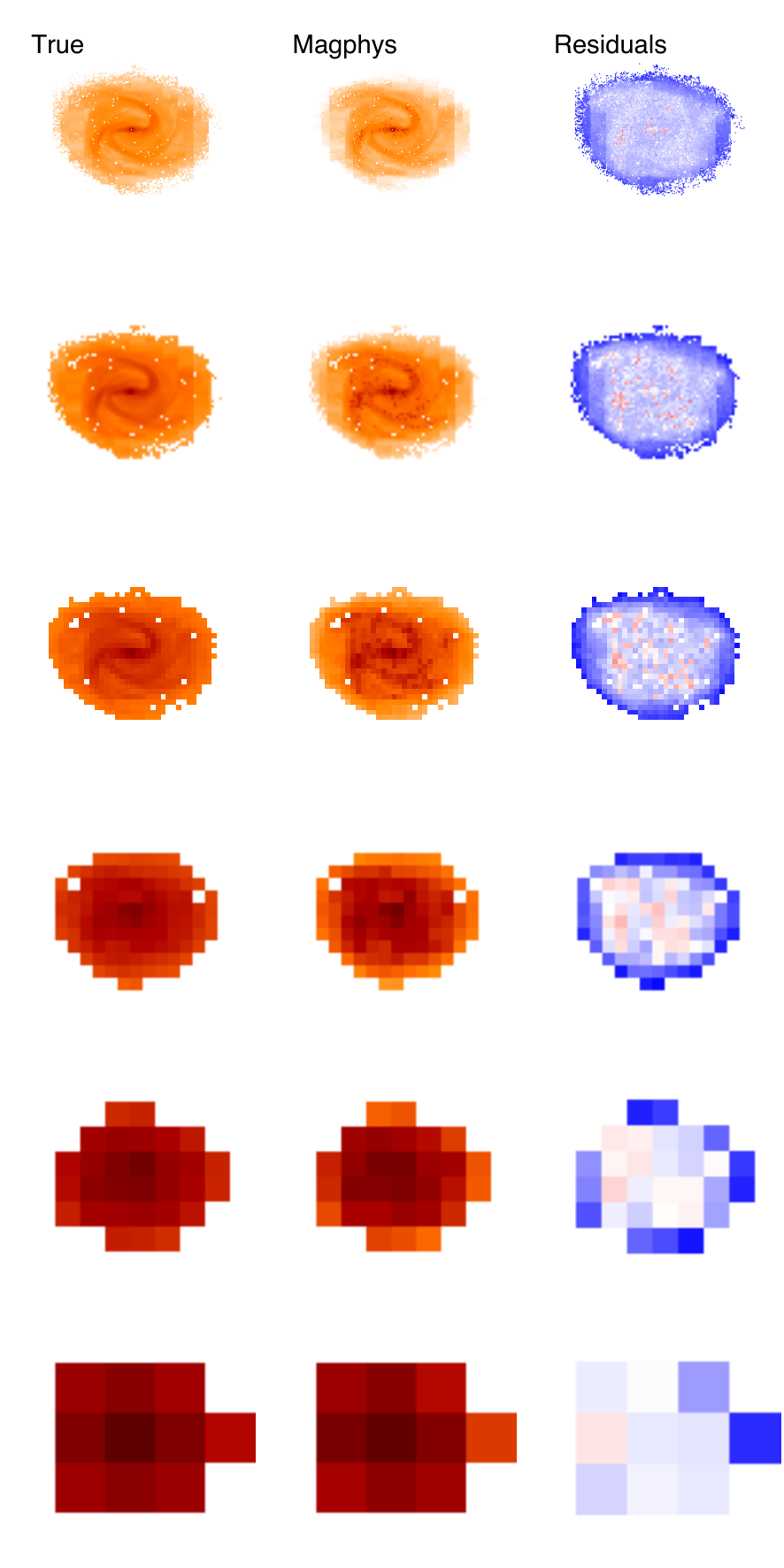}\\
		\includegraphics[width=\textwidth,trim=0cm 1.2cm 0cm 2.4cm]{figures/res_cbar_mdustmaps.pdf}
		\end{minipage}
		}
		\vline
		\subfloat[Metallicity]{
		\begin{minipage}{0.25\textwidth}
		\includegraphics[width=\textwidth,trim=0cm 0cm 0cm 0cm,clip]{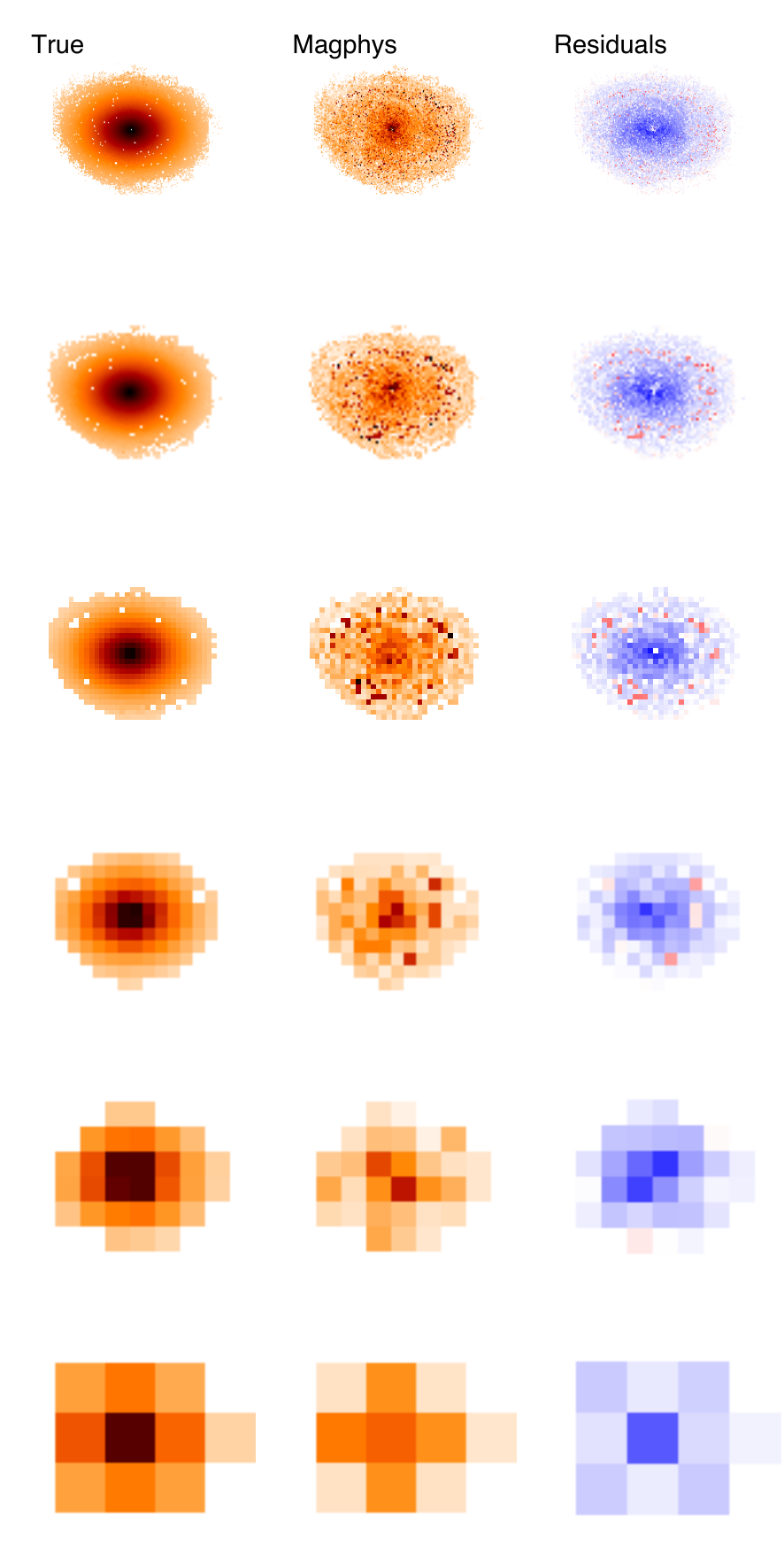}\\
		\includegraphics[width=\textwidth,trim=0cm 1.2cm 0cm 2.4cm]{figures/res_cbar_zmaps.pdf}
		\end{minipage}		
		}
		\vline
		\subfloat[$A_V$]{
		\begin{minipage}{0.25\textwidth}
		\includegraphics[width=\textwidth,trim=0cm 0cm 0cm 0cm,clip]{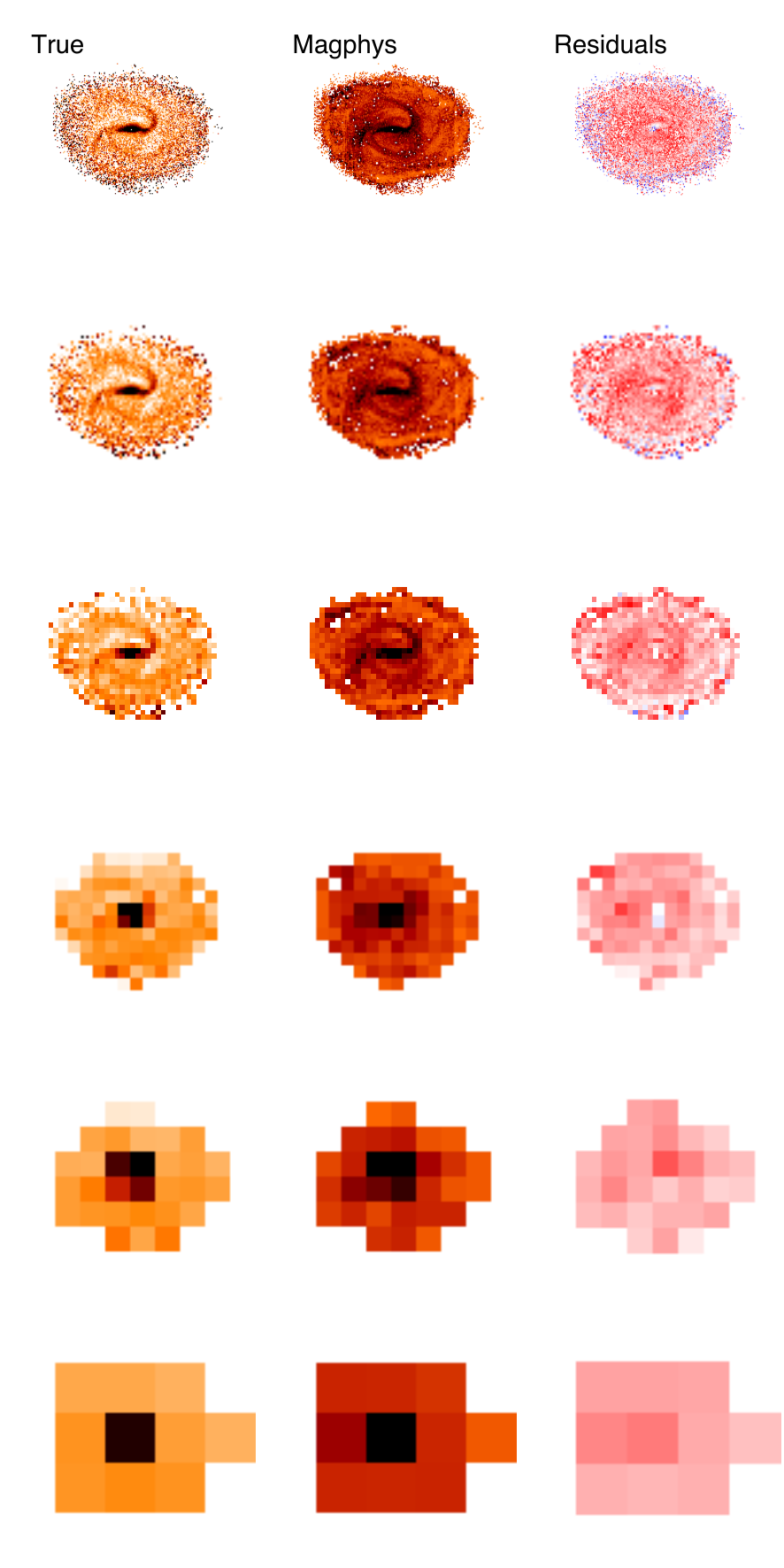}\\
		\includegraphics[width=\textwidth,trim=0cm 1.2cm 0cm 2.4cm]{figures/res_cbar_avmaps.pdf}
		\end{minipage}
		}
		\vline
		\subfloat[Mass-weighted Age]{
		\begin{minipage}{0.25\textwidth}
		\includegraphics[width=\textwidth,trim=0cm 0cm 0cm 0cm,clip]{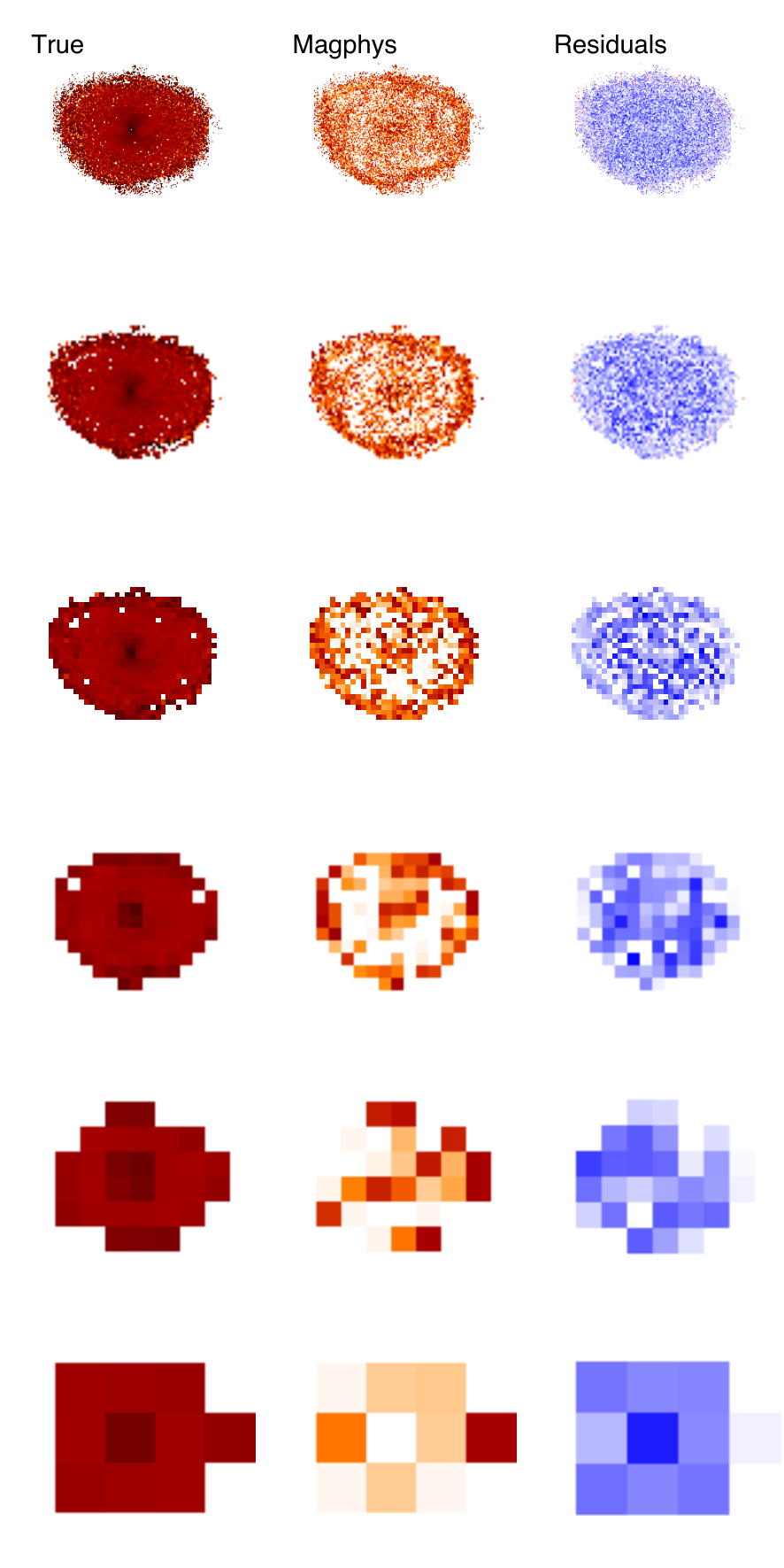}\\
		\includegraphics[width=\textwidth,trim=0cm 1.2cm 0cm 2.4cm]{figures/res_cbar_agemaps.pdf}		
		\end{minipage}
		}
	\caption{Each sub-panel shows comparisons between the true (left) and \magphys -derived values of (a) stellar mass, (b) dust luminosity, (c) SFR, (d) specific-SFR, (e) dust mass, (f) Metallicity, (g) $A_V$, and (h) mass-weighted age for our simulated galaxy snapshot viewed from camera 3, at a range of spatial resolutions from 0.2\,kpc pixels (top) to 10\,kpc pixels (bottom).}
	\label{fig:recovery_images_c3-part1}
\end{figure*}

Here we include figures showing comparisons of projected maps of (a) stellar mass, (b) dust luminosity, (c) SFR, (d) sSFR, (e) dust-mass, (f) metallicity, (g) A$_{V}$, and (h) mass-weighted age for the different viewing angles, with camera 3 shown in Figure \ref{fig:recovery_images_c3-part1} and camera 5 in Figure \ref{fig:recovery_images_c5-part1}.

\setcounter{figure}{1}
\begin{figure*}
	\centering 	
		\subfloat[Mass]{
		\begin{minipage}{0.25\textwidth}
		\includegraphics[width=\textwidth,trim=0cm 0cm 0cm 0cm,clip]{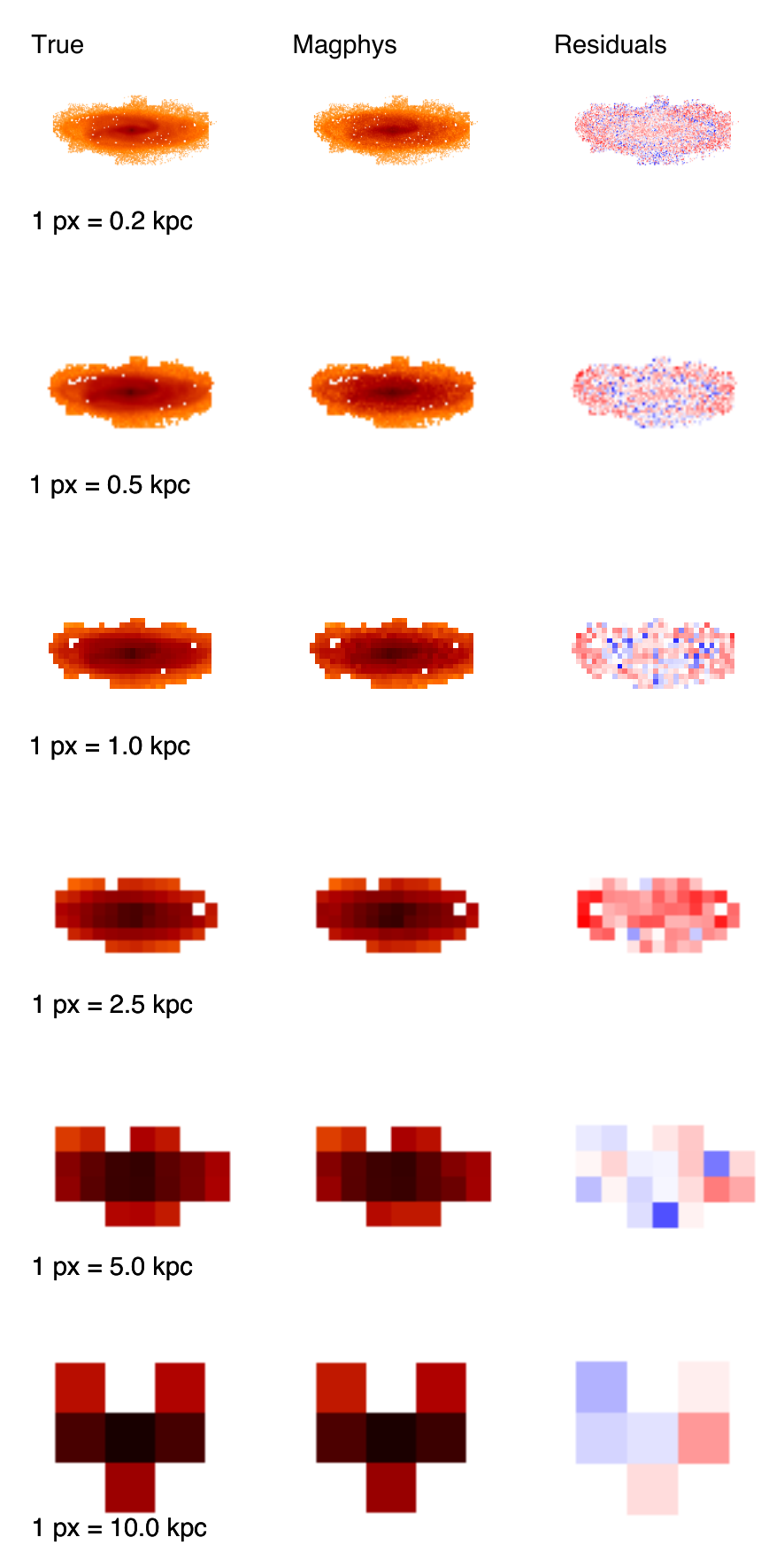} \\
		\includegraphics[width=\textwidth,trim=0.0cm 1.2cm 0.0cm 2.4cm,clip=true]{figures/res_cbar_massmaps.pdf}
		\end{minipage}
		}
		\vline
		\subfloat[Dust luminosity]{
		\begin{minipage}{0.25\textwidth}
		\includegraphics[width=\textwidth,trim=0cm 0cm 0cm 0cm,clip]{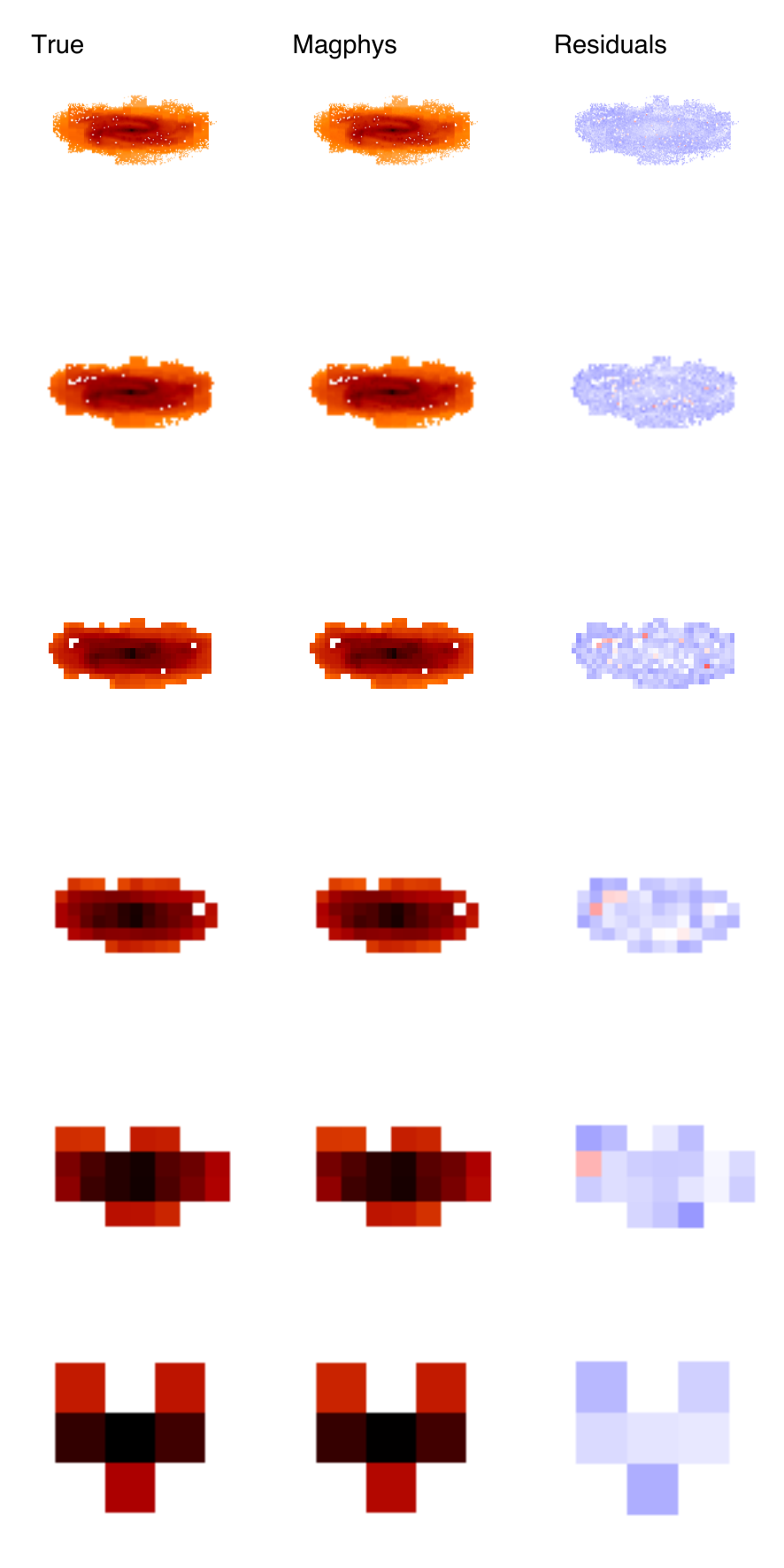}\\
		\includegraphics[width=\textwidth,trim=0cm 1.2cm 0cm 2.4cm]{figures/res_cbar_ldustmaps.pdf}
		\end{minipage}
		}
		\vline
		\subfloat[SFR]{
		\begin{minipage}{0.25\textwidth}
		\includegraphics[width=\textwidth,trim=0cm 0cm 0cm 0cm,clip]{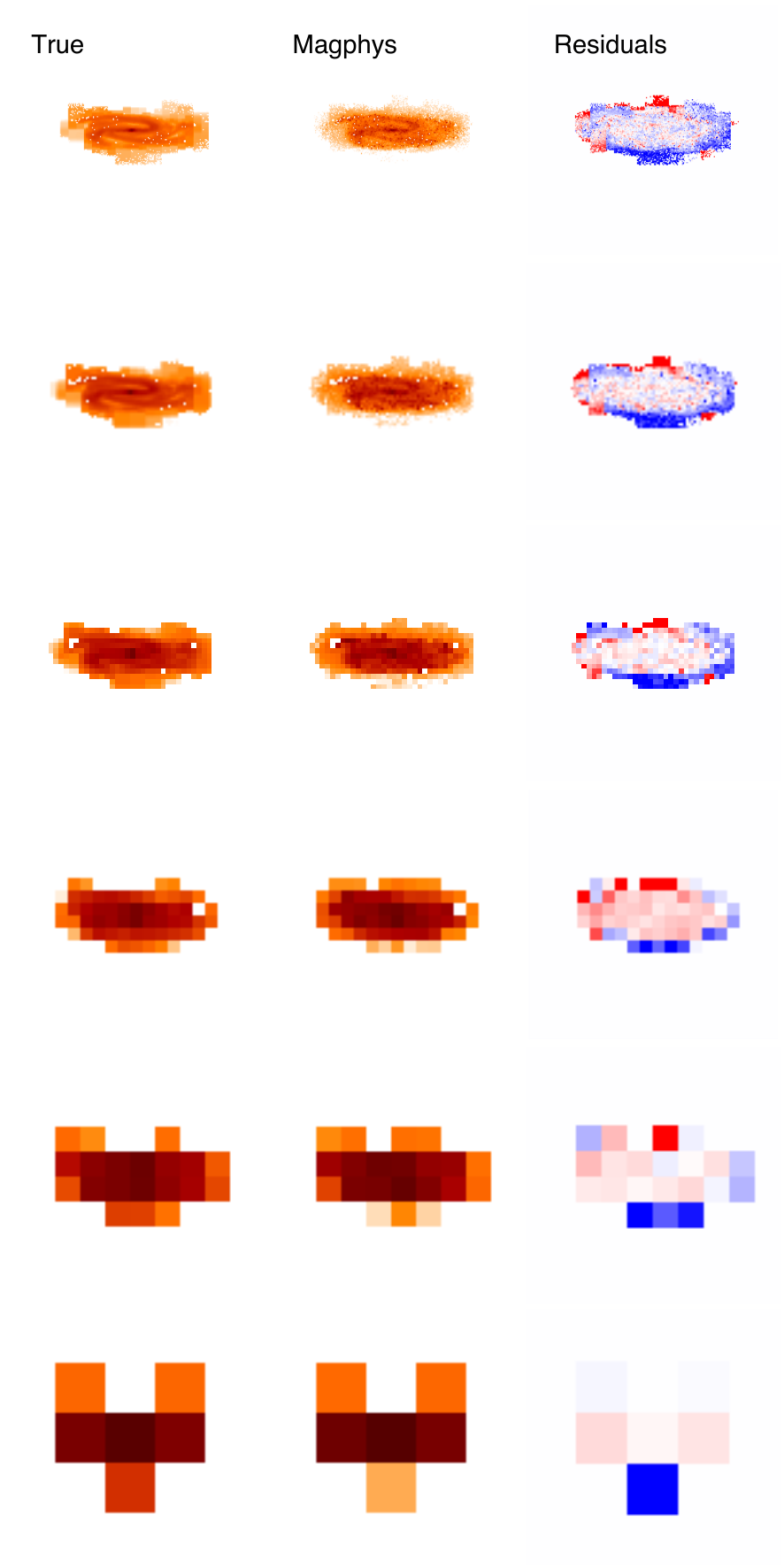} \\
		\includegraphics[width=\textwidth,trim=0cm 1.2cm 0cm 2.4cm]{figures/res_cbar_sfrmaps.pdf}
		\end{minipage}
		}
		\vline
		\subfloat[sSFR]{
		\begin{minipage}{0.25\textwidth}	
		\includegraphics[width=\textwidth,trim=0cm 0cm 0cm 0cm,clip]{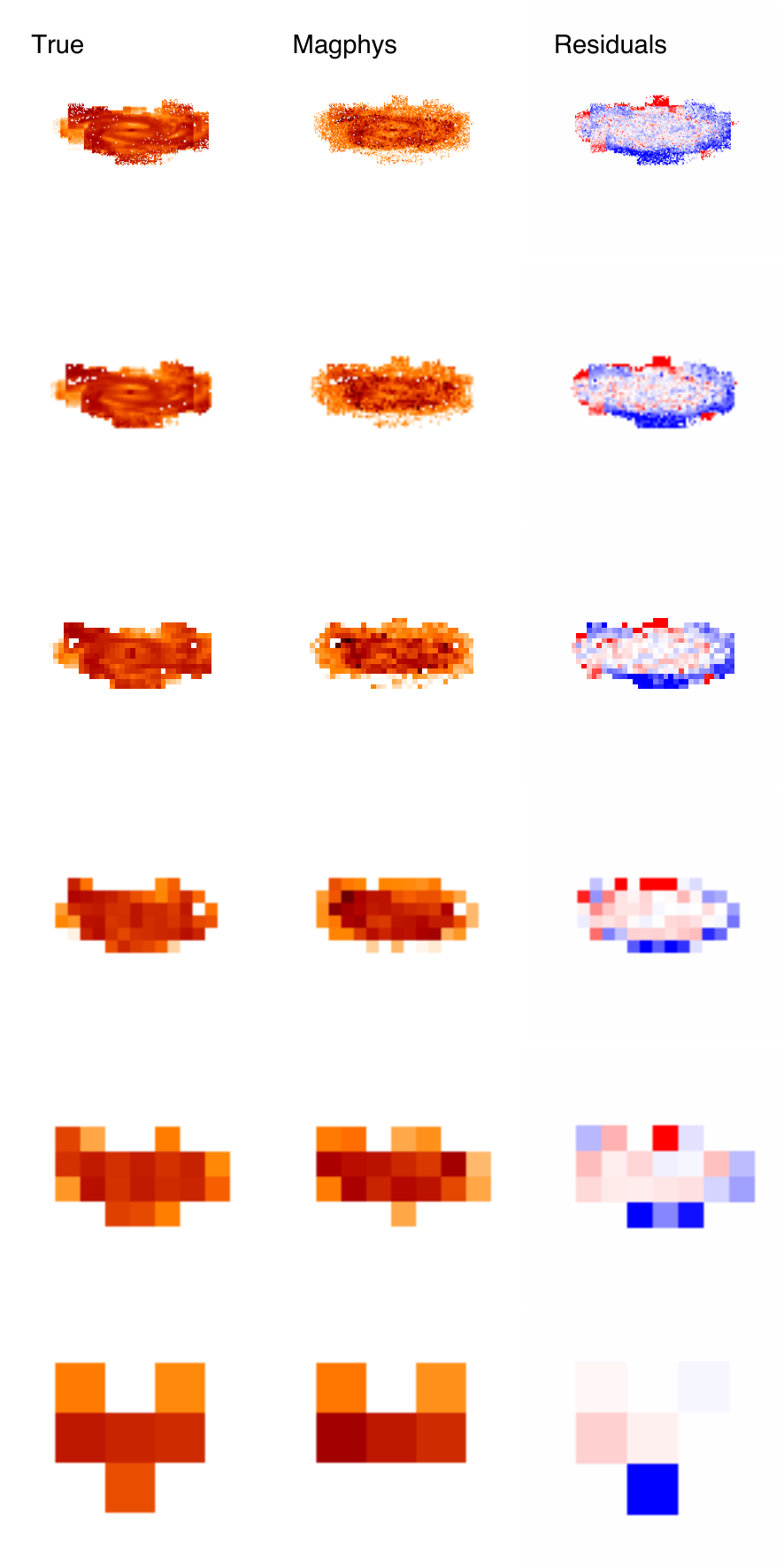} \\
		\includegraphics[width=\textwidth,trim=0cm 1.2cm 0cm 2.4cm]{figures/res_cbar_ssfrmaps.pdf}
		\end{minipage}
		} \\
		\hrule
		\subfloat[Dust mass]{
		\begin{minipage}{0.25\textwidth}
		\includegraphics[width=\textwidth,trim=0cm 0cm 0cm 0cm,clip]{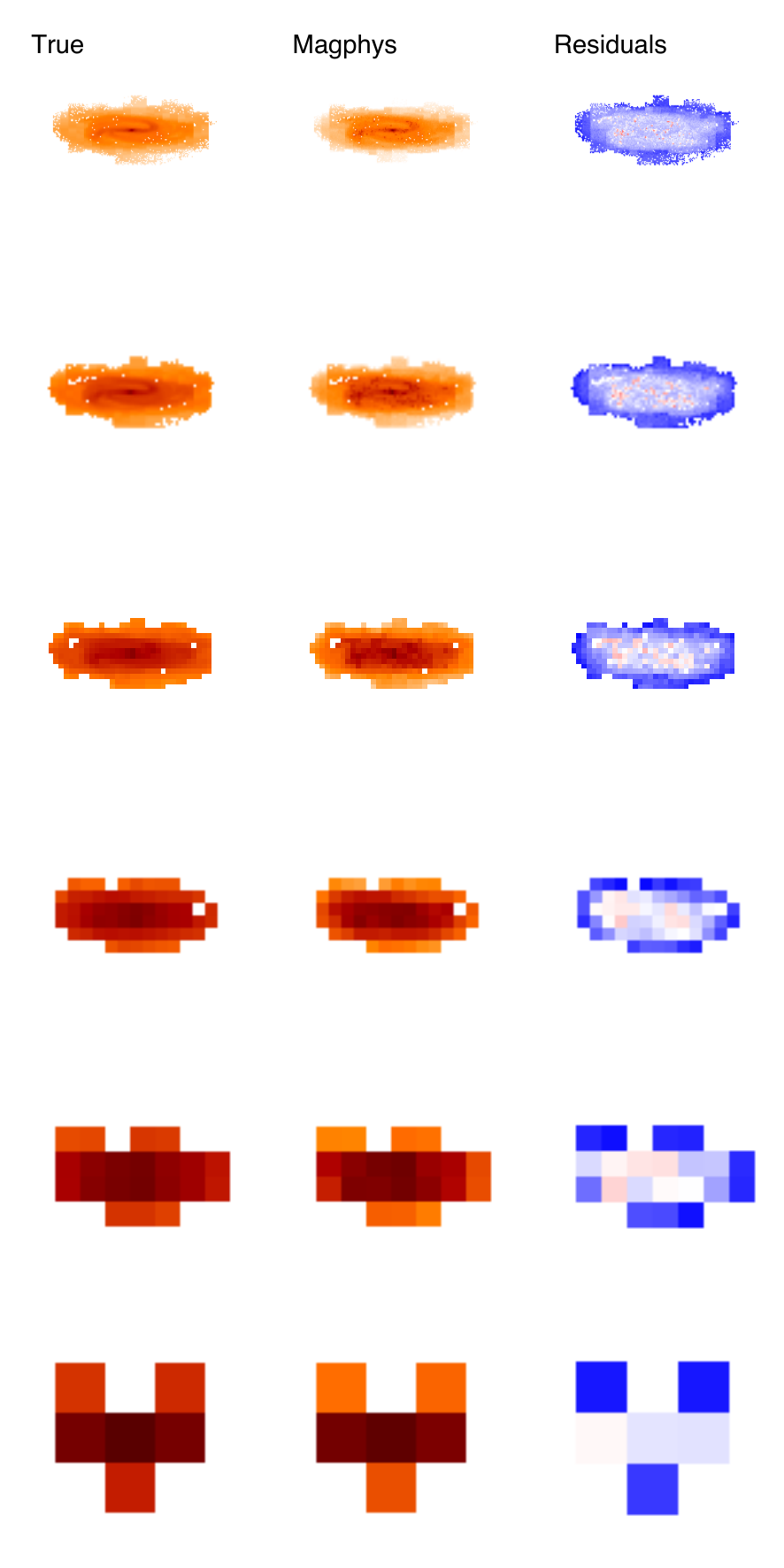}\\
		\includegraphics[width=\textwidth,trim=0cm 1.2cm 0cm 2.4cm]{figures/res_cbar_mdustmaps.pdf}
		\end{minipage}
		}
		\vline
		\subfloat[Metallicity]{
		\begin{minipage}{0.25\textwidth}
		\includegraphics[width=\textwidth,trim=0cm 0cm 0cm 0cm,clip]{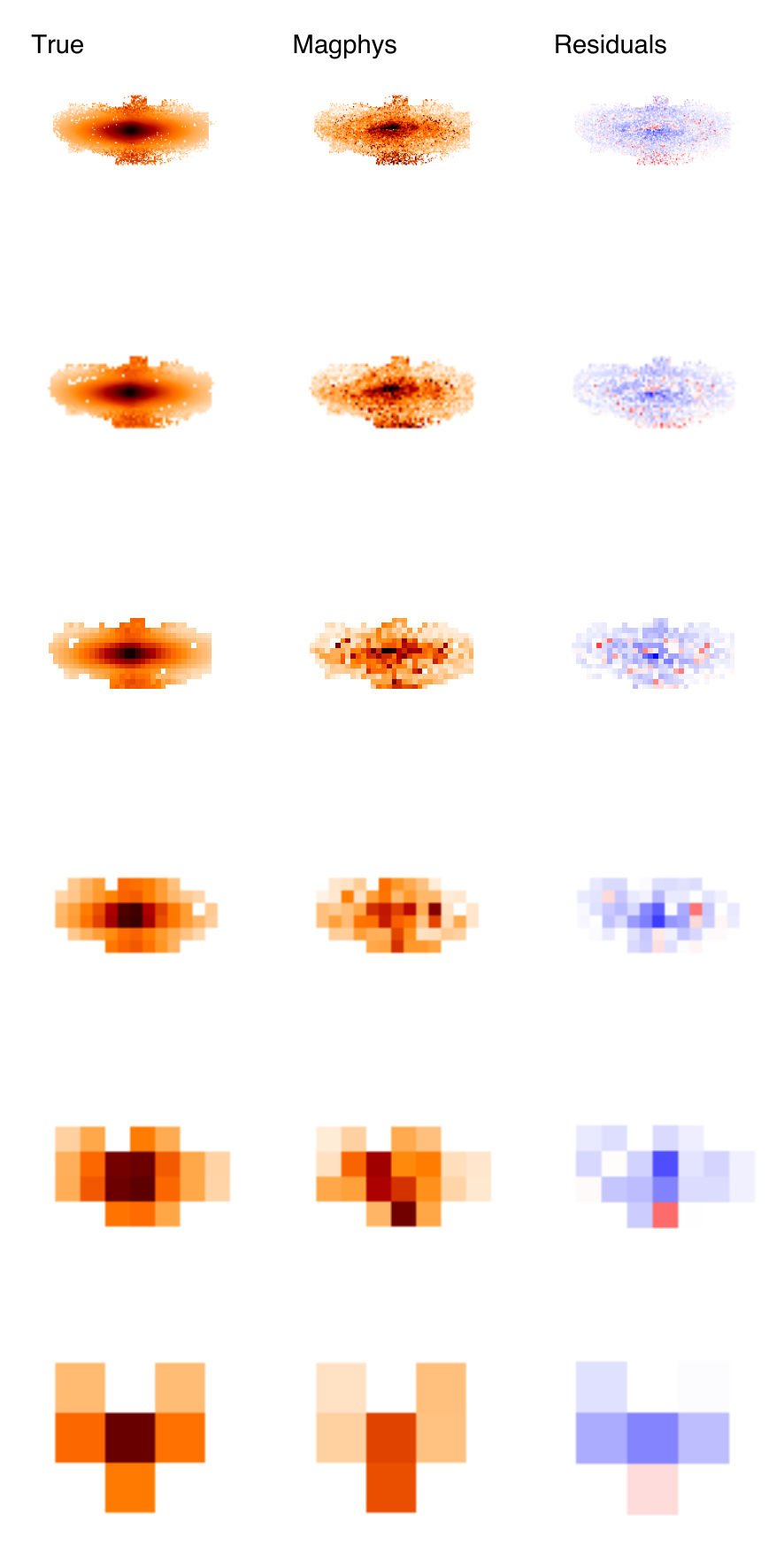}\\
		\includegraphics[width=\textwidth,trim=0cm 1.2cm 0cm 2.4cm]{figures/res_cbar_zmaps.pdf}
		\end{minipage}		
		}
		\vline
		\subfloat[$A_V$]{
		\begin{minipage}{0.25\textwidth}
		\includegraphics[width=\textwidth,trim=0cm 0cm 0cm 0cm,clip]{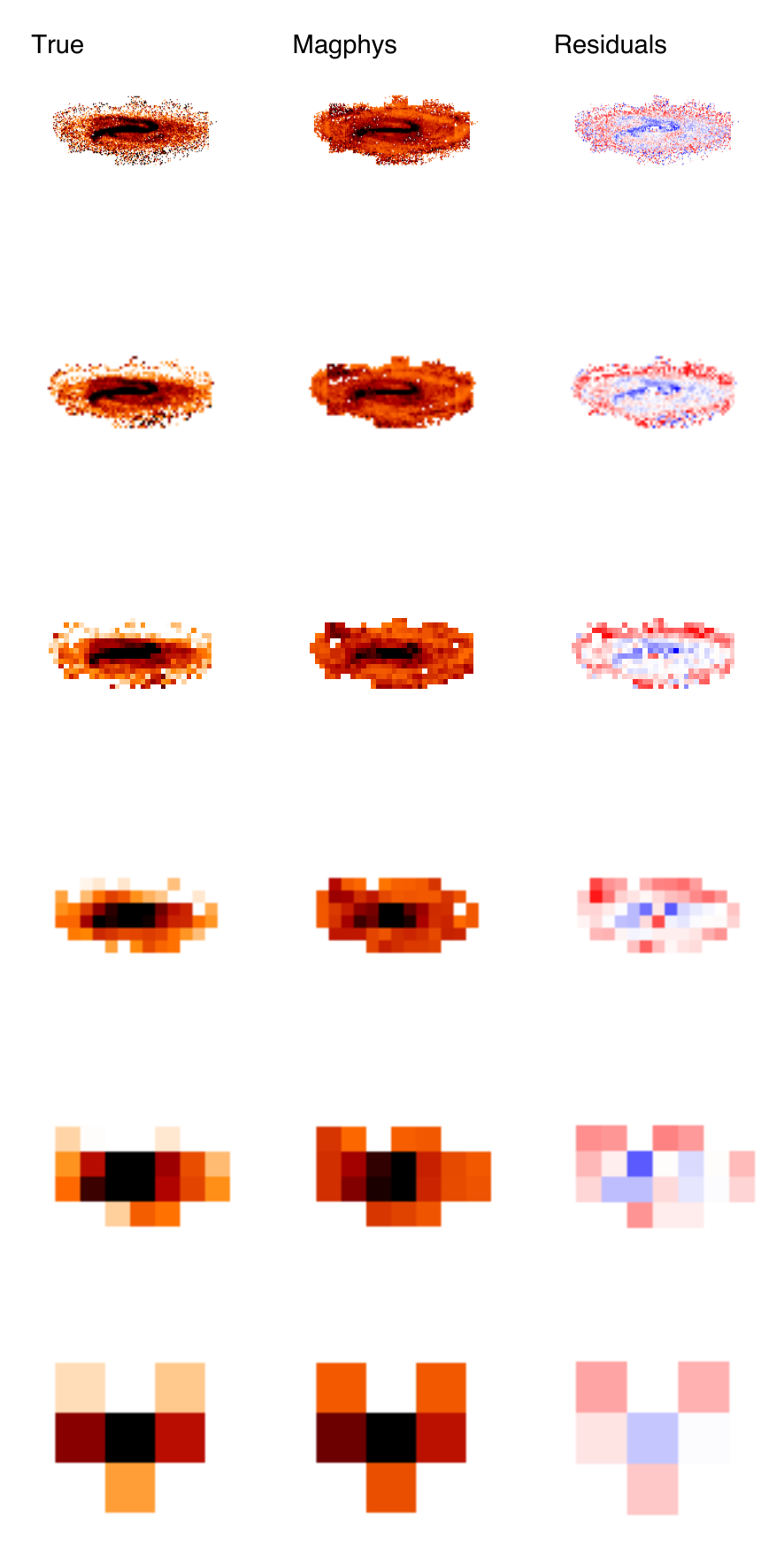}\\
		\includegraphics[width=\textwidth,trim=0cm 1.2cm 0cm 2.4cm]{figures/res_cbar_avmaps.pdf}
		\end{minipage}
		}
		\vline
		\subfloat[Mass-weighted Age]{
		\begin{minipage}{0.25\textwidth}
		\includegraphics[width=\textwidth,trim=0cm 0cm 0cm 0cm,clip]{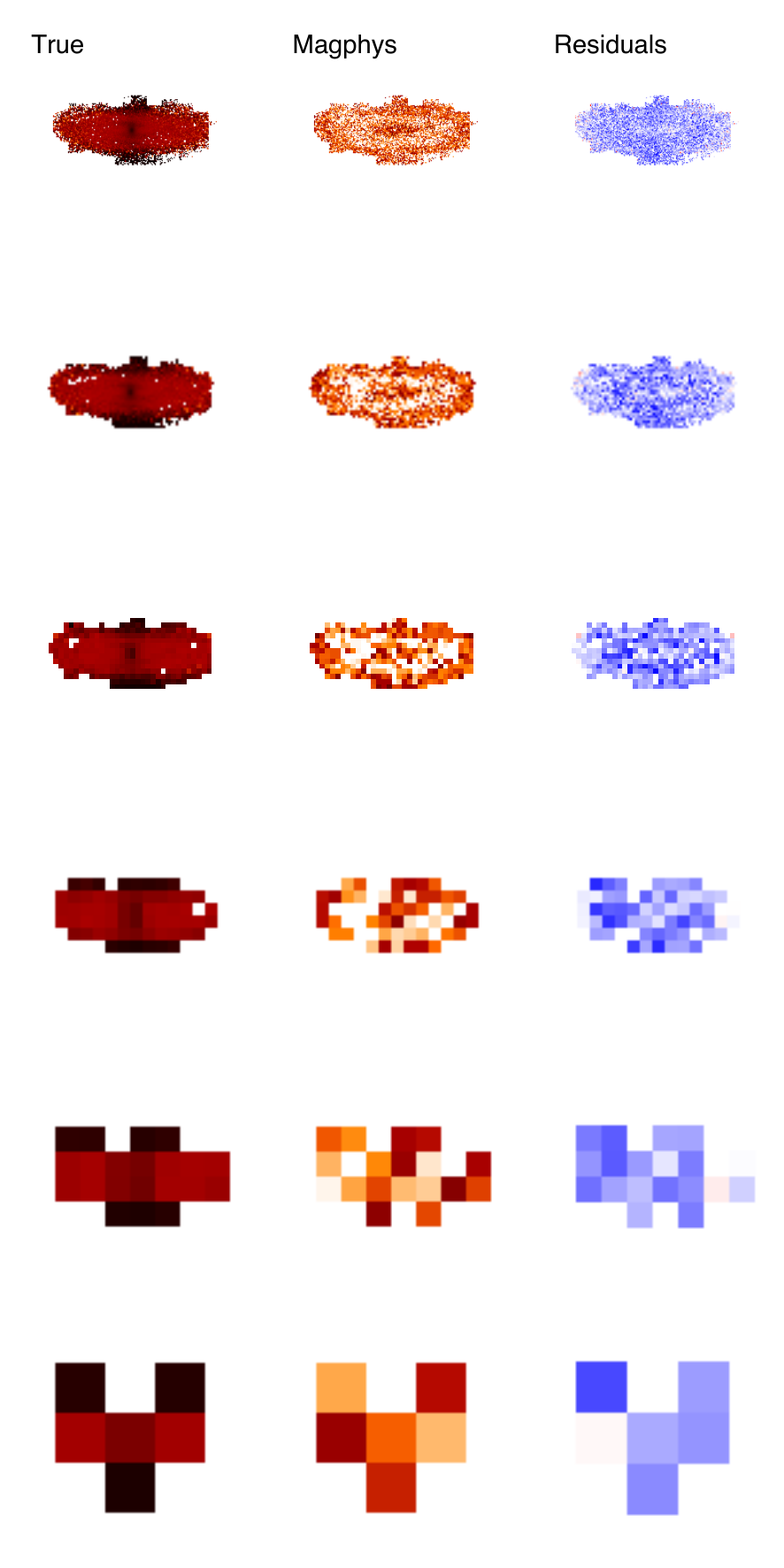}\\
		\includegraphics[width=\textwidth,trim=0cm 1.2cm 0cm 2.4cm]{figures/res_cbar_agemaps.pdf}		
		\end{minipage}
		}
	\caption{Similar to Figure \ref{fig:recovery_images_c5-part1}, but for camera 5 (the nearly edge-on view).}
	\label{fig:recovery_images_c5-part1}
\end{figure*}


\end{appendix}

\end{document}